\documentclass[12pt]{article}
\usepackage[dvips]{graphicx}
\usepackage{epsfig}
\psfigdriver{dvips}
\begin{document}
\setcounter{page}{1}
\renewcommand{\thefootnote}{\fnsymbol{footnote}}
\renewcommand{\theenumi}{(\roman{enumi})}
\renewcommand{\thefigure}{\arabic{figure}}
\renewcommand{\theequation}{\thesection.\arabic{equation}}
\newcommand{\clean}{\setcounter{equation}{0}}
\def\beq{\begin{equation}}
\def\eeq{\end{equation}}
\def\bea{\begin{eqnarray}}
\def\eea{\end{eqnarray}}
\def\bseq{\begin{subequations}}
\def\eseq{\end{subequations}}
\def\nn{\nonumber}
\def\dfrac{\displaystyle\frac}
\def\tfrac{\textstyle\frac}
\def\numt#1#2{#1 \times 10^{#2}}
\def\etal{{\it et al.}}
\def\etc{{\it etc.~}}
\def\ie{{\it i.e.,~}}
\def\eg{{\it e.g.~}}
\def\bs{\bigskip}
\def\ms{\medskip}
\def\ss{\smallskip}
\def\st{{\it s.t.,~}}
\def\id{{\mit I}}

\def\btiny{\begin{tiny}}
\def\etiny{\end{tiny}}
\def\bsc{\begin{scriptsize}}
\def\esc{\end{scriptsize}}
\def\bfoot{\begin{footnotesize}}
\def\efoot{\end{footnotesize}}
\def\bsm{\begin{small}}
\def\esm{\end{small}}
\def\bno{\begin{normalsize}}
\def\eno{\end{normalsize}}
\def\bla{\begin{large}}
\def\ela{\end{large}}
\def\bLa{\begin{Large}}
\def\eLa{\end{Large}}
\def\bLA{\begin{LARGE}}
\def\eLA{\end{LARGE}}
\def\bhu{\begin{huge}}
\def\ehu{\end{huge}}
\def\bHu{\begin{Huge}}
\def\eHu{\end{Huge}}

\def\bCe{\begin{center}}
\def\eCe{\end{center}}
\def\bFR{\begin{flushright}}
\def\eFR{\end{flushright}}
\def\bFL{\begin{flushleft}}
\def\eFL{\end{flushleft}}

\def\UL{\underline}

\def\PR#1#2#3{Phys. Rev. {\bf #1}, #2 (#3)}
\def\PRL#1#2#3{Phys. Rev. Lett. {\bf #1}, #2 (#3)}
\def\PL#1#2#3{Phys. Lett. {\bf #1}, #2 (#3)}
\def\NL#1#2#3{Nucl. Phys. {\bf #1}, #2 (#3)}
\def\NP#1#2#3{Nucl. Phys. {\bf #1}, #2 (#3)}
\def\PREP#1#2#3{Phys. Report {\bf #1}, #2 (#3)}
\def\Mod#1#2#3{Mod. Phys. Lett. {\bf #1}, #2 (#3)}
\def\PTP#1#2#3{Prog. Theor. Phys. {\bf #1}, #2 (#3)}
\def\EPJ#1#2#3{Eur. Phys. J. {\bf #1}, #2 (#3)}
\def\MPLA#1#2#3{Mod. Phys. Lett. {\bf A#1} #2 (#3)}
\def\PRD#1#2#3{Phys. Rev. {\bf D#1} #2 (#3)}
\def\NPB#1#2#3{Nucl. Phys. {\bf B#1} #2 (#3)}
\def\ZPC#1#2#3{Z. Phys. {\bf C#1} #2 (#3)}
\def\EPJC#1#2#3{Eur. Phys. J. {\bf C#1} #2 (#3)}
\def\PLB#1#2#3{Phys. Lett. {\bf B#1} #2 (#3)}

\def\eqref#1{eq.(\ref{eqn:#1})}
\def\Eqref#1{Equation~(\ref{eqn:#1})}
\def\eqsref#1{eqs.(\ref{eqn:#1})}
\def\Eqsref#1{Equations~(\ref{eqn:#1})}
\def\eqvref#1{(\ref{eqn:#1})}
\def\Eqvref#1{(\ref{eqn:#1})}
\def\eqlab#1{\label{eqn:#1}}

\def\tbref#1{table~\ref{tbl:#1}}
\def\Tbref#1{Table~\ref{tbl:#1}}
\def\tbsref#1{tables~\ref{tbl:#1}}
\def\Tbsref#1{Tables~\ref{tbl:#1}}
\def\tbvref#1{\ref{tbl:#1}}
\def\Tbvref#1{\ref{tbl:#1}}
\def\tblab#1{\label{tbl:#1}}

\def\Fgref#1{Fig.~\ref{fig:#1}}
\def\Figref#1{Figure~\ref{fig:#1}}
\def\Fgsref#1{Figs.~\ref{fig:#1}}
\def\Figsref#1{Figures~\ref{fig:#1}}
\def\fgvref#1{(\ref{fig:#1})}
\def\Fgvref#1{\ref{fig:#1}}
\def\Fglab#1{\label{fig:#1}}

\def\scref#1{section~\ref{sec:#1}}
\def\Scref#1{Section~\ref{sec:#1}}
\def\sclab#1{\label{sec:#1}}

\def\bmaT{\left(\begin{array}{ccc}}
\def\emaT{\end{array}\right)}
\def\bma{\left( \begin{array} }
\def\ema{\end{array} \right)}
\def\vev#1{\langle #1 \rangle}
\newcommand{\VEV}[1]{{\langle {#1} \rangle}}

\def\ov{\overline}
\def\wt{\widetilde}
\def\l{\left}
\def\r{\right}
\def\gsim{~{\rlap{\lower 3.5pt\hbox{$\mathchar\sim$}}\raise 1pt\hbox{$>$}}\,}
\def\lsim{~{\rlap{\lower 3.5pt\hbox{$\mathchar\sim$}}\raise 1pt\hbox{$<$}}\,}
\makeatletter
\newtoks\@stequation

\def\subequations{\refstepcounter{equation}%
  \edef\@savedequation{\the\c@equation}%
  \@stequation=\expandafter{\theequation}
  \edef\@savedtheequation{\the\@stequation}
  \edef\oldtheequation{\theequation}%
  \setcounter{equation}{0}%
  \def\theequation{\oldtheequation\alph{equation}}}

\def\endsubequations{%
  \ifnum\c@equation < 2 \@warning{Only \the\c@equation\space subequation
    used in equation \@savedequation}\fi
  \setcounter{equation}{\@savedequation}%
  \@stequation=\expandafter{\@savedtheequation}%
  \edef\theequation{\the\@stequation}%
  \global\@ignoretrue}

\def\eqnarray{\stepcounter{equation}\let\@currentlabel\theequation
\global\@eqnswtrue\m@th
\global\@eqcnt\z@\tabskip\@centering\let\\\@eqncr
$$\halign to\displaywidth\bgroup\@eqnsel\hskip\@centering
     $\displaystyle\tabskip\z@{##}$&\global\@eqcnt\@ne
      \hfil$\;{##}\;$\hfil
     &\global\@eqcnt\tw@ $\displaystyle\tabskip\z@{##}$\hfil
   \tabskip\@centering&\llap{##}\tabskip\z@\cr}

\makeatother
\begin{titlepage}
\thispagestyle{empty}
\begin{flushright}
\begin{tabular}{l}
{KEK-TH-798}\\
{VPI-IPPAP-01-03}\\
{hep-ph/0112338} \\
{December, 2001}
\end{tabular}
\end{flushright}
\baselineskip 24pt 
\begin{center}
{\Large\bf
Prospects of Very Long Base-Line Neutrino Oscillation Experiments
with the KEK-JAERI High Intensity Proton Accelerator
}
\vspace{5mm}

\baselineskip 18pt 
\renewcommand{\thefootnote}{\fnsymbol{footnote}}
\setcounter{footnote}{0}

{M. Aoki$^1$\footnote{{e-mail:mayumi.aoki@kek.jp}},
 K. Hagiwara$^1$,
 Y. Hayato$^2$\footnote{{e-mail:hayato@neutrino.kek.jp}},
 T. Kobayashi$^2$\footnote{{e-mail:takashi.kobayashi@kek.jp}},}\\
{T. Nakaya$^3$\footnote{{e-mail:nakaya@scphys.kyoto-u.ac.jp}}, 
 K. Nishikawa$^3$\footnote{{e-mail:nishikaw@neutrino.kek.jp}},
 and
 N. Okamura$^{4}\footnote{{e-mail:nokamura@vt.edu}}$}\\
\bs
$^1${\it Theory Group, KEK, Tsukuba, Ibaraki 305-0801, Japan}\\
$^2${\it Inst.\,of Particle and Nuclear Studies, High Energy Accelerator
Research Org. (KEK), Tsukuba, Ibaraki 305-0801, Japan}\\
$^3${\it Department of Physics, Kyoto University, Kyoto 606-8502, Japan}\\
$^4${\it IPPAP, Physics Department, Virginia Tech. Blacksburg, VA 24061, USA}
\\
\end{center}
\begin{abstract}
We study physics potential of Very Long
Base-Line (VLBL) Neutrino-Oscillation Experiments with the
High Intensity Proton Accelerator (HIPA), which will be completed by the
year 2007 in Tokai-village, Japan, as a joint project of KEK and JAERI
(Japan Atomic Energy Research Institute). 
The HIPA 50 GeV proton beam will deliver neutrino beams of a few GeV
range with the intensity about two orders of magnitude higher than the present
KEK beam for K2K experiment.
As a sequel to the proposed HIPA-to-Super-Kamiokande experiment, we study
impacts of experiments with a 100 kton-level detector and the base-line 
length of a few-thousand km. 
The pulsed narrow-band $\nu_\mu$ beams (NBB) allow us to measure 
the $\nu_\mu \to \nu_e$ transition probability
and the $\nu_\mu$ survival probability through counting experiments at 
large water-$\check {\rm C}$erenkov detector.
We study sensitivity of such experiments to the neutrino
mass hierarchy, the mass-squared differences, the three angles, and one CP
phase of the three-generation lepton-flavor-mixing matrix.
We find that experiments at a distance between 1,000 and 2,000 km can
determine the sign of the larger mass-squared difference
$(m_3^2-m_1^2)$
if the mixing between $\nu_e$ and
$\nu_3$ (the heaviest-or-lightest neutrino) is not too small;
$2|U_{e3}|^2(1-|U_{e3}|^2)\gsim 0.03$.
The CP phase can be constrained if the $|U_{e3}|$ element is sufficiently 
large, $2|U_{e3}|^2(1-|U_{e3}|^2)\gsim 0.06$,
and if the smaller mass-squared difference $(m_2^2-m_1^2)$
and the $U_{e2}$ element are in the 
prefered range of the large-mixing-angle
solution of the solar-neutrino deficit.
The magunitude $\left| m_3^2-m_1^2\right|$
and the matrix element $U_{\mu 3}$ can
be precisely measured, but we find little sensitivity to
$m_2^2-m_1^2$
and the matrix element $U_{e2}$. 
 
\end{abstract}
\bs
\bs
{
 \small 
 \begin{flushleft}
 \end{flushleft}
}
 
\end{titlepage}

\newpage
\renewcommand{\thepage}{- \roman{page} -}
\tableofcontents
\newpage
\setcounter{page}{1}
\renewcommand{\thepage}{\arabic{page}}
\baselineskip 18pt 
\renewcommand{\thefootnote}{\arabic{footnote}}
\setcounter{footnote}{0}
\section{Introduction}
\clean
 Many neutrino experiments \cite{atm}-\cite{SNO}
strongly suggest that there are 
flavor mixings in the lepton sector, and 
that neutrinos are massive.
 According to the atmospheric-neutrino observations \cite{atm},
the lepton-flavor-mixing matrix
(the Maki-Nakagawa-Sakata (MNS) matrix \cite{MNS})
has a large mixing angle.
Especially, the Super-Kamiokande (SK) collaboration \cite{atm_SK} reported
that $\nu^{}_{\mu}$ oscillates into the other species
with maximal mixing.
 The K2K \cite{K2K} experiment,
the current long-base-line (LBL) neutrino oscillation
experiment from KEK to SK with $L=250$ km  
and $\langle E_\nu \rangle=1.3$ GeV,
obtained results which are consistent with 
the neutrino oscillation in the atmospheric-neutrino anomaly with
$\delta m^2 \sim 3 \times 10^{-3}$ and $\sin^2 2\theta \sim 1$. 
 The two reactor neutrino experiments CHOOZ \cite{CHOOZ}
and Palo Verde \cite{PaloVarde} reported that no oscillation
is found from $\ov{\nu}_e^{}$, and they exclude significant mixings
between $\ov{\nu}_e^{}$ and the other neutrinos.
 An important conclusion from these observations is
that the atmospheric neutrino oscillation cannot be due to
$\nu_\mu^{}$-$\nu_e^{}$ oscillation.
 Recently, the SK collaboration showed an evidence
that $\nu^{}_{\mu}$ oscillates into an active neutrino rather than
sterile neutrinos \cite{atm_mutau}.
 According to the results of solar-neutrino observations \cite{sun0},
the $\nu_{e}^{}$ flux from the sun is less than
that of the prediction of the standard solar model \cite{SSM}
and the reduction factor depends on neutrino energies.
 The most convincing explanation for this deficit is
the $\nu^{}_{e}$ oscillation to the other neutrinos.
 Four possible scenarios of the solar-neutrino oscillation have been
identified :
the MSW \cite{MSW1,MSW2} large-mixing-angle (LMA) solution,
the MSW small-mixing-angle (SMA) solution,
the vacuum oscillation (VO) solution \cite{VO} and
the MSW low-$\delta m^2$ (LOW) solution.
 Recently, the SK collaboration reported 
the energy spectrum and the day-night asymmetry data and showed that
the LMA solution
is more favorable than the other scenarios \cite{SK-solar}.
The SNO experiment \cite{SNO},
which observes the solar neutrino flux with heavy water,
showed conclusively, when combined with the SK flux data, that
$\nu_e^{}$ oscillates into the other active neutrinos \cite{SNO}.
A consistent picture of three-neutrino oscillations with two large mixing
angles and two hierarchically different mass-squared differences 
emerges from those observations,
with the exception of the LSND experiment \cite{LSND} which may
indicate the existence of the fourth and non-standard (sterile) neutrino.

 Several LBL neutrino oscillation experiments \cite{MINOS,ICARUS,OPERA,H2SK}
and a short-base-line experiment \cite{MiniBooNe}
have been proposed to confirm the results of these experiments and
to measure the neutrino oscillation parameters
more precisely.
 The MINOS experiments \cite{MINOS}, from Fermilab to the Soudan mine, 
with the base-line length of $L=730$ km and $\langle E_\nu \rangle=3.5$ GeV,
will start producing data in 2005. 
 The observation of the survival probability
$P_{\nu_\mu \to \nu_\mu}$ will 
allow us to measure the larger mass-squared difference
and the mixing angle with about $10 \%$ accuracy.
 Two LBL experiments, ICARUS \cite{ICARUS} and OPERA \cite{OPERA},
from CERN to Gran-Sasso with the base-line length of $L=730$ km
and at higher energies with $\langle E_\nu \rangle \sim 20$ GeV
have been proposed, and they may begin operation in 2005. 
 The CERN experiments expect to observe the $\nu_\mu \to \nu_\tau$
appearance.
 Physics discover potential of these LBL experiments have
been studied extensively \cite{LBL_th}.

 In Japan, as a sequel to the K2K experiment, a new
LBL neutrino oscillation
experiment between the
High Intensity Proton Accelerator (HIPA) \cite{HIPA}
and SK has been proposed \cite{H2SK}.
 The facility, HIPA, has a 50 GeV proton accelerator, 
which will be completed by the year 2007
in the site of JAERI (Japan Atomic Energy Research Institute),
as the joint project of KEK and JAERI.
 The proton beam of HIPA can deliver high intensity neutrino beams
in the $\sim 1$ GeV range, whose intensity is two-orders-of-magnitude
higher than that of the KEK 12 GeV proton synchrotron
beam for the K2K experiment.
 The HIPA-to-SK experiment with $L=295$ km and
$\langle E_\nu \rangle = 1.3$ GeV will measure
the larger mass-squared difference
with 3 $\%$ accuracy and
the mixing angle at about 1 $\%$ accuracy.

 All these experiments use conventional neutrino beams,
which are made from decaying pions and Kaons
that are produced by high-energy proton beams.
 The possibility of a neutrino factory \cite{NuFac} has been
discussed as the
next generation of LBL neutrino-oscillation experiments \cite{NuFac2}.
 Here the neutrino beam is delivered from 
decaying muons in a muon-storage ring, where the stored muon
energy may be in the several 10 GeV range.
A neutrino factory can deliver very high intensity
neutrino beams that are consist of 
the same amount of $\nu_\mu$ and $\ov\nu_e^{}$
($\ov \nu_\mu$ and $\nu_e$ from $\mu^+$)
with precisely known spectra.
Possibility of constructing a neutrino factory in the JAERI site
by upgrading the HIPA is now being extensively studied \cite{HIPA_NuFact}.

 In this paper we examine an alternative possibility of using 
conventional neutrino beams from the HIPA for Very-Long-Base-Line
(VLBL) neutrino oscillation experiments, 
whose base-line length exceeding a thousand km \cite{BCP4}.
A possible 100 kton detector in Beijing \cite{BAND} can be a target 
at about $L=2,100$ km away.
Physics capability of such experiments should be seriously studied because 
a neutrino factory may turn out to be too difficult 
or too expensive to realize in the near future.
In order to take full advantage of conventional $\nu$ beams, we examine
the case of using
pulsed narrow-band $\nu^{}_{\mu}$ beams (NBB)
and as a target we consider a large water-$\check {\rm C}$erenkov detector
{\it a la} SK which is capable of 
measuring the $\nu^{}_{\mu}$-to-$\nu^{}_{e}$ transition probability
and the $\nu^{}_{\mu}$ survival probability.
 We study sensitivity of such experiments to the
signs and the magnitudes of the two mass-squared differences,
the three angles and one CP phase of the three-flavor MNS matrix.

This article is organized as follows.
In section 2, we fix our notation and
review the present status of the neutrino-oscillation
experiments in the three-neutrino model. 
In section 3, we study the properties of the narrow-band neutrino beams
that can be delivered by the HIPA 50 GeV proton synchrotron.
In section 4, we study the signals, the backgrounds and systematic errors
of the VLBL experiments and present our findings for the prospects of 
experiments at the base-line length of 2,100 km and 1,200 km.
Our results are summarized in section 5.

\section{Neutrino oscillation in the three-neutrino model }
In this section, we give the definition and
useful parameterization 
of the 3$\times$3 Maki-Nakagawa-Sakata
(MNS) lepton-flavor-mixing matrix \cite{MNS}, and give constraints on 
its matrix elements and the neutrino mass-squared differences.
\clean
\subsection{The MNS matrix}
The MNS matrix is defined analogously to
the CKM matrix \cite{CKM} through the
charged-current (CC) weak interactions, where the charged-current can be
expressed as  
\bea
 J^{\mu}_{\rm cc} =
(\overline{d},\overline{s},\overline{b})
{V_{_{\rm CKM}}^{\dagger}}
\gamma^{\mu}(1-\gamma_5)
(u,c,t)^T 
+
(\overline{e},\overline{\mu},\overline{\tau})
\gamma^{\mu}(1-\gamma_5)
{V_{_{\rm MNS}}^{}}
(\nu_1^{},\nu_2^{},\nu_3^{})^T\,.
\eqlab{current}
\eea
Here
$\nu_i^{}~(i=1,2,3)$ denotes the neutrino
mass-eigenstates.
The flavor-eigenstates of the neutrinos are then expressed as
\begin{eqnarray}
 \nu_\alpha^{}=
\sum_{i=1}^3 \l({{V}_{_{\rm MNS}}}\r)_{{\alpha}{i}}^{}
~{\nu_i^{}}\,,
\eqlab{massdef}		
\end{eqnarray}
where $\alpha=e,\mu,\tau$ are the lepton-flavor indices.

The $3\times3$ MNS matrix has three mixing angles
and three phases in general.
We adopt the following parameterization \cite{HO1}
\beq
V_{_{\rm MNS}}^{} =
\bmaT
U_{e 1}    & U_{e 2}    & U_{e 3} \\
U_{\mu 1}  & U_{\mu 2}  & U_{\mu 3} \\
U_{\tau 1} & U_{\tau 2} & U_{\tau 3}  
\emaT
\bmaT
1 & 0 & 0 \\
0 & e^{i \varphi_2^{}} & 0 \\
0 & 0 & e^{i \varphi_3^{}} 
\emaT
\equiv U{\cal P}
\,,
\eeq
where ${\cal P}$ is the diagonal phase matrix with two Majorana phases, 
$\varphi_2^{}$ and $\varphi_3^{}$.
The matrix $U$, which has three mixing angles and one phase,
can be parameterized in the same way as the CKM matrix.
Because the present neutrino oscillation experiments
constrain directly the elements,
$U_{e2}$, $U_{e3}$, and $U_{\mu 3}$,
we find it most convenient to adopt the parameterization \cite{HO1}
where these three matrix elements in the upper-right
corner of the $U$ matrix are the independent parameters.
Without losing generality, we can take 
$U_{e2}$ and $U_{\mu 3}$ to be real and non-negative.
By allowing $U_{e3}$ to have the complex phase 
\bea
U_{e2},U_{\mu 3} \geq 0,~~~ 
U_{e3}\equiv \left|U_{e3}\right| e^{-i\delta_{_{\rm MNS}}} 
~~(0 \leq \delta_{_{\rm MNS}} < 2\pi )\,,
\eea
the four independent parameters are $U_{e2}, U_{\mu 3}, |U_{e3}|$
and $\delta_{_{\rm MNS}}$.
All the other matrix elements of the $U$
are then determined by the unitary conditions :
\bseq
\bea
U_{e1} &=& \sqrt{1-|U_{e3}|^2-|U_{e2}|^2}\,, 
~~~~~U_{\tau 3}
=
\sqrt{1-|U_{e3}|^2-|U_{\mu 3}|^2}\,, 
\\ 
U_{\mu 1} &=& - \frac{U_{e2}U_{\tau 3} + U_{\mu 3}U_{e1}U_{e3}^{\ast} }
                      {1-|U_{e3}|^2}\,, 
~~U_{\mu 2} = \frac{U_{e1}U_{\tau 3} - U_{\mu 3}U_{e2}U_{e3}^{\ast} }
                    {1-|U_{e3}|^2}\,,\\
U_{\tau 1} &=& \frac{U_{e2}U_{\mu 3} - U_{\tau 3}U_{e1}U_{e3}^{\ast} }
                     {1-|U_{e3}|^2}\,,
~~~~U_{\tau 2} = - \frac{U_{\mu 3}U_{e1} + U_{e2}U_{\tau 3}U_{e3}^{\ast} }
                       {1-|U_{e3}|^2}\,.
\eea
\eseq
\hspace*{-1ex}
In this phase convention,
$U_{e1}$, $U_{e2}$, $U_{\mu 3}$, and $U_{\tau 3}$
are all real and non-negative numbers,
and the other five elements are complex numbers.

The Jarlskog parameter \cite{Jar} of the MNS matrix
is defined as
\bea
J_{_{\rm MNS}}^{}
\equiv {\rm Im} \left({V_{\alpha i} V_{\beta i}^{\ast} V_{\beta j} 
V_{\alpha j}^{\ast}}\right) 
= {\rm Im} \left({U_{\alpha i} U_{\beta i}^{\ast} U_{\beta j} 
U_{\alpha j}^{\ast}}\right)
= - \frac{U_{e1}U_{e2}U_{\mu 3}U_{\tau 3}}{1-\left|U_{e3}\right|^2}
              {\rm Im}\left({U_{e3}}\right)\,,
\eqlab{Jmns}
\eea
where $(\alpha, \beta)=(e, \mu),(\mu, \tau),(\tau, e) $ and
$(i,j)=(1,2),(2,3),(3,1)$.
The last expression above is obtained in our phase convention.
The two Majorana phases $\varphi_2^{}$ and $\varphi_3^{}$
do not contribute to the Jarlskog parameter.

\subsection{Constraints on the MNS matrix and the mass-squared differences}
\def\cU#1#2{U_{#1}^{#2}}
\def\cUm#1#2{\wt{U}_{#1}^{#2}}
The probability of finding the flavor-eigenstate $\beta$
from the original flavor-eigenstate $\alpha$ at the base-line length $L$
in the vacuum is given by
\bea
P_{\nu_\alpha \to \nu_\beta} 
&=&
\left|\sum_{j=1}^3 (V_{_{\rm MNS}})_{\beta j}^{} 
\exp\left(-i\frac{m_j^2}{2E_\nu}L\right) (V_{_{\rm MNS}})_{j \alpha} 
\right|^2\nn \\
\nn \\
&=&
\left| \cU{\beta 1}{} \cU{\alpha 1}{\ast}+\cU{\beta 2}{}
  e^{-i\Delta_{12}}
  \cU{\alpha 2}{\ast} 
+\cU{\beta 3}{}
  e^{-i\Delta_{13}}
  \cU{\alpha 3}{\ast}
\right|^2\nn\\
&=&
\delta_{\alpha\beta}
-4{\rm Re} 
\left\{
\cU{\alpha 1}{} \cU{\beta 1}{\ast} \cU{\beta 2}{} \cU{\alpha 2}{\ast}
\sin^2\frac{\Delta_{12}}{2}
+
\cU{\alpha 2}{} \cU{\beta 2}{\ast} \cU{\beta 3}{} \cU{\alpha 3}{\ast}
\sin^2\frac{\Delta_{23}}{2}
+
\cU{\alpha 3}{} \cU{\beta 3}{\ast} \cU{\beta 1}{} \cU{\alpha 1}{\ast}
\sin^2 \frac{\Delta_{31}}{2}
\right\}
\nn \\
& &~~~~~~+
2{\rm Im}\left[
\cU{\alpha 1}{} \cU{\beta 1}{\ast} \cU{\beta 2}{} \cU{\alpha 2}{\ast}
\right]
\left[
\sin \Delta_{12}+\sin \Delta_{23}+\sin \Delta_{31}
\right],
\eqlab{P_ex_0}
\eea
where $\Delta_{ij}$ is
\begin{equation}
\Delta_{ij} \equiv \frac{\delta m_{ij}^2}{2E_\nu}L
\simeq 2.534 \frac{\delta m_{ij}^2 ({\rm eV}^2)}{E_\nu({\rm GeV})}L({\rm km}
)\,,
\end{equation}
with the neutrino energy $E_\nu$ and the mass-squared differences
$\delta m_{ij}^2=m_j^2 - m_i^2$.
In particular,
the survival probability
in the vacuum is
\begin{eqnarray}
 P_{\nu_\alpha \to \nu_\alpha} &=&
1
-4\left\{\left|
\cU{\alpha 1}{} \cU{\alpha 2}{}
\right|^2
\sin^2\frac{\Delta_{12}}{2}
+
\left|
\cU{\alpha 2}{}\cU{\alpha 3}{}
\right|^2
\sin^2\frac{\Delta_{23}}{2}
+
\left|
\cU{\alpha 3}{} \cU{\alpha 1}{}
\right|^2
\sin^2 \frac{\Delta_{31}}{2}
\right\}\,.
\eqlab{P_surv_0}
\end{eqnarray}
The oscillation probabilities of anti-neutrinos in the vacuum are
obtained from those of neutrinos simply by reversing the sign of the Jarlskog 
parameter; 
\bseq
\bea
P_{\ov \nu_{\alpha}\to \ov \nu_{\alpha}}&=&
P_{\nu_{\alpha}\to \nu_{\alpha}}\,, \\
P_{\ov \nu_{\alpha}\to \ov \nu_{\beta}}&=&
P_{\nu_{\alpha}\to \nu_{\beta}}(J_{_{\rm MNS}}\to -J_{_{\rm MNS}})\,.
\eea
\eseq
For instance,
\bea
\hspace*{-7ex}
\left(
\begin{array}{c}
P_{\nu_\mu \to \nu_e}\\
P_{\ov \nu_\mu \to \ov \nu_e}
\end{array}
  \right)&=&-4{\rm Re}\l\{U_{\mu 1}U_{e1}^\ast U_{e2}U_{\mu 2}^\ast
\sin^2\frac{\Delta_{12}}{2}
+U_{\mu 2}U_{e2}^\ast U_{e3}U_{\mu 3}^\ast \sin^2\frac{\Delta_{23}}{2}
\right.\nn \\
&&+
\left.
U_{\mu 3}U_{e3}^\ast U_{e1}U_{\mu 1}^\ast \sin^2\frac{\Delta_{31}}{2}\r\} 
\mp 2J_{_{\rm MNS}}\l\{\sin\Delta_{12}+\sin\Delta_{23}
+\sin\Delta_{31}\r\}\,.~~~~~
\eea
The unitarity leads to the identities 
\bea
\sum_{\beta =e,\mu,\tau} P_{\nu_\alpha \to \nu_\beta}
=\sum_{\beta =e,\mu,\tau} P_{\nu_\beta \to \nu_\alpha}
=\sum_{\beta =e,\mu,\tau} P_{\ov \nu_\alpha \to \ov \nu_\beta}
=\sum_{\beta =e,\mu,\tau} P_{\ov \nu_\beta \to \ov \nu_\alpha}
=1\,.
\eea

The following approximations for the oscillation probabilities are
useful in our study.
When $\left|\Delta_{12}^{}\right|\ll\left|\Delta_{13}^{}\right|\sim 1$,
the oscillation probabilities can be expressed as
\begin{eqnarray}
P_{\nu_\alpha^{} \to \nu_\beta^{}} &=&
\delta_{\alpha\beta}-4{\rm Re}\left\{
\cU{\alpha 3}{}\cU{\beta 3}{\ast}\cU{\beta 1}{}\cU{\alpha 1}{\ast}
+
\cU{\alpha 2}{}\cU{\beta 2}{\ast}\cU{\beta 3}{}\cU{\alpha 3}{\ast}
\right\}
\sin^2\dfrac{\Delta_{13}^{}}{2} \nn \\
&&+
\left\{
2{\rm Re}
 \left(
 \cU{\alpha 2}{}\cU{\beta 2}{\ast}\cU{\beta 3}{}\cU{\alpha 3}{\ast} 
 \right)
 \sin\Delta_{13}^{}
 \pm
 4J_{\rm MNS}^{} \sin^2\dfrac{\Delta_{13^{}}}{2}
\right\}
\Delta_{12}^{} \nn \\
&&+
{\rm Re}\left(
 \cU{\alpha 1}{}\cU{\beta 1}{\ast}\cU{\beta 2}{}\cU{\alpha 2}{\ast} 
+
 \cU{\alpha 2}{}\cU{\beta 2}{\ast}\cU{\beta 3}{}\cU{\alpha 3}{\ast} 
\cos \Delta_{13}^{}
\right)
\Delta_{12}^2
+{\cal O}(\Delta_{12}^3)\,.
\eqlab{approx1}
\end{eqnarray}
When $1 \ll \left| \Delta_{12}\right| \sim
\left|\Delta_{13}\right|$,
we may take into account finite resolution of 
$\Delta_{ij}^{}$
$(\Delta_{ij}^{}\to\Delta_{ij}^{}\pm\delta\Delta_{ij}^{})$,
and find
\begin{eqnarray}
P_{\nu_\alpha^{} \to \nu_\beta^{}}
&\to&
\dfrac{1}{2\delta \Delta_{ij}^{}}
\int_{_{\Delta_{ij}^{}-\delta \Delta_{ij}^{}}}
    ^{^{\Delta_{ij}^{}+\delta \Delta_{ij}^{}}}
d \Delta_{ij}^{}
P_{\nu_\alpha^{} \to \nu_\beta^{}}
\left(\Delta_{ij}^{}\right)
\nn \\
&=&
\delta_{\alpha \beta}^{}
-2{\rm Re}
\left\{
\cU{\alpha 1}{}\cU{\beta 1}{\ast}\cU{\beta 2}{}\cU{\alpha 2}{\ast}
+
\cU{\alpha 2}{}\cU{\beta 2}{\ast}\cU{\beta 3}{}\cU{\alpha 3}{\ast} 
+
\cU{\alpha 3}{}\cU{\beta 3}{\ast}\cU{\beta 1}{}\cU{\alpha 1}{\ast} 
\right\}\nn\\
&&
+2{\rm Re}
\left\{
\cU{\alpha 1}{}\cU{\beta 1}{\ast}\cU{\beta 2}{}\cU{\alpha 2}{\ast}
\cos{\Delta_{12}^{}}\dfrac{\sin{\delta \Delta_{12}}}{\delta \Delta_{12}}
+
\cU{\alpha 2}{}\cU{\beta 2}{\ast}\cU{\beta 3}{}\cU{\alpha 3}{\ast} 
\cos{\Delta_{23}^{}}\dfrac{\sin{\delta \Delta_{23}}}{\delta \Delta_{23}}
\right.
\nn\\
&&~~~~~~~~~~
\left.
+
\cU{\alpha 3}{}\cU{\beta 3}{\ast}\cU{\beta 1}{}\cU{\alpha 1}{\ast} 
\cos{\Delta_{31}^{}}\dfrac{\sin{\delta \Delta_{31}}}{\delta \Delta_{31}}
\right\}\nn\\
&&
\pm 2J_{_{\rm MNS}}
\left\{
\sin{\Delta_{12}^{}}\dfrac{\sin{\delta \Delta_{12}}}{\delta \Delta_{12}}
+
\sin{\Delta_{23}^{}}\dfrac{\sin{\delta \Delta_{23}}}{\delta \Delta_{23}}
+
\sin{\Delta_{31}^{}}\dfrac{\sin{\delta \Delta_{31}}}{\delta \Delta_{31}}
\right\}\,.
\eqlab{large_13}
\end{eqnarray}
In LBL experiments the uncertainty in $E_\nu^{}$,
$\delta E_{\nu}^{}$, dictates $\Delta_{ij}^{}$,
\begin{equation}
 \delta \Delta_{ij}^{} \simeq -\Delta_{ij}\dfrac{\delta E_\nu^{}}{E_\nu^{}}\,,
\end{equation}
and the following two cases are relevant.
When
\begin{equation}
\left|\delta \Delta_{12}^{}\right|
\simeq
\left|\Delta_{12}^{}\dfrac{\delta E_\nu^{}}{E_\nu^{}} \right|
\ll 1
\ll
\left|\Delta_{13}^{}\dfrac{\delta E_\nu^{}}{E_\nu^{}} \right|
 \simeq
\left|\delta \Delta_{13}^{}\right|
 \simeq
\left|\delta \Delta_{23}^{}\right|\,,
\end{equation} 
\eqref{large_13} can be expressed as
\begin{eqnarray}
P_{\nu_\alpha^{} \to \nu_\beta^{}} \to
\delta_{\alpha \beta}^{}
&-&2{\rm Re}
\left\{
\cU{\alpha 1}{}\cU{\beta 1}{\ast}\cU{\beta 2}{}\cU{\alpha 2}{\ast}
+
\cU{\alpha 2}{}\cU{\beta 2}{\ast}\cU{\beta 3}{}\cU{\alpha 3}{\ast} 
+
\cU{\alpha 3}{}\cU{\beta 3}{\ast}\cU{\beta 1}{}\cU{\alpha 1}{\ast} 
\right\}\nn\\
&&
+2{\rm Re}
\left\{
\cU{\alpha 1}{}\cU{\beta 1}{\ast}\cU{\beta 2}{}\cU{\alpha 2}{\ast}
\cos{\Delta_{12}^{}}
\right\}
\pm 2J_{_{\rm MNS}}
\left\{
\sin{\Delta_{12}^{}}
\right\}
\nn\\
&&+{\cal O}\left(
\left(
\Delta_{12}^{}\dfrac{\delta E_\nu^{}}{E_\nu^{}}
\right)^2
\right)
+{\cal O}\left(
\left(
\Delta_{13}^{}\dfrac{\delta E_\nu^{}}{E_\nu^{}}
\right)^{-1}
\right)
\,.
\eqlab{high_frq}
\end{eqnarray}
On the other hand, when
\begin{equation}
1\ll
\left|\Delta_{12}^{}\dfrac{\delta E_\nu^{}}{E_\nu^{}} \right|
\ll
\left|\Delta_{13}^{}\dfrac{\delta E_\nu^{}}{E_\nu^{}} \right|
\end{equation} 
\eqref{large_13} is simplified to
\begin{eqnarray}
P_{\nu_\alpha^{} \to \nu_\beta^{}} \to
\delta_{\alpha \beta}^{}
&-&2{\rm Re}
\left\{
\cU{\alpha 1}{}\cU{\beta 1}{\ast}\cU{\beta 2}{}\cU{\alpha 2}{\ast}
+
\cU{\alpha 2}{}\cU{\beta 2}{\ast}\cU{\beta 3}{}\cU{\alpha 3}{\ast} 
+
\cU{\alpha 3}{}\cU{\beta 3}{\ast}\cU{\beta 1}{}\cU{\alpha 1}{\ast} 
\right\}
\nn\\
&&+{\cal O}\left(
\left(
\Delta_{12}^{}\dfrac{\delta E_\nu^{}}{E_\nu^{}}
\right)^{-1}
\right)
+{\cal O}\left(
\left(
\Delta_{13}^{}\dfrac{\delta E_\nu^{}}{E_\nu^{}}
\right)^{-1}
\right)
\,.
\end{eqnarray}
Note that in \eqsref{approx1}, \eqvref{large_13},
and \eqvref{high_frq}, $\pm J_{_{\rm MNS}}^{}$ stands for
${\rm Im}\left[
\cU{\alpha 1}{} \cU{\beta 1}{\ast} \cU{\beta 2}{} \cU{\alpha 2}{\ast}
\right]$;
see \eqref{Jmns}.

All the above formulas remain valid for the
neutrino oscillation probabilities
in the matter, by replacing the mass-squared differences and the MNS 
matrix elements with the effective ones in the matter,
\bea
\Delta_{ij} \to \wt \Delta_{ij}\,,~~U_{\alpha i} \to \wt U_{\alpha i}\,,
~~J_{_{\rm MNS}} \to \wt J_{_{\rm MNS}}\,,
\eea
as long as the matter density remains the same along the base-line.
The definitions of the effective parameters
$\wt \Delta_{ij}$ and $\wt U_{\alpha i}$ are
given in section \ref{sec:matter}, \
and $\wt J_{_{\rm MNS}}$ is obtained from
\eqref{Jmns} by replacing all
$U_{\alpha i}$'s by $\wt U_{\alpha i}$'s.

In the following,
we summarize the constraints on the neutrino mass-squared differences
and the MNS matrix elements from the recent neutrino-oscillation
experiments; the atmospheric-neutrino anomaly \cite{atm,atm_SK},
the CHOOZ reactor experiment \cite{CHOOZ},
and the
solar-neutrino deficit observations \cite{sun0,SK-solar,SNO}.

\subsubsection{Atmospheric-neutrino anomaly}
\label{sec:atm}
A recent analysis of the atmospheric-neutrino data
from the Super-Kamiokande (SK) experiment \cite{atm_SK} finds
\bseq
\begin{eqnarray}
0.88 < &\sin^2 2\theta_{_{\rm ATM}} & < 1.0 \,, \\
1.6 \times 10^{-3}<&\delta m^2_{_{\rm ATM}} ({\rm eV}^2) &
< 4.0 \times 10^{-3}\,,
\eqlab{atmex}
\end{eqnarray}
\eseq	
from the $\nu_{\mu}^{}\to\nu_\mu$ survival probability in
the two-flavor oscillation model:
\bea
P_{\nu_\mu \to \nu_\mu}=1-\sin^22\theta_{_{\rm ATM}}\sin^2
\left(\frac{\delta m^2_{_{\rm ATM}}}{4E_\nu}L \right).
\eea

The base-line of this observation is less than about $10^4$ km 
for the Earth diameter,
and the typical neutrino energy is one to a few GeV.
The survival probability \eqref{P_surv_0} may then be expanded as 
\begin{eqnarray}
P_{\nu_\mu \to \nu_\mu}=
1-4\left|\cU{\mu 3}{}\right|^2 \left(1-\left|\cU{\mu 3}{}\right|^2\right)
\sin^2\frac{\Delta_{13}}{2}
+
2\left|\cU{\mu 2}{}\right|^2\left|\cU{\mu3}{}\right|^2
\Delta_{12}\sin\Delta_{13}
+{\cal O}(\Delta_{12}^2)\,.~~~~~~~
\end{eqnarray}
When $|\Delta_{12}|\ll 1$, we may
neglect terms of order $\Delta_{12}$ and obtain the following 
identification :
\bseq
\eqlab{atmdef2}
\begin{eqnarray}
4|U_{\mu 3}|^2(1-|U_{\mu 3}|^2) &=&
\sin^22\theta_{_{\rm ATM}} \,, \\
\l|\delta m^2_{13}\r| &=&
\delta m^2_{_{\rm ATM}}\,.
\eqlab{atmdef}
\end{eqnarray}
\eseq
The independent parameter $\cU{\mu 3}{}$
($\geq 0$ in our convention) is then
\begin{eqnarray}
 \cU{\mu 3}{} = 
\sqrt{
1-\sqrt{1-\sin^2 2\theta_{_{\rm ATM}}}
}\Bigg/\sqrt{2}\,.
\end{eqnarray}

The magnitude of the neglected terms in the above approximation is largest
when the large-mixing-angle solution of the solar-neutrino deficit is taken. 
For $\delta m^2_{_{\rm SOL}}=10^{-4}$ eV$^2$, $E_\nu=1$ GeV, and
$L=10^4$ km, we have $|\Delta_{12}|\simeq 
\delta m^2_{_{\rm SOL}}(L/2E_\nu)\sim 1$,
and a more careful analyses
in the three-neutrino model are required
to constrain the model parameters. 
The results of such analyses \cite{3gene}
show that the identifications 
\eqsref{atmdef2} remain valid approximately even for 
the large-mixing-angle solution. 

\subsubsection{Reactor neutrino experiments}
\label{sec:chooz}
The CHOOZ experiment \cite{CHOOZ} 
measured the survival probability of $\ov{\nu}_e^{}$,
\bea
P_{\ov \nu_e \to \ov \nu_e}=1-\sin^22\theta_{_{\rm CHOOZ}}\sin^2
\left(\frac{\delta m^2_{_{\rm CHOOZ}}}{4E_\nu}L \right)\,,
\eea
and it was found that
\begin{eqnarray}
\sin^2 2\theta_{_{\rm CHOOZ}} <
\left\{
\begin{array}{lcl}
0.10~~&{\mbox{{  for  }}}
  & \delta m^2_{_{\rm CHOOZ}} > \numt{3.5}{-3}\,{\rm eV}^2\,,\\
\\
0.18~~&{\mbox{{  for  }}}
  & \delta m^2_{_{\rm CHOOZ}} > \numt{2.0}{-3}\,{\rm eV}^2\,,\\
\\
0.52~~&{\mbox{{  for  }}}
  & \delta m^2_{_{\rm CHOOZ}} > \numt{1.0}{-3}\,{\rm eV}^2\,.
\end{array}
\right.
\eqlab{chooz}
\end{eqnarray}
The base-line length of this experiment is about 1 km and
the typical anti-neutrino energy is 1 MeV.
For those $L$ and $E_\nu$, the Earth matter effects are negligible, and
$|\Delta_{12}|=\delta m^2_{_{\rm SOL}}(L/2E_\nu)$ can be safely neglected even
for the large-mixing-angle solution.
The survival probability of the three-neutrino model is
then approximated by
\begin{equation}
 P_{\ov{\nu}_e \to \ov{\nu}_e} =
1-4\l|\cU{e3}{}\r|^2\l(1-\l|\cU{e3}{}\r|^2\r)
\sin^2\dfrac{ \Delta_{13}}{2}+{\cal O}(\Delta_{12})\,,
\end{equation}
and we obtain the identifications :
\bseq
\begin{eqnarray}
4\l|\cU{e3}{}\r|^2\l(1-\l|\cU{e3}{}\r|^2\r) &=& 
\sin^2 2\theta_{_{\rm CHOOZ}}\,, \eqlab{anglecho}\\
|\delta m^2_{13}| &=&\delta m^2_{_{\rm CHOOZ}}.
\end{eqnarray}
\eseq
With the above identifications, we find that the element
$\left|U_{e3}^{}\right|$
\eqref{anglecho} is constrained by \eqref{chooz} in the region of 
$|\delta m^2_{13}|$ allowed by the atmospheric-neutrino oscillation data 
through \eqref{atmex} and \eqref{atmdef}.
The independent parameter $\cU{e3}{}$
is now constrained by \eqref{chooz} through the identification
\begin{eqnarray}
 \l|\cU{e3}{}\r| = 
\sqrt{
1-\sqrt{1-\sin^2 2\theta_{_{\rm CHOOZ}}}
}\Bigg/\sqrt{2}\,.
\eqlab{boundcho}
\end{eqnarray}

\subsubsection{Solar-neutrino deficit}
Deficit of the solar neutrinos observed at several
terrestrial experiments \cite{sun0,SK-solar,SNO}
have been successfully interpreted
in terms of the $\nu_e^{} \to \nu_{X}^{}$
($\nu_{X}^{} \neq \nu_e^{}$ or $\ov{\nu}_e$) oscillation
\bea
P_{\nu_e \to \nu_e}=1-\sin^22\theta_{_{\rm SOL}}\sin^2
\left(\frac{\delta m^2_{_{\rm SOL}}}{4E_\nu}L \right)
\eea
in the following four scenarios \cite{sun0,SK-solar,SNO}.

\noindent
MSW large-mixing-angle solution (LMA)~:
\bseq
\begin{eqnarray}
0.7 < &\sin^2 2\theta_{_{\rm SOL}}& < 0.9\,, 
\eqlab{MSW_L_sin} \\
\numt{3}{-5} < &\delta m^2_{_{\rm SOL}}(\mbox{eV}^2) & < \numt{15}{-5}\,.
\eqlab{MSW_L_mass}
\end{eqnarray}
\eqlab{MSW_L}
\eseq

\noindent
MSW small-mixing-angle solution (SMA)~:
\bseq
\begin{eqnarray}
\numt{1.2}{-3}< &\sin^2 2\theta_{_{\rm SOL}}& < \numt{12}{-3}\,,
\eqlab{MSW_S_sin} \\
\numt{0.3}{-5}< &\delta m^2_{_{\rm SOL}}(\mbox{eV}^2) & < \numt{1}{-5}\,.
\eqlab{MSW_S_mass}
\end{eqnarray}
\eseq

\noindent
MSW low-$\delta m^2$ solution (LOW)~:
\bseq
\begin{eqnarray}
0.8< &\sin^2 2\theta_{_{\rm SOL}}& \le 1\,,
\eqlab{LOW_sin} \\
10^{-8}< &\delta m^2_{_{\rm SOL}}(\mbox{eV}^2) & < \numt{2.5}{-7}\,.
\eqlab{LOW_mass}
\end{eqnarray}
\eseq

\noindent
Vacuum Oscillation solution (VO)~:
\bseq
\begin{eqnarray}
&\sin^2 2\theta_{_{\rm SOL}}& \sim 0.9 \,,
\eqlab{VO_sin}\\
&\delta m^2_{_{\rm SOL}}(\mbox{eV}^2) & \sim 10^{-9} \,.
\eqlab{VO_mass}
\end{eqnarray}
\eseq

The SK collaboration reported that
their data on the energy spectrum and the day-night
asymmetry disfavor the SMA solution at $95\%$ C.L..
Recently the SNO collaboration 
gives us the first direct indication of a non-electron
and active flavor component in the solar neutrino flux.
Because $\delta m_{_{\rm ATM}}^{2} \gg \delta m_{_{\rm SOL}}^{2}$
the $\nu_e^{}$ survival probability in the three-neutrino model
can be expressed as
\begin{equation}
P_{\nu_e \to \nu_e} =
1-2\left|U_{e 3}\right|^2
\left(
1-\left|U_{e 3}\right|^2
\right)
-4
\left|U_{e 1}\right|^2
\left|U_{e 2}\right|^2
\sin^2
\left(
\dfrac{\Delta_{12}^{}}{2}
\right)+{\cal O}\left(
\dfrac{E_{\nu}^{}}{\delta E_{\nu}^{}\Delta_{13}}
\right)\,,
\eqlab{surv2}
\end{equation}
where terms of order
$\left[
\Delta_{13}
\left(
\delta E_{\nu}^{}/E_{\nu}^{}
\right)\right]^{-1}$ in \eqref{large_13} are 
safely neglected.
The energy-independent deficit factor,
$2|\cU{e3}{}|^2(1-|\cU{e3}{}|^2)$,
should be smaller than $5\%$ by the CHOOZ
constraint \eqref{chooz}
if $\left|\delta m_{13}^2\right|>\numt{3.5}{-3}$eV$^2$.
Because we need only rough estimates of the allowed
ranges of the MNS matrix elements\footnote{{%
See for example, more detail discussions in \cite{3flavor}
}},
we ignore the small energy-independent deficit factor
and interpret the results of the two-flavor analysis
\eqref{MSW_L_sin}-\eqref{VO_sin} by using the
following identifications
\bseq
\begin{eqnarray}
4\l|\cU{e1}{}\r|^2\l|\cU{e2}{}\r|^2 &=& \sin^2 2\theta_{_{\rm SOL}}\,,\\
|\delta m^2_{12}| &=&  \delta m^2_{_{\rm SOL}}\,.
\end{eqnarray}
\eseq
By using unitary condition, the independent parameter $\cU{e2}{}$
is obtained as
\begin{equation}
 \cU{e2}{} = \left[
\left(
1-|\cU{e3}{}|^2
-
\sqrt{\left(1-|\cU{e3}{}|^2\right)^2-\sin^22\theta_{_{\rm SOL}}}
\right)\Bigg/2\right]^{1/2}
\,.
\end{equation}

\subsection{Neutrino mass hierarchy}
All the above constraints on the three-neutrino model parameters
are obtained from
the survival probabilities
which are  even-functions
of $\delta m^2_{ij}$.
We have made the identification
\begin{equation}
\delta m^2_{_{\rm SOL}} =
 \left|\delta m^2_{12}\right| 
\ll
 \left|\delta m^2_{13}\right| = \delta m^2_{_{\rm ATM}}\,,
\end{equation}
which is valid for all the four scenarios of
the solar neutrino oscillation.

There are four mass hierarchy cases corresponding to the
sign of the $\delta m^2_{ij}$,
as shown in \Fgref{cases} and \Tbref{cases}.
We name them the neutrino mass hierarchy I, II, III and IV, respectively.

\begin{table}[htbp]
\begin{center}
\begin{tabular}{|c|c|c|c|c|}
\hline
 &I &II &III & IV \\
\hline 
 $\delta  m_{12}^{2^{\,}}$ &
 $+\delta m_{\rm SOL}^2$&
 $-\delta m_{\rm SOL}^2$&
 $+\delta m_{\rm SOL}^2$&
 $-\delta m_{\rm SOL}^2$ \\[1mm]
\hline 
 $\delta  m_{13}^{2^{\,}}$ &
 $+\delta m_{\rm ATM}^2$&
 $+\delta m_{\rm ATM}^2$&
 $-\delta m_{\rm ATM}^2$&
 $-\delta m_{\rm ATM}^2$\\[1mm]
\hline
\end{tabular}
\end{center}
\caption{%
The four neutrino mass hierarchies and the corresponding
sign assignments for $\delta m^2_{12}$ and $\delta m^2_{13}$.
}
\tblab{cases}
\end{table}

 If the MSW effect is relevant for
the solar neutrino oscillation, then the neutrino mass hierarchy cases
II and IV are 
not favored, especially for the LMA and SMA solutions.
The hierarchy I may be called `normal' and
the hierarchy III may be called `inverted'.
Within the three-neutrino model, there is an indication that the normal
hierarchy I is favored against the inverted one III from the Super-Nova
1987A observation \cite{SN1987A}.
Nevertheless a terrestrial experiment is needed to determine 
the neutrino mass hierarchy. 
\begin{figure}[htbp]
\begin{center}
\rotatebox{-90}{\scalebox{0.5}{
\includegraphics{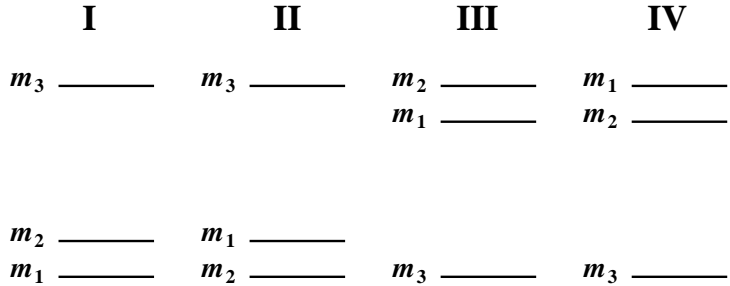}
}}
\end{center} 
\caption{Schematical view of the four-types of neutrino mass hierarchy.
}
\Fglab{cases}
\end{figure}

We notice here that there are two types of mass eigenstate for the neutrinos.
The states $\nu_i$ with the mass $m_i~(i=1,2,3)$ appear
in the definition \eqref{massdef} of the MNS matrix, whose elements are 
constrained by the existing experiments
that measure essentially the neutrino-flavor survival probabilities.
Since the survival probabilities 
\eqref{P_surv_0} do not depend on the sign of 
the mass-squared differences, these constraints do not depend on 
the neutrino mass hierarchy.
Because the MNS matrix elements are constrained uniquely by
the neutrino-flavor survival probabilities,
we may call these states as `current-based' mass-eigenstates.
We find this basis most convenient for our study in this paper.
On the other hand, 
the `mass-ordered' mass-eigenstates $\nu'_i$, whose masses satisfy
\begin{equation}
m'_1 < m'_2 < m'_3 \,,
\end{equation}
are useful when studying the high-energy behavior of the neutrino mass matrix
\cite{okamura}, the matter effects on the neutrino-flavor oscillation 
\cite{MSW1,MSW2}, and when studying the lepton-number violation effects which 
are proportional to the magnitudes of the Majorana masses.
The relation between the current-eigenstates and the two mass-eigenstates
is
\begin{equation}
\left(
\begin{array}{c}
\nu_e  \\
\nu_\mu  \\
\nu_\tau 
\end{array}
\right)
=
V_{_{\rm MNS}}
\left(
\begin{array}{c}
\nu_1  \\
\nu_2  \\
\nu_3 
\end{array}
\right)
=
V'_{_{\rm MNS}}
\left(
\begin{array}{c}
\nu'_1 \\
\nu'_2 \\
\nu'_3 
\end{array}
\right)\,,
\end{equation}
where $V'_{_{\rm MNS}}$ is the MNS matrix in the
mass-ordered mass-eigenstate base.
It can be obtained from $V_{_{\rm MNS}}$ by
\begin{equation}
 V'_{_{\rm MNS}}=V_{_{\rm MNS}}O^{\rm X}
\mbox{{~~~~~(X = I, II, III, IV)}}\,,
\end{equation}
where the permutation matrices $O^{\rm X}$ 
\bea
O^{\rm I} =
\bmaT
1 & 0 & 0 \\
0 & 1 & 0 \\
0 & 0 & 1 
\emaT\,, 
O^{\rm II} =
\bmaT
0 & 1 & 0 \\
1 & 0 & 0 \\
0 & 0 & 1 
\emaT\,,
O^{\rm III} =
\bmaT
0 & 1 & 0 \\
0 & 0 & 1 \\
1 & 0 & 0 
\emaT ,
O^{\rm IV} =
\bmaT
0 & 0 & 1 \\
0 & 1 & 0 \\
1 & 0 & 0 
\emaT\,~~~
\eea
relate the two mass-eigenstates
\beq
\bma{c}
\nu_1  \\
\nu_2  \\
\nu_3 
\ema 
=O^{\rm X}
\bma{c}
\nu'_1 \\
\nu'_2 \\
\nu'_3 
\ema\,,
\eeq
for the neutrino mass hierarchy I, II, III and IV, respectively.

\subsection{Neutrino oscillation in the Earth matter}
\label{sec:matter}
Neutrino-flavor oscillation inside matter is governed by the
Schr{\" o}dinger equation
\begin{eqnarray}
i\frac{\partial}{\partial t}
\bma{c}
\nu_e \\
\nu_\mu \\
\nu_\tau\\
\ema
=\frac{1}{2E_\nu}
\left[
{H_0} + 
\bmaT
a & 0 & 0 \\
0 & 0 & 0 \\
0 & 0 & 0
\emaT
\right]
\bma{c}
\nu_e \\
\nu_\mu \\
\nu_\tau\\
\ema
=H
\bma{c}
\nu_e \\
\nu_\mu \\
\nu_\tau\\
\ema,
\end{eqnarray}
where $H_0$ is the Hamiltonian in the vacuum,
\begin{eqnarray}
H_0=
{\cU{}{}}
\bmaT
 0 & 0 & 0 \\
 0 & \delta m^2_{21} & 0 \\
 0 & 0 & \delta m^2_{31}
\emaT
{\cU{}{\dagger}}\,,
\eqlab{H_0}
\end{eqnarray}
and $a$ is the matter effect term \cite{MSW1},
\begin{eqnarray}
a=2\sqrt{2}G_F n_e E_\nu^{} 
={7.56}\times 10^{-5}({\rm eV}^2)\left(\frac{\rho}{{\rm g/cm}^{3}}\right)
  \left(\frac{E_\nu}{\rm GeV}\right)\,.
\eqlab{matter_a}
\end{eqnarray}
Here
$n_e$ is the electron density of the matter,
$G_F$ is the Fermi constant,
and $\rho$ is the matter density.
In our analysis,
we assume that the density of the earth's crust 
relevant for the VLBL experiments up to about 2,000 km
is a constant\footnote{{more detail discussion of the matter
profile in Ref.\cite{matter}.}},
$\rho=3$, with an uncertainty of $\Delta \rho=0.1$;
\bea
\rho ~({\rm g/cm}^3)=3.0\pm 0.1\,.
\eea
The Hamiltonian in the matter $H$ is diagonalized as
\begin{eqnarray}
{H} = \dfrac{1}{2E_\nu}
{{\cUm{}{}}}
\bmaT
 {\lambda_1} & 0 & 0 \\
 0 & {\lambda_2} & 0 \\
 0 & 0 & {\lambda_3}
\emaT
{{\cUm{}{\dagger}}}
,
\eqlab{H_w_mat}
\end{eqnarray}
by the MNS matrix in the matter $\wt U$.
The neutrino-flavor oscillation probabilities in the matter 
\begin{eqnarray}
 P_{\nu_\alpha \to \nu_\beta} &=&
\l|\cUm{\beta 1}{} \cUm{\alpha 1}{\ast}
+\cUm{\beta 2}{}
  e^{-i{{\wt{\Delta}_{12}}}}
  \cUm{\alpha 2}{\ast}
+\cUm{\beta 3}{}
  e^{-i{{\wt{\Delta}_{13}}}}
  \cUm{\alpha 3}{\ast}
\r|^2\,
\end{eqnarray}
takes the same
form as those in the vacuum \eqref{P_ex_0},
where 
the elements $U_{\alpha i}^{}$ are replaced by 
$\wt U_{\alpha i}^{}$ and the terms
$\Delta_{ij}^{}$ are replaced by
\begin{equation}
\wt{\Delta}_{ij} = 
\dfrac{{\lambda_j}-{\lambda_i}}{2E_\nu}L
\equiv
\dfrac{\delta \wt{m}_{ij}^{2}}{2E_\nu}L\,.
\end{equation}

In \Fgref{MNSelements} we show the $E_\nu$-dependence of the effective 
mass-squared
differences and the MNS matrix elements inside of the matter at 
$\rho=3.0$ g/cm$^3$.
The curves are obtained for $|\delta m^2_{13}|=3.5\times 10^{-3}$ eV$^2$,
$|\delta m^2_{12}|=15\times 10^{-5}$ eV$^2$,
$\sin^22\theta_{_{\rm ATM}}=1.0$, 
$\sin^22\theta_{_{\rm SOL}}=0.8$,
$\sin^22\theta_{_{\rm CHOOZ}}=0.1$,
and $\delta_{_{\rm MNS}}=0^\circ$.
For the mass hierarchy III and IV, $\delta m_{13}^2 < 0$, and for II and IV,
$\delta m_{12}^2 < 0$.
The elements $|\wt U_{\mu i}|^2$ affect the survival probability 
$P_{\nu_\mu \to \nu_\mu}$ in the matter, whereas the terms 
$\wt U_{\mu i}\wt U_{ei}^\ast$ affect the transition probability
$P_{\nu_\mu \to \nu_e}$ ; see \eqref{P_surv_0} and \eqref{P_ex_0},
whose expressions remain valid in the matter by replacing $U_{\alpha i}$
and $\Delta_{ij}$ by $\wt U_{\alpha i}$
and $\wt \Delta_{ij}$, respectively.

Between the hierarchy I and II (III and IV), the sign of the smaller 
mass squared difference $\delta m_{12}^2$ is different.
Even though the individual terms behave quite differently in the matter,
we find that neither $P_{\nu_\mu \to \nu_\mu}$ nor $P_{\nu_\mu \to \nu_e}$
depend strongly on the sign of $\delta m^2_{12}$ in the range of $E_\nu$
(1 $\sim$ 6 GeV) and $L$ (300 $\sim$ 2,100 km) that we study.
On the other hand, we find strong dependence of the transition probability
$P_{\nu_\mu \to \nu_e}$ on the sign of $\delta m^2_{13}$,
between the hierarchy I and III (II and IV).
We show in \Fgref{amplitude} the $\nu_{\mu} \to \nu_e$ amplitude
\bea
\wt U_{\mu 1}\wt U^\ast_{e 1}+\wt U_{\mu 2}e^{-i\wt \Delta_{12}}
\wt U^\ast_{e 2}
+\wt U_{\mu 3}e^{-i\wt \Delta_{13}}\wt U^\ast_{e 3}
\eea 
in the complex plane 
for the same parameter set at four $\delta_{_{\rm MNS}}^{}$ points.
The three terms in the above sum are shown in the complex plane
as two-vectors,
whose sum is chosen to lie along the horizontal axis.
The absolute value squared of the sum
gives the probability $P_{\nu_\mu \to \nu_e}$.
The left figures are for the normal hierarchy I and the right figures for
the inverted hierarchy III.
The double circle shows the origin, and the solid-circle (with solid line),
solid-square (with long-dased line), open-circle (with short-dashed line), 
open-square (with dotted line)
shows the amplitude for $\delta_{_{\rm MNS}}=0^\circ,~
90^\circ,~180^\circ$ and $270^\circ$, respectively.
 The four figures in each column show the amplitudes at the
same $L/E_\nu^{}=420$km$/$GeV;
from the top to the bottom, we show the amplitudes in
vacuum, in the earth crust $(\rho=3.0$ g/cm$^3)$
at $L=295$ km, $L=1,200$ km, and $L=2,100$ km.
We can clearly see from these figures that the $\nu_\mu$ to $\nu_e$ 
transition amplitude increases by the matter effect at higher energies
and hence at a larger distance in case of the normal hierarchy I,
whereas the amplitude decreases significantly in case of the inverted hierarchy
III.
The difference in the transition probability is more striking after
the amplitude is squared.

\begin{figure}[htb]
\begin{center}
{\scalebox{0.75}{\includegraphics{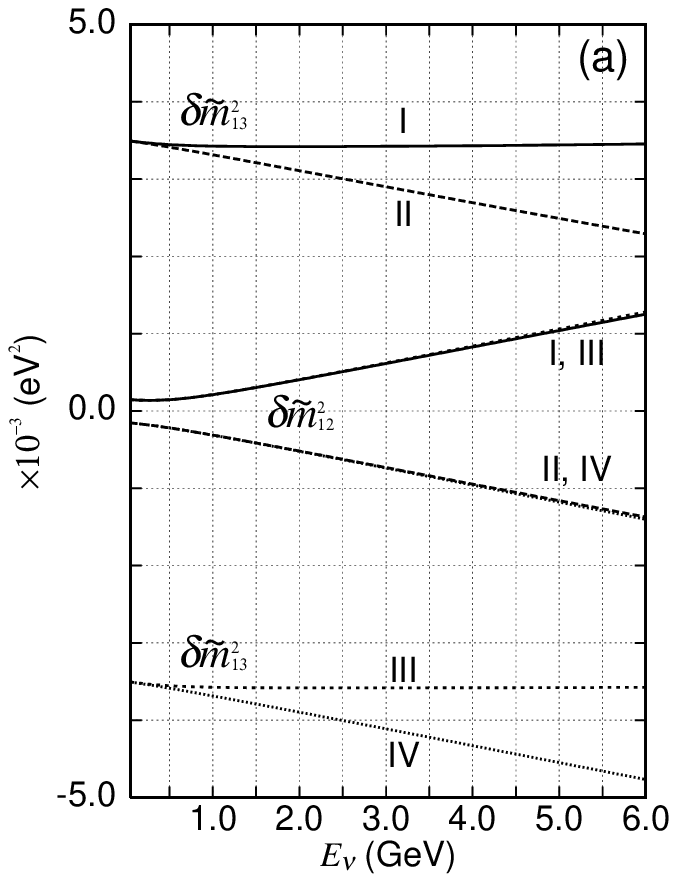}}}
{\scalebox{0.75}{\includegraphics{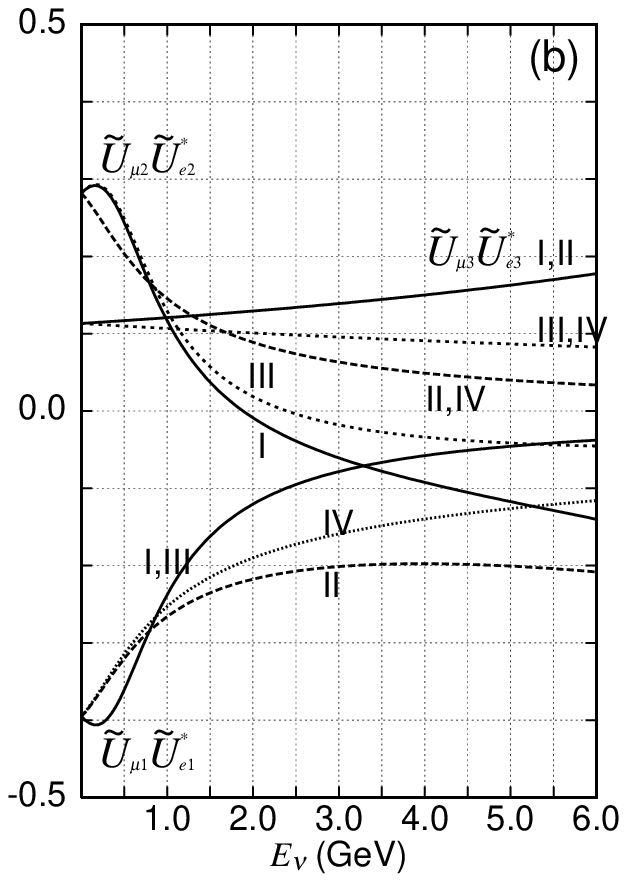}}}
{\scalebox{0.75}{\includegraphics{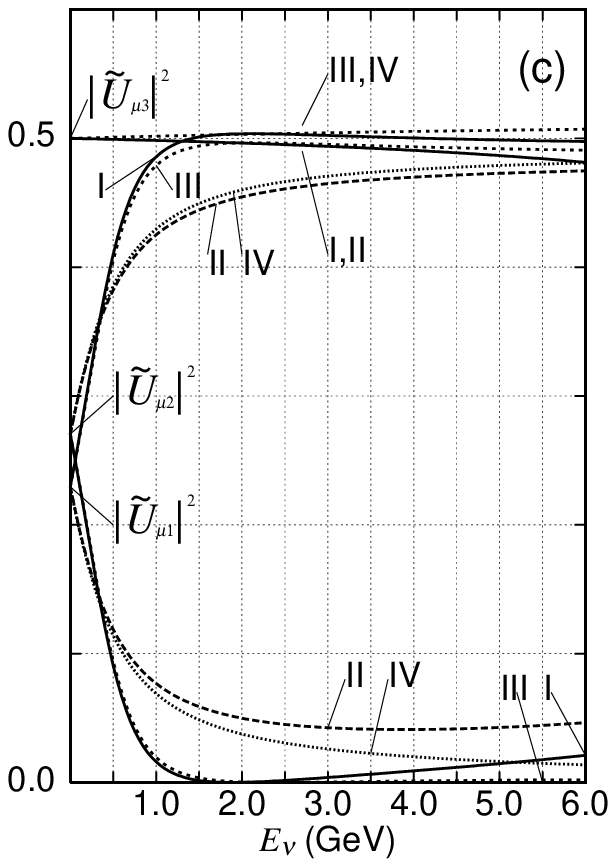}}} 
\end{center}
\caption{Mass-squared differences and the MNS matrix elements inside of 
the matter ($\rho=$3.0 g/cm$^3$) for the four neutrino mass hierarchies, 
I (solid lines), II (long-dashed lines), III (short-dashed lines) 
and IV (dotted lines).
The curves are obtained for $|\delta m^2_{13}|=3.5\times 10^{-3}$ (eV$^2$),
$|\delta m^2_{12}|=15\times 10^{-5}$ (eV$^2$),
$\sin^22\theta_{_{\rm ATM}}=1.0$, 
$\sin^22\theta_{_{\rm SOL}}=0.8$, and
$\sin^22\theta_{_{\rm CHOOZ}}=0.1$.
}
\Fglab{MNSelements}
\end{figure}
\begin{figure}[htbp]
\begin{center}
{\scalebox{0.53}{\includegraphics{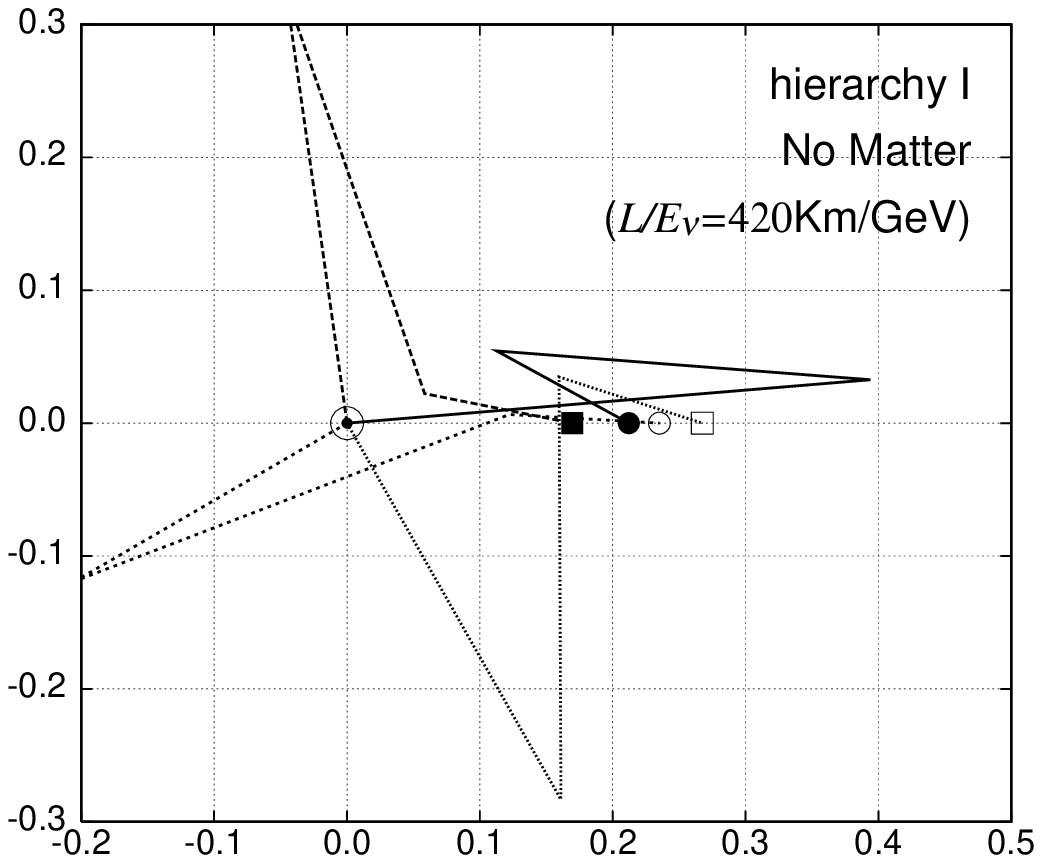}}}
{\scalebox{0.53}{\includegraphics{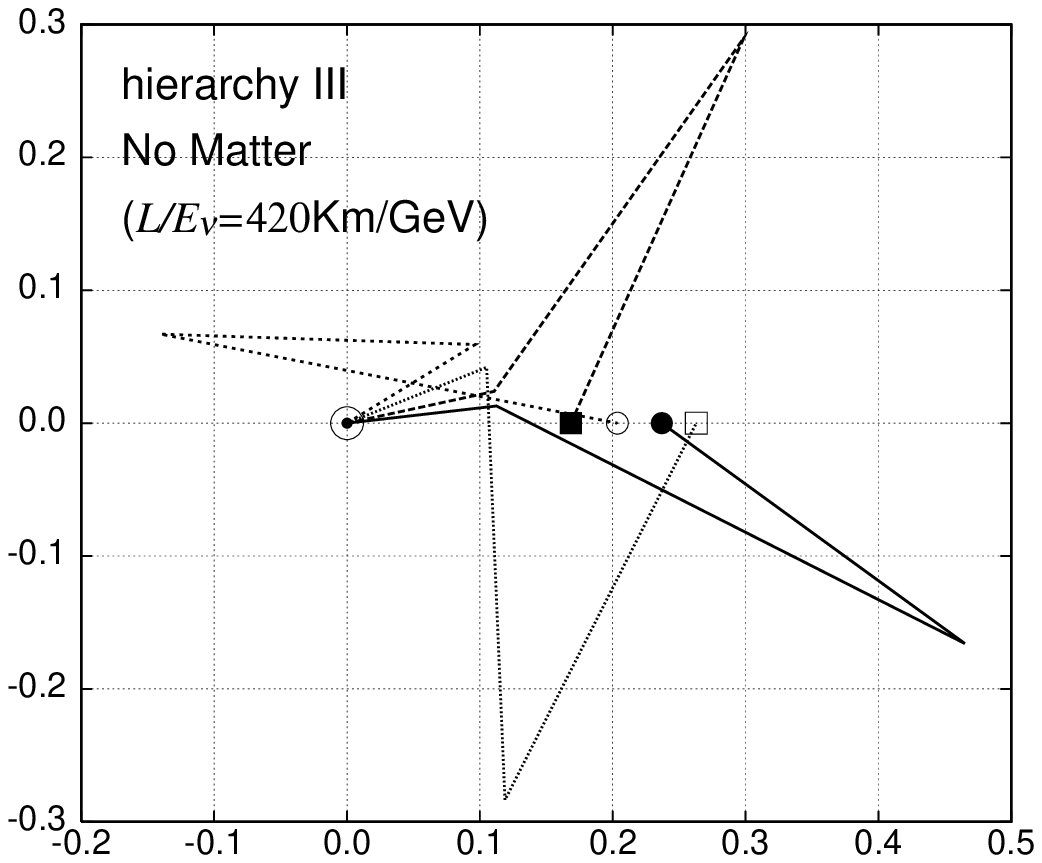}}}
{\scalebox{0.53}{\includegraphics{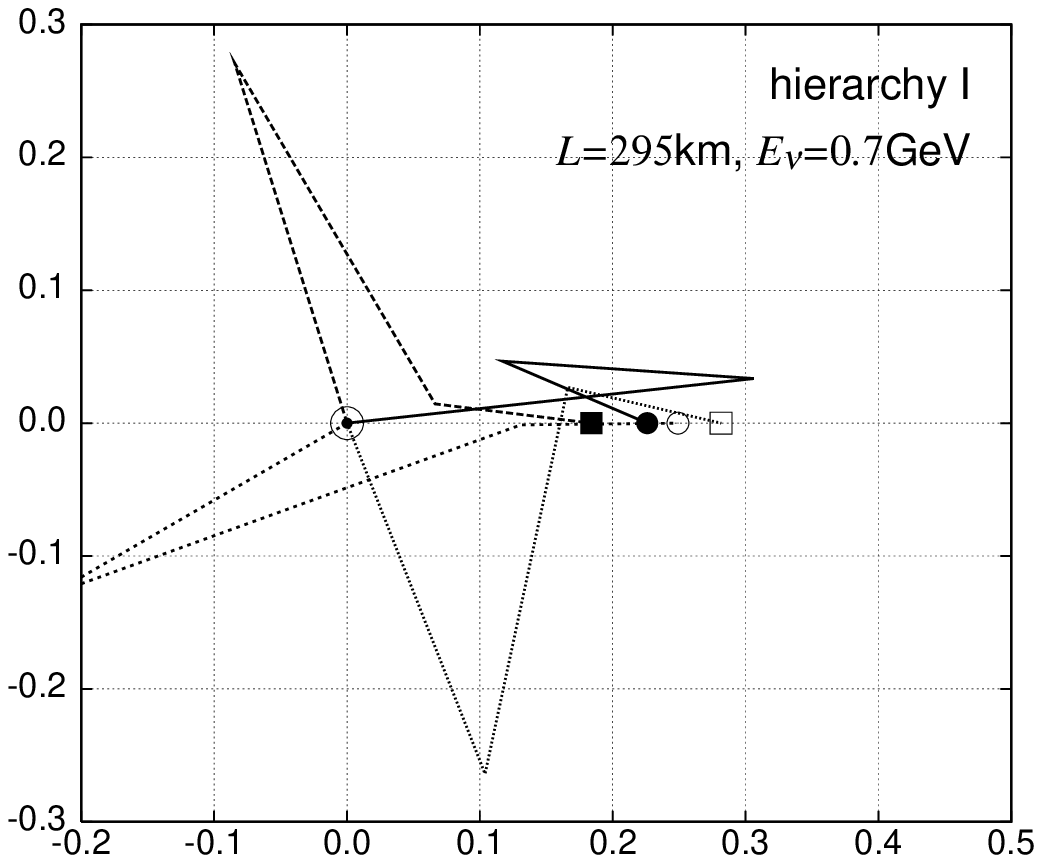}}}
{\scalebox{0.53}{\includegraphics{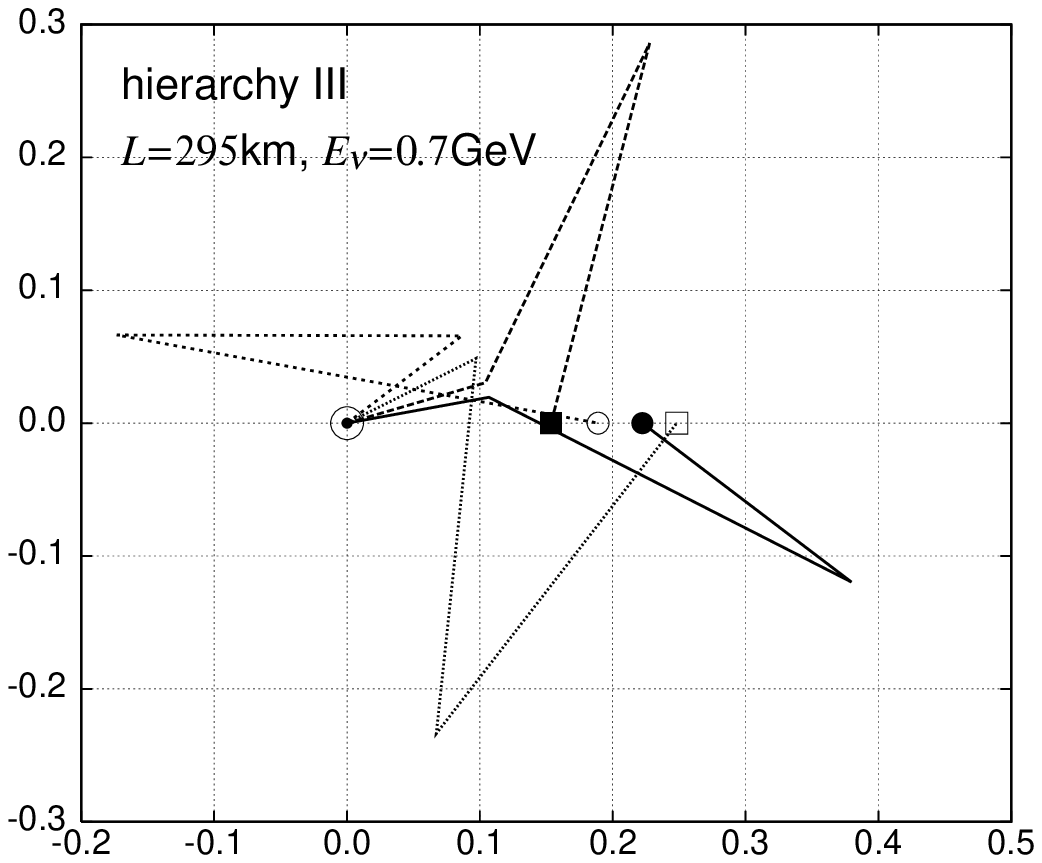}}}
{\scalebox{0.53}{\includegraphics{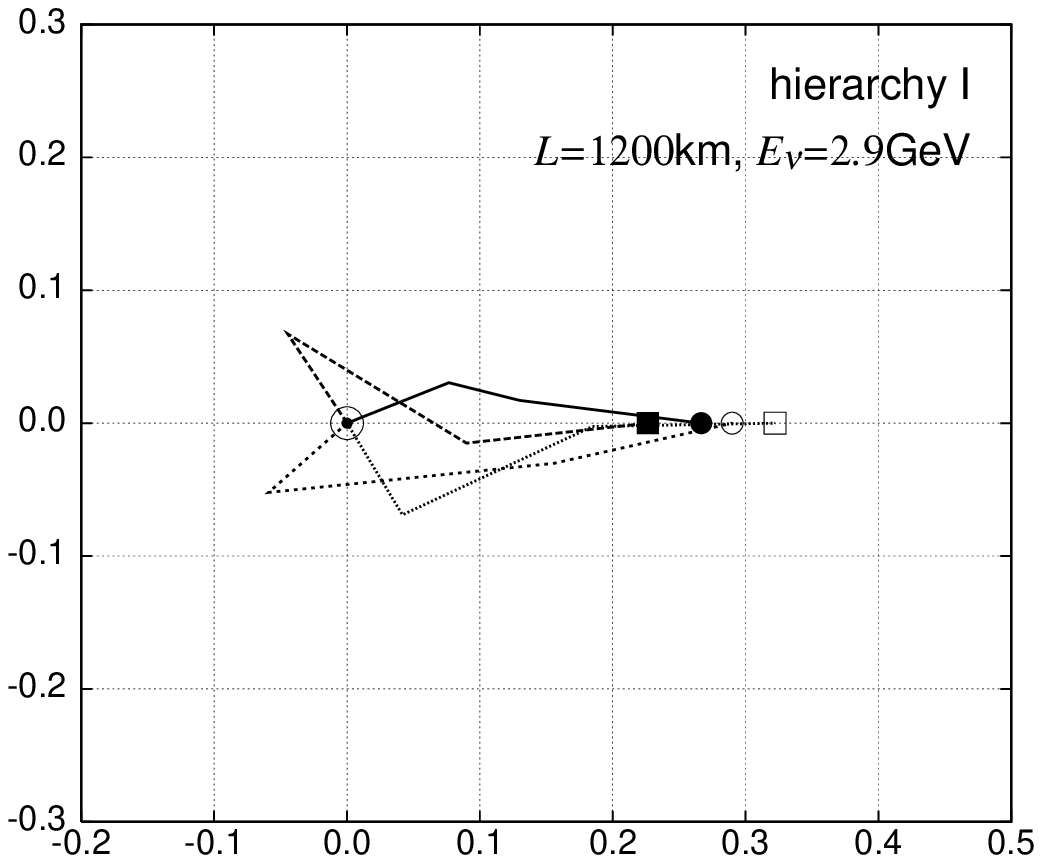}}}
{\scalebox{0.53}{\includegraphics{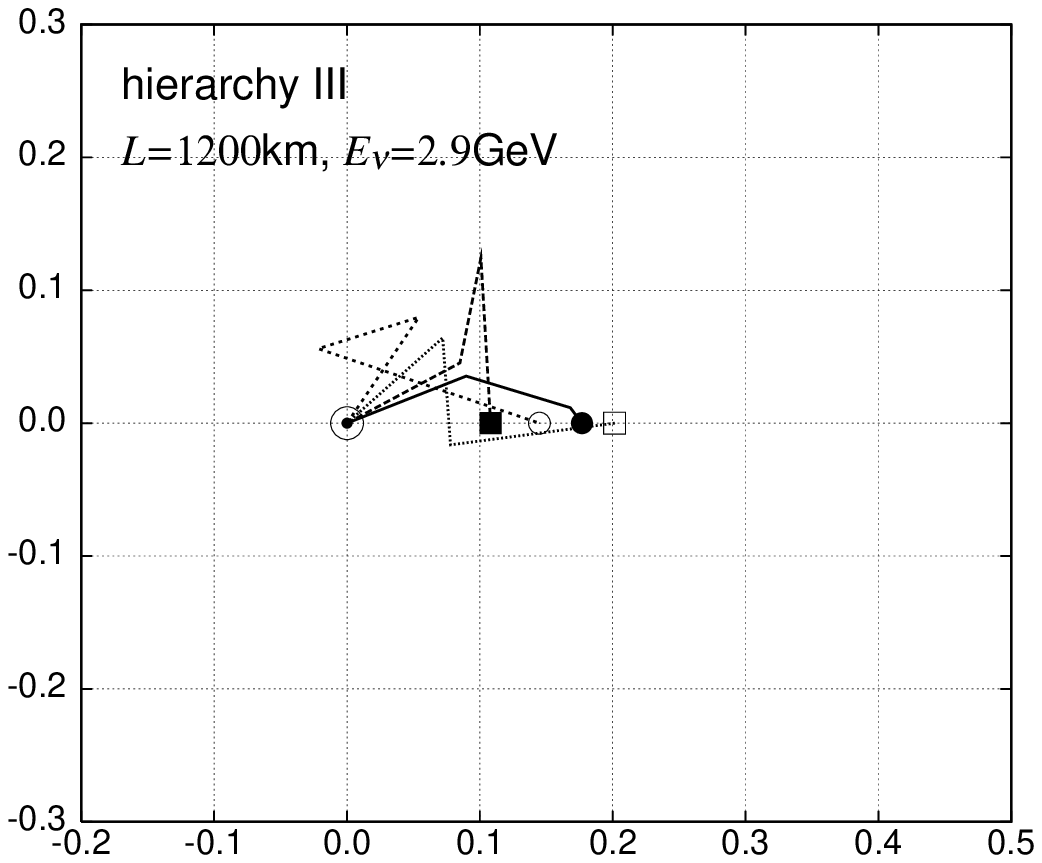}}}
{\scalebox{0.53}{\includegraphics{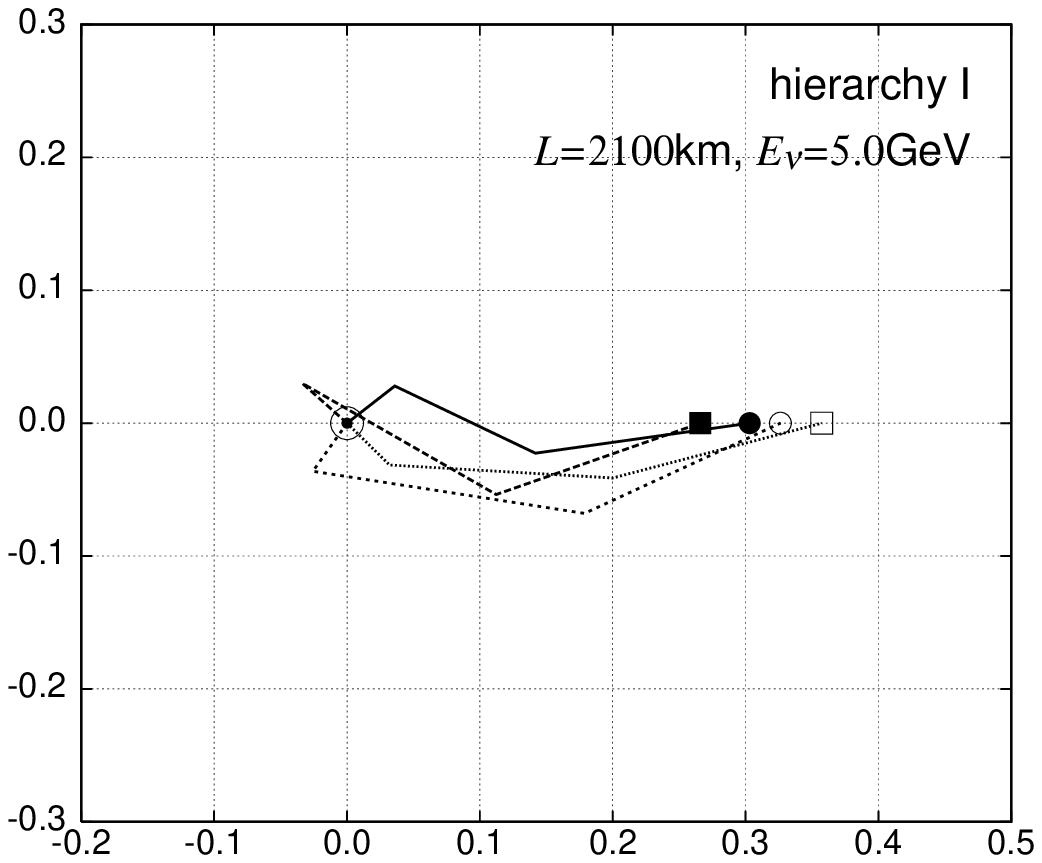}}}
{\scalebox{0.53}{\includegraphics{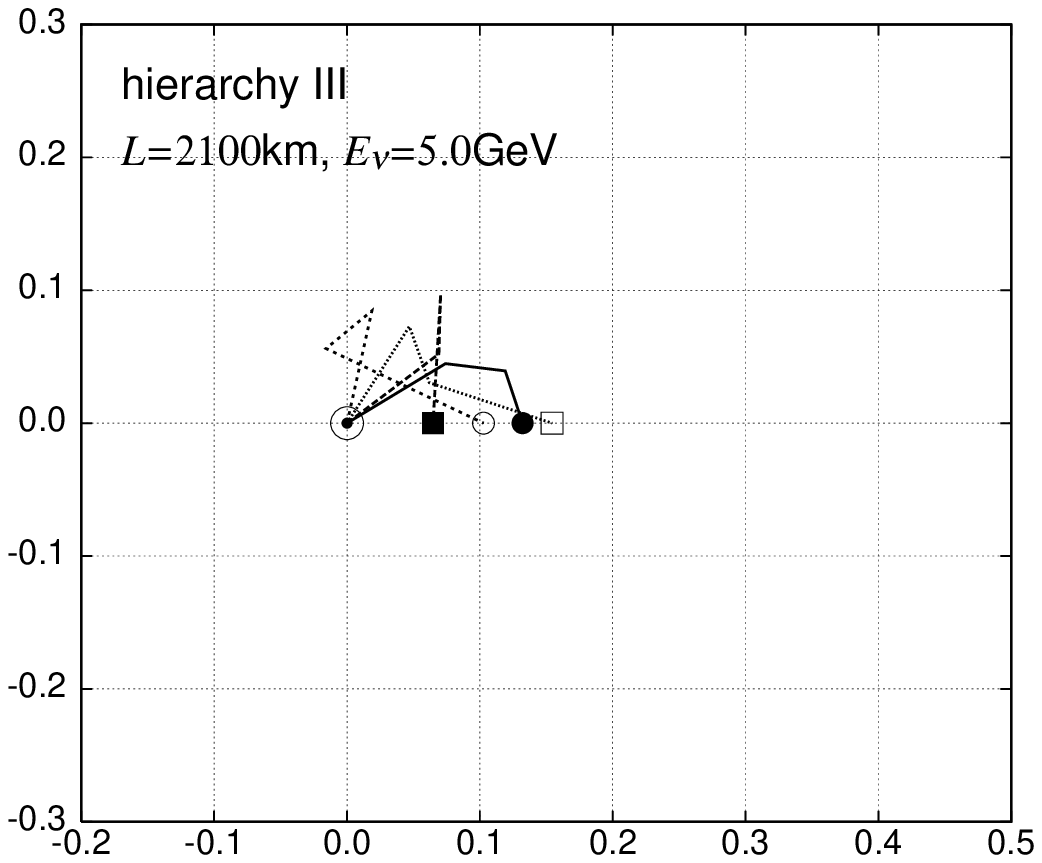}}}
\end{center}
\caption{{%
The $\nu_\mu\to\nu_e$ amplitude 
$
\wt U_{\mu 1}\wt U^\ast_{e 1}
+\wt U_{\mu 2}e^{-i\wt \Delta_{12}}\wt U^\ast_{e 2}
+\wt U_{\mu 3}e^{-i\wt \Delta_{13}}\wt U^\ast_{e 3}
$ 
in the mater ($\rho=3.0$ g/cm$^3$) at $L/E_\nu=2,100$ km/5 GeV
for $|\delta m^2_{13}|=3.5\times 10^{-3}$ eV$^2$, 
$|\delta m^2_{12}|=15\times 10^{-5}$ eV$^2$,
$\sin^22\theta_{_{\rm ATM}}=1.0$, 
$\sin^22\theta_{_{\rm SOL}}=0.8$, and
$\sin^22\theta_{_{\rm CHOOZ}}=0.1$.
The three terms in the sum are shown in the complex plane, where
the real part of the sum gives
$\sqrt{P_{\nu_\mu \to \nu_e}}$.
The four figures in the left are for the mass hierarchy I and 
the figures in the right
are for the hierarchy III.
The solid-circle, solid-square, open-circle, open-square are for
$\delta_{_{\rm MNS}}= 0^\circ, 90^\circ, 180^\circ$ and $270^\circ$,
respectively.
The double circle shows the origin.
}}
\Fglab{amplitude}
\end{figure}

Before closing this section, we give a useful relationship
between the $\nu_\alpha \to \nu_\beta$ transition and the
$\ov \nu_\alpha \to \ov \nu_\beta$ transition
which may be valid
in  the terrestrial LBL experiments where both the accelerator and 
the detectors are near the earth surface.
The oscillation of anti-neutrinos in matter
is given by the 
Schr{\" o}dinger equation:
\begin{eqnarray}
 i\frac{\partial}{\partial t}
\bma{c}
\ov{\nu}_e \\
\ov{\nu}_\mu \\
\ov{\nu}_\tau\\
\ema
=
\left[
\overline{H_0} + 
\bmaT
-a & 0 & 0 \\
0 & 0 & 0 \\
0 & 0 & 0
\emaT
\right]
\bma{c}
\ov{\nu}_e \\
\ov{\nu}_\mu \\
\ov{\nu}_\tau\\
\ema
=\overline{H}
\bma{c}
\ov{\nu}_e \\
\ov{\nu}_\mu \\
\ov{\nu}_\tau\\
\ema\,.
\eqlab{osc_nu}
\end{eqnarray}
The Hamiltonian in the vacuum is identical to the one governing 
the neutrino-flavor oscillation \eqref{H_0},
\begin{eqnarray}
\overline{H_0}=
\frac{1}{2E}U^\dag
\bmaT
0 & 0 & 0 \\
0 & \delta m^2_{21} & 0 \\
0 & 0 & \delta m^2_{31}
\emaT
U
=H^\dag_0=H_0\,.
\eqlab{osc_antinu}
\end{eqnarray}
By comparing the total Hamiltonians $H$ and $\overline{H}$;
\bseq
\bea
H&=&
\frac{\delta m_{13}^2}{2E}
\left[
\bmaT
\frac{2aE}{\delta m_{13}^2}+|U_{e 3}|^2 & U_{e 3} U_{\mu 3}^\ast
& U_{e 3} U_{\tau 3}^\ast\\
U_{\mu 3} U_{e 3}^\ast  & |U_{\mu 3}|^2  & U_{\mu 3} U_{\tau 3}^\ast \\
U_{\tau 3} U_{e 3}^\ast & U_{\tau 3} U_{\mu 3}^\ast & |U_{\tau 3}|^2  
\emaT
\right.
\nn \\
&&
\left.
~~~~~~~~~~~~~~~~~~~~~~~~~~~~~~~~~~~~~~~
+\frac{\delta m_{12}^2}{\delta m_{13}^2}
\bmaT
|U_{e 2}|^2            & U_{e 2} U_{\mu 2}^\ast  & U_{e 2} U_{\tau 2}^\ast \\
U_{\mu 2} U_{e 2}^\ast  & |U_{\mu 2}|^2  & U_{\mu 2} U_{\tau 2}^\ast \\
U_{\tau 2} U_{e 2}^\ast & U_{\tau 2} U_{\mu 2}^\ast & |U_{\tau 3}|^2  
\emaT
\right], 
\eqlab{hamiltonian1}
~~~~~~~
\\
\overline H &=&
\frac{\delta m_{13}^2}{2E}
\left[
\bmaT
-\frac{2aE}{\delta m_{13}^2}+|U_{e 3}|^2 & U_{e 3}^\ast U_{\mu 3}
& U_{e 3}^\ast U_{\tau 3}\\
U_{\mu 3}^\ast U_{e 3}  & |U_{\mu 3}|^2  & U_{\mu 3}^\ast U_{\tau 3} \\
U_{\tau 3}^\ast U_{e 3} & U_{\tau 3}^\ast U_{\mu 3} & |U_{\tau 3}|^2  
\emaT
\right.
\nn\\
&&
\left.
~~~~~~~~~~~~~~~~~~~~~~~~~~~~~~~~~~~~~~~
+\frac{\delta m_{12}^2}{\delta m_{13}^2}
\bmaT
|U_{e 2}|^2            & U_{e 2}^\ast U_{\mu 2}  & U_{e 2}^\ast U_{\tau 2} \\
U_{\mu 2}^\ast U_{e 2}  & |U_{\mu 2}|^2  & U_{\mu 2}^\ast U_{\tau 2} \\
U_{\tau 2}^\ast U_{e 2} & U_{\tau 2}^\ast U_{\mu 2} & |U_{\tau 3}|^2  
\emaT
\right],
\eqlab{hamiltonian2}
~~~~~~~
\eea
\eseq
we find
\bea
-\overline{H}^\ast(\delta m^2_{12},\delta m^2_{13})
 =H(-\delta m^2_{12},-\delta m^2_{13}).
\eea
Because the Hamiltonian $-\overline H^\ast$ governs the oscillation in the
reversed time
direction, we find that the following identities hold
\bseq
\eqlab{anti-P}
\begin{eqnarray}
P_{\ov{\nu}_\alpha \to \ov{\nu}_\beta}^{~\rm I}
 &=&
P_{\nu_\alpha \to \nu_\beta}^{~\rm IV}, \\
P_{\ov{\nu}_\alpha \to \ov{\nu}_\beta}^{~\rm II}
 &=&
P_{\nu_\alpha \to \nu_\beta}^{~\rm III}, \\
P_{\ov{\nu}_\alpha \to \ov{\nu}_\beta}^{~\rm III}
 &=&
P_{\nu_\alpha \to \nu_\beta}^{~\rm II}, \\
P_{\ov{\nu}_\alpha \to \ov{\nu}_\beta}^{~\rm IV}
 &=&
P_{\nu_\alpha \to \nu_\beta}^{~\rm I},
\end{eqnarray}
\eseq
if the matter density along the baseline is symmetric under the reversal of
the beam direction, $\it{i.e}$ under the exchange of the injector and the
detector.
 This condition is met approximately for all terrestrial LBL experiments where
both the accelerator and the detector are on or near the earth surface.

In the following, we therefore give results for all the four hierarchy
patterns
but only for the neutrino beam.
Oscillation probabilities for anti-neutrino beams are then obtained according
to the rule \eqsref{anti-P}, while the CC and 
neutral-current (NC) event rates are obtained after
multiplying the ratio of the anti-neutrino and neutrino cross sections on the
target.

\section{Narrow-Band Neutrino Beams with HIPA}
\clean
\def\Journal#1#2#3#4{{#1}{\bf#2}, #3 (#4)}
\def\ZPC{{\em Z. Phys.} C}
\def\PRD{{\em Phys. } C}
\def\PLB{{\em Z. Phys.} C}
\def\to{\rightarrow}
\def\50ps{\mbox{50 GeV-PS}}
\def\12ps{\mbox{12 GeV-PS}}
\def\nue{\nu_{e}}
\def\num{\nu_{\mu}}
\def\nut{\nu_{\tau}}
\def\dm{\Delta m^2}
\def\dme{\Delta m_{12}^2}
\def\dmm{\Delta m_{23}^2}
\def\dmt{\Delta m_{13}^2}
\def\sint{\sin^22 \theta}
\def\sine{\sin^22 \theta_{e}}
\def\sinm{\sin^22 \theta_{\mu}}
\def\ct13{\cos^22 \theta_{13}}
\def\st23{\sin^22 \theta_{23}}
\def\stt13{\sin^22 \theta_{13}}
\def\c13{\cos^4 \theta_{13}}
\def\s23{\sin^2 \theta_{23}}
\subsection{The KEK-JAERI joint project}
\hspace*{\parindent}
The KEK-JAERI joint project on HIPA 
is a proton accelerator
complex and associated experimental facilities which will be
constructed in the site of JAERI,
Tokai-village, 60 km north-east of KEK.
The project consists of 400 MeV Linac, 3 GeV and
50 GeV synchrotrons.
The design parameters of 50 GeV machine are listed
in Table \ref{PScomp} together with some other proton
machines for LBL experiments. The intensity is 3.3$\times
10^{14}$ protons/pulse (ppp) and the repetition rate is
0.275 Hz. The power reaches 0.75 MW which is 2 orders
of magnitudes
higher than the KEK 12 GeV proton synchrotron (PS).
The accelerators in the facility will be
in the power frontier in the  world.
The facility is approved by the Japanese government in
December, 2000 and the construction will take 6 years from 2001.

In the following discussion, $10^{21}$ protons on target
(POT) is adopted as a typical 1 year
operation. This corresponds to about 100 days of operation
with the design intensity.

\begin{table}[htbp]
\caption{Comparison of accelerator parameters used for LBL projects}
\label{PScomp}
\begin{center}
\begin{tabular}{lcccc}
\hline\hline
        & Energy & Intensity & Rep. rate & Power \\
        & (GeV)  & (10$^{12}$ppp) & (Hz) & (MW) \\
\hline
HIPA     & 50  & 330 & 0.275 & 0.75 \\
NuMI    & 120 &  40 & 0.53  & 0.41 \\
KEK-PS  & 12  &   6 & 0.45  & 0.0052 \\
\hline\hline
\end{tabular}
\end{center}
\end{table}

\subsection{Neutrino Beams for LBL experiment 
        between HIPA and SK}
\hspace*{\parindent}
As a first stage neutrino experiment at this new facility, long baseline
experiment from HIPA to SK has been
planned and discussed seriously by the JHF neutrino working
group \cite{H2SK}.
Before going into the description of higher energy beam for
VLBL experiment, we briefly introduce the beam for the LBL
experiment.

Major purposes of the HIPA-to-SK experiment are 1) precise measurement of
oscillation parameters in $\nu_\mu$ disappearance, 2) to
discover $\nu_e$ appearance.
The principles of the experiment are
\begin{itemize}
\item Use of low-energy narrow band beam (NBB) whose peak energy
is tuned at the oscillation maximum.
Since the distance between
HIPA and SK is 295 km, the peak energy should be around $E_\nu = 0.7\sim
1.2$ GeV for the region of $\delta m_{_{\rm ATM}}^2$
allowed by the SK observation \cite{atm_mutau}.

\item Neutrino-energy is kinematically reconstructed event-by-event
from the measured lepton momentum 
by assuming the charged-current quasi-elastic (CCqe) scattering.
Inelastic scattering with invisible secondary hadrons mimic
the CCqe interactions and smears the $E_\nu$ measurement.
Below $E_\nu \sim 1$ GeV, $\nu_\mu$ interaction is
dominated by the CCqe interaction. So the low-energy beam
with small high-energy tail is favorable for this method.
\end{itemize}

Currently, three beam options are being considered, namely
the wide-band beam (WBB), the NBB and the off-axis
beam (OAB);
\begin{description}
\item{WBB:}
Secondary charged pions from production target
are focused by two electromagnetic horns \cite{K2Khorns}.
Since the momentum and angular acceptances of the horns are wide,
resulting neutrino spectrum are also wide.
The advantage of the WBB is the wide sensitivity in $\delta
m^2$. But backgrounds from inelastic scattering of neutrinos
from high-energy neutrinos limit the precision of the oscillation-parameter
measurements.
\item{NBB:}
Two electromagnetic horns have their axis
displaced by about
10$^\circ$, and an dipole magnet is
placed between them to select pion momentum. 
Resulting spectrum has a sharp peak and much less high energy
tail than WBB.
\item{OAB:}
The arrangement of beam optics is almost the same as WBB,
{\it i.e.} coaxially aligned two horns.
The axis of OAB is intensionally displaced from the SK direction
by a few degrees. Pions with various momenta but with a finite angle
from the SK direction contribute to a narrow energy region in the $\nu_\mu$
spectrum \cite{BNL-E889}. The OAB can produce a factor of 2 or 3
more intense beam than NBB. But unwanted high-energy component is 
larger than NBB.
\end{description}
The length of the decay pipe is chosen to be relatively short,
80 m for all the configurations.
This is because 1) high-energy neutrinos do not improve
the measurements,
2) longer pipe costs very much due to the heavy shielding
required by the Japanese radiation regulation.
In \cite{H2SK}, the WBB is used only 
in the early stage of the project 
in order to pin down $\delta m_{_{\rm ATM}}^2$ at about $\sim 10 \%$
accuracy\footnote{{
Very recently the JHF-SK neutrino working group has modified its
strategy and the beam configuration.
In the most recent plan, the OAB will be adapted for the
LBL experiment and the decay pipe length will be $130$ m.
For more details see Ref.\cite{KEKTC5}.}}.
Typical expected spectra of those options are plotted in
\Fgref{bmfig:SKspec} and the flux and number of interactions
are summarized in \Tbref{bmtbl:SKnint}.

\subsection{High Energy Narrow Band Beam for VLBL
Experiments}
\hspace*{\parindent}
In order to
explore the physics potential of the VLBL
experiment with HIPA, we need to estimate the neutrino flux whose
spectrum has a peak at higher energies.
 We study the profile of such beams at 
distances of 1,200 km and 2,100 km by using Monte
Carlo simulation.
 In this subsection, we describe the beam in detail.

First, we chose the decay pipe length
to be 350 m.
For the baseline length currently under consideration, 1,200
km to Seoul and 2,100 km to Beijing, oscillation maximum
lies at $E_\nu = 2\sim 4$ GeV and $4\sim 7$ GeV,
respectively, for $\delta m^2_{23}=(2\sim 4) \times 10^{-3}$ eV$^2$.
In order to make 5 GeV neutrinos for example, we need
pions of momentum about 10 GeV.
The 10 GeV pions run about 560 m during
their life. Therefore we need a long decay pipe of several 100
m for efficient neutrino production.
Considering the site boundary of JAERI and layout of accelerators,
maximum decay pipe length is about 350 m.
\begin{figure}[ht]
\centerline{\epsfig{file=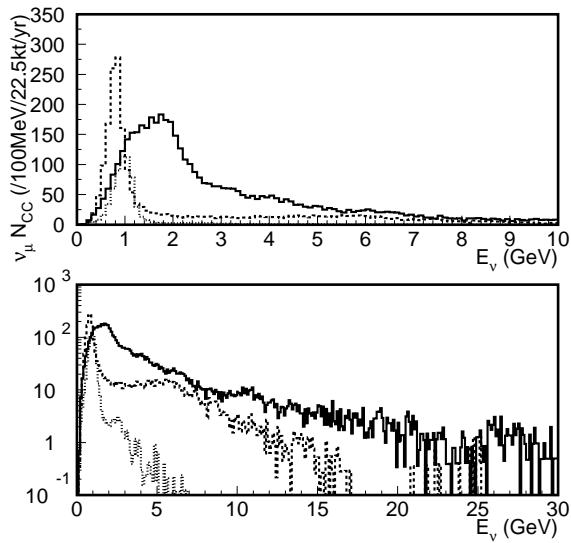,width=8cm}}
\caption{Typical spectra of the $\nu_\mu$ CC interactions in the absence of
neutrino oscillation in the HIPA-to-SK LBL experiment. 
The solid, dashed and dotted
histograms are spectra for WBB, OAB(2$^\circ$) and NBB
($\langle p_{\pi}\rangle=2$ GeV),
respectively. 
}
\Fglab{bmfig:SKspec}
\end{figure}
\begin{table}
\begin{center}
\begin{tabular}{l|rrrr}
\hline\hline
Beam & $\nu_\mu \Phi$ &  $\nu_\mu N_{\rm tot}$ &  $\nu_\mu N_{\rm CC}$ &  
$\nu_e N_{\rm tot}$ \\
\hline
NBB($\langle p_{\pi}\rangle=2$ GeV) & 7.0 &  870 &  620 & 6.8 \\
OAB(2$^\circ$) & 19  & 3100 & 2200 & 60 \\
WBB   & 26  & 7000 & 5200 & 78 \\
\hline\hline
\end{tabular}
\end{center}
\caption{Summary of beam simulation of the HIPA-to-SK LBL
experiment. The neutrino flux $\Phi$
at SK is in the unit of 10$^6$/cm$^2$/year,
$N_{\rm tot}$ and $N_{\rm CC}$ are the
number of total and CC interactions, respectively 
in SK's fiducial volume of 22.5kton for
1 year (10$^{21}$POT) in the absence of neutrino oscillation.
}
\tblab{bmtbl:SKnint}
\end{table}

Secondly, we adopt the NBB configuration.
Use of WBB at these high energies has at least two
disadvantages;
\begin{enumerate}
\item
The reconstruction of neutrino energy is difficult.
At multi GeV region, $\nu_\mu$ interactions are dominated by 
deep-inelastic scattering with multi-pion production.
It is difficult for a water $\check {\rm C}$erenkov detector to make
such measurements.
It is a non-trivial exercise to construct a 100 kton-level detector
of a reasonable cost
which has the capability of reconstructing the neutrino energy at several 
GeV range.
See \eg a proposal in ref \cite{BAND}.

\item 
The construction of the beam line costs very much.
In the WBB configuration, the proton beam which passes
through the production target goes through the decay pipe
all the way down. The intensity is
still of the order of $10^{14}$ ppp even beyond the target.
In order to shield the extremely high radiation, we need a
considerable amount of shielding around the decay
pipe. Constructing a decay pipe of several 100 m with heavy
shielding is unrealistic.
\end{enumerate}
The OAB also has the second disadvantage.
Therefore we choose high energy NBB for the present study,
since it does not suffer from the above disadvantages.
With an ideal NBB, neutrino energy reconstruction is not
necessarily be done by the detector.
It is possible to design a NBB beam line where the 50 GeV proton beam
does not enter the decay pipe.
Simulation of high energy WBB is done only for comparison.

For the purpose of the focusing secondary pions, we adopt the
quadrupole (Q) magnets instead of horns.
In general, focusing by Q magnets has smaller
angular and momentum acceptance than the horn focusing.
The reasons why we choose the Q focusing are
\begin{itemize}
\item Q focusing gives narrower neutrino-energy spectrum
\item High energy pions of $\sim$ 10 GeV are emitted at 
smaller angles and hence reasonable acceptance for those pions
can be obtained by the Q optics.
\item The Q-magnet can be operated at low DC current of several kA.
Compared with the horn magnets which require pulsed operation
with a few 100 kA, much more stable operation can be expected.
\end{itemize}

We used GEANT \cite{GEANT} for the beam-line simulation to
estimate the neutrino flux. 
A target, Q magnets, a dipole magnet, and a decay pipe are
put into the geometry.
The target is Cu rod of 1 cm diameter and 30 cm length.
The length corresponds to about 2 nuclear interaction length.
GCALOR code \cite{GCALOR} is used for hadron production in the
target.
Every secondary particles produced in the target are tracked.
The beam line optics assumed for the present study is drawn
in \Fgref{bmfig:optics}.
\begin{figure}[htbp]
\centerline{\epsfig{file=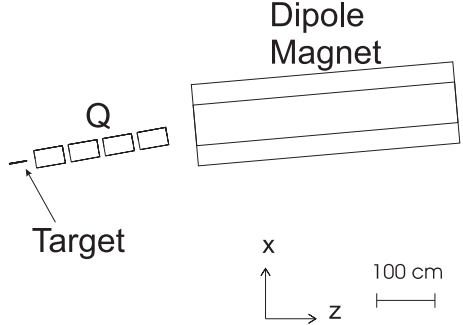,width=6cm}}
\caption{Beam line optics for high energy narrow band beam}
\Fglab{bmfig:optics}
\end{figure}
The secondary pions from the target are focused by the following 
4 Q magnets and bent by 10$^\circ$ by a dipole magnet.
The optics is not fully optimized. Several tens of $\%$
increase in flux could be expected by tuning the size or
field  of the Q and bending magnets, the position of the target 
\etc

In \Fgref{bmfig:numuspec}, some typical spectra obtained
by the MC simulations are plotted. Spectra with a narrow peak
structure are generated.
The peak has a sharper edge in the high-energy side than in the 
low-energy side. The trailing edge in the low-energy side comes
from pions with finite angle.
We observe a small secondary peak at an energy about twice the
primary peak position. The second peak comes from Kaon decays.
In the left-hand-side figure,
we show the flux of the neutrino at 2,100 km away from
HIPA in units of $10^4$/400MeV/cm$^2$/year, where $10^{21}$ POT is assumed
for one-year operation.
Three types of the NBB where peak energy is at about 3, 5, 8 GeV, and
the high-energy WBB spectra are shown.
The right-side figures show the expected number of $\nu_\mu$ CC events
par year, in the absence of neutrino oscillation, for a 100 kton detector.
The cross section has been obtained by assuming that the target detector
made of water.

\begin{figure}[tbp]
\centerline{
\epsfig{file=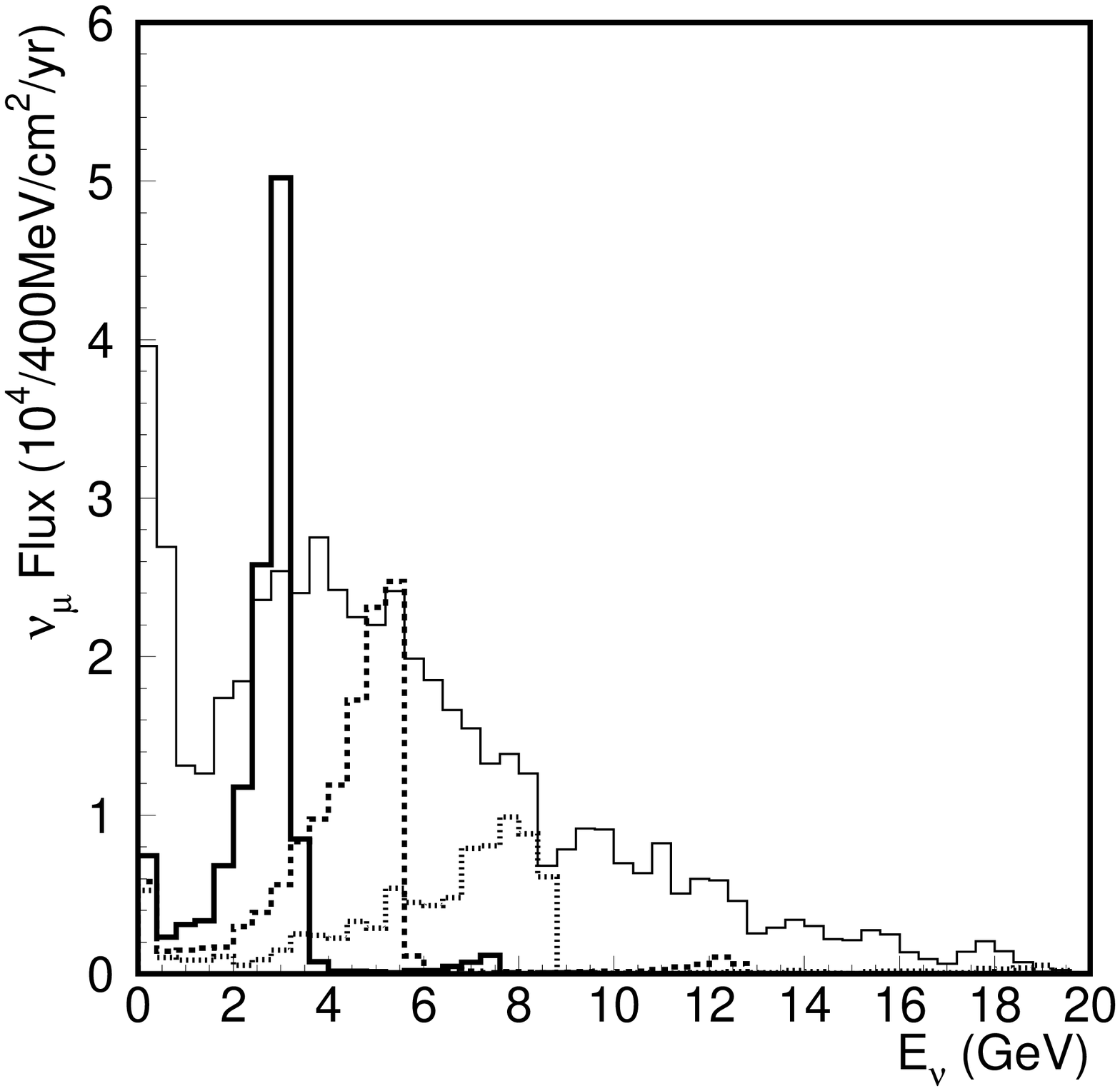,width=7cm}
\epsfig{file=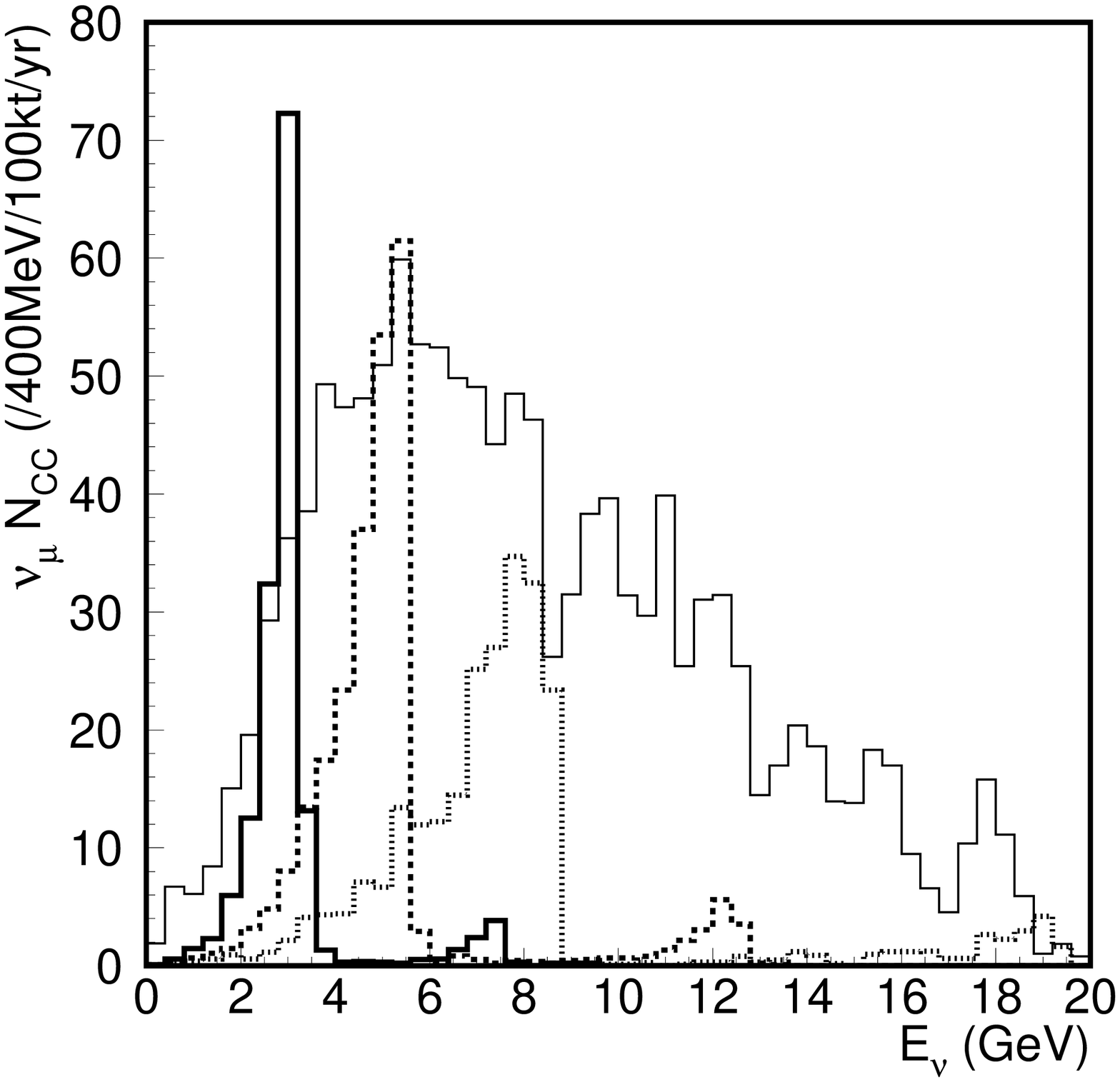,width=7cm}
}
\caption{Typical spectra of narrow band beams.
The left figure shows the $\nu_\mu$ flux and the right one gives the
number of
$\nu_\mu$ charged current interactions in the absence of oscillation
at $L=2,100$ km.
The solid, dashed and dotted histograms are 3, 5 and 8 GeV
NBB, respectively.
The thin solid histograms show the WBB spectra for comparison.}
\Fglab{bmfig:numuspec}
\end{figure}
\begin{figure}[htbp]
\centerline{
\epsfig{file=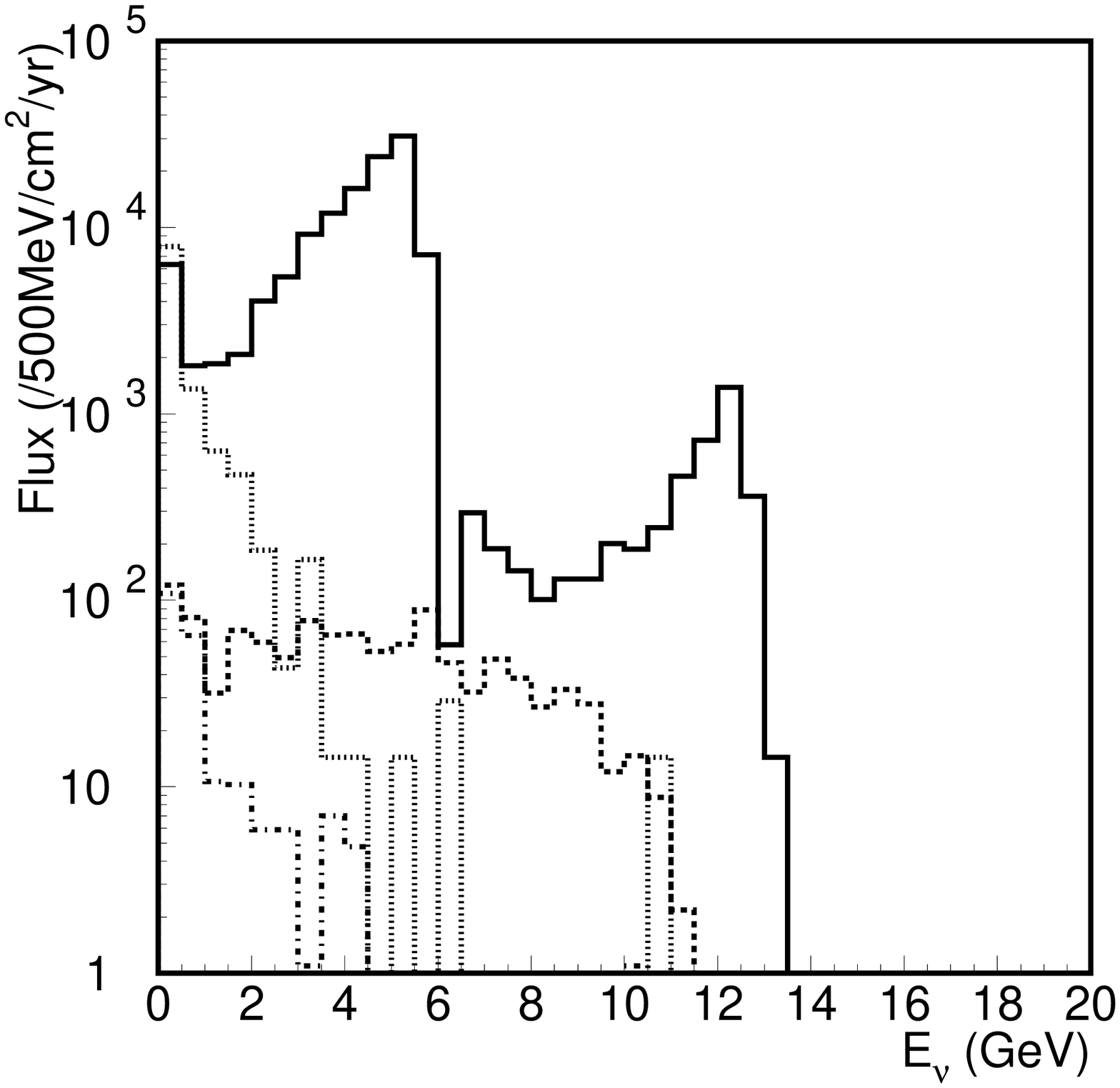,width=7cm}
\epsfig{file=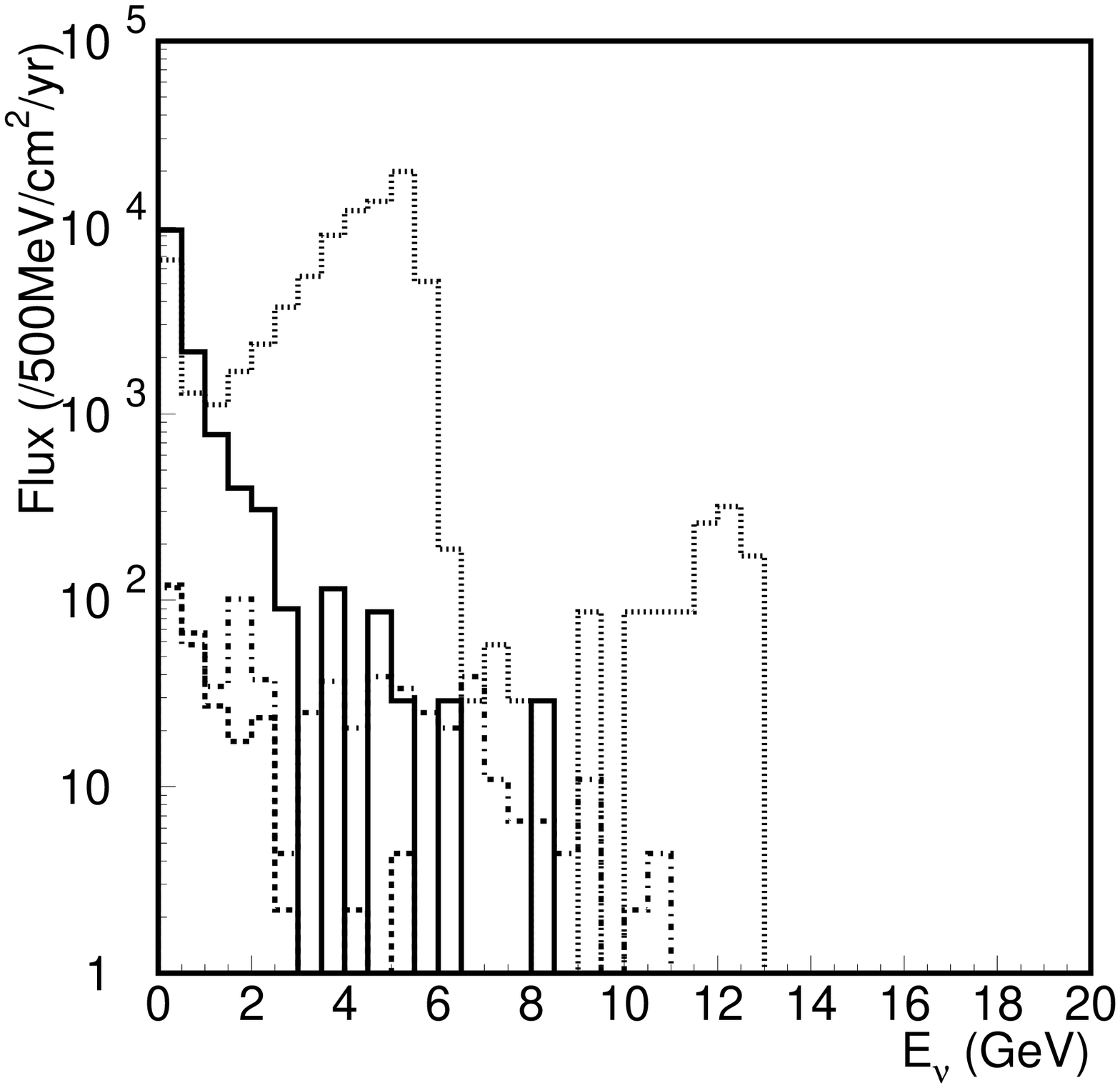,width=7cm}
}
\caption{Composition of neutrino species for a typical NBB. 
Solid, dashed, dotted and dot-dashed 
histograms correspond to $\nu_\mu$, $\nu_e$, $\ov{\nu}_\mu$ 
and $\ov{\nu}_e$, respectively.
The left and right figures shows 5 GeV $\nu_\mu$ and $\ov{\nu}_\mu$
beam, respectively.}
\Fglab{bmfig:nucomp}
\end{figure}

Spectra for each neutrino species in the 5 GeV $\nu_\mu$ and
$\ov{\nu}_\mu$ beams are shown in \Fgref{bmfig:nucomp}.
The flux ratio of $\nu_e$($\ov{\nu}_e$) to
$\nu_\mu (\ov{\nu}_\mu$) is $0.9 \%$ ($0.8 \%$) in total,
and $0.2 \%$ ($0.1 \%$) at the peak energy for $\nu_\mu$
($\ov{\nu}_\mu$) beam.
In the left figure, we show the $\nu_\nu, \nu_e, \ov \nu_\mu$ and
$\ov \nu_e$ spectrum of the NBB $\nu_\mu$ beam, while in the right figure
the corresponding spectrum are shown for the NBB $\ov \nu_\mu$ beam.
The number of wrong-sign $\ov{\nu}_\mu$ ($\nu_\mu$) CC
interactions in $\nu_\mu$ ($\ov{\nu}_\mu$) beam is
about $0.4 \%$ ($4 \%$) of the right-sign interactions.
Although the flux of the $\nu_\mu$ beam and the $\ov \nu_\mu$ beam are
almost the same, as well as the fraction of contaminated neutrino flux,
the $\ov \nu_\mu$ beam suffers from 10 times higher wrong-sign events
because of a factor of 3 smaller $\ov \nu_\mu$ CC interactions than
the $\nu_\mu$ CC interactions off water target at these energies.

\begin{table}[thbp]
\footnotesize
\caption{Expected number of interactions of $\nu_\mu$ NBB.
Assumed parameters are hierarchy I,
$\delta m^2_{_{\rm ATM}} =  3.5\times 10^{-3}$ eV$^2$, 
$\sin^22\theta_{_{\rm ATM}}  = 1.0$,
$\delta m^2_{_{\rm SOL}} =  10 \times 10^{-5}$ eV$^2$,  
$\sin^22\theta_{_{\rm SOL}}= 0.8$,
$\sin^22\theta_{_{\rm CHOOZ}} = 0.06$,
$\delta_{_{\rm MNS}}=0^\circ$ and $\rho=3$ g/cm$^3$. 
First and second lines for each set of baseline length 
and the peak energy
indicate the number of interactions without and with
oscillation, respectively.}
\tblab{bmtbl:nuflux}
\begin{center}
\begin{tabular}{l|l|r|r|r|r|r}
\hline\hline
  $L$    &  $E_{\rm peak}$     &
$\nu_\mu$ CC &
$\ov{\nu}_\mu$ CC &
$\nu_e$ CC &
$\ov{\nu}_e$ CC &
$N_{\rm NC}$ \\
\hline
300km  &  3GeV & 7495. &  43.0     &  55.0   &   0.90    & 2540.9 \\
       &       & 5903. &  22.0     & 105.0   &   1.20    & 2540.9 \\
\hline
       &  6GeV & 13321.&    44.0   &  82.0   &   1.90    & 4457.4 \\
       &       & 12400.&    21.0   & 110.0   &   1.70    & 4457.4 \\
\hline\hline
700km  &  3GeV & 1376. &  7.9     &  10.1  &   0.17    &   466.7 \\
       &       &  382. &  3.8     &  43.9  &   0.28    &   466.7 \\
\hline
       &  6GeV &  2446.&  8.1     &  15.0   &   0.35    & 818.7 \\
       &       &  1699.&  3.5     &  40.6   &   0.36    & 818.7 \\
\hline\hline
1,200km &  3GeV &   468.&     2.7  &    3.4 &    0.05    & 158.8 \\
       &       &    85.&     0.8  &   18.1 &    0.13    & 158.8 \\
\hline
       &  6GeV &   833.&     2.7  &    5.1 &     0.12   & 278.6 \\
       &       &   297.&     1.1  &   25.3 &     0.13   & 278.6 \\
\hline\hline
2,100km &  3GeV &   153.&  0.9     &    1.1 &    0.02    &  51.9 \\
       &       &   119.&  0.5     &    2.4 &    0.05    &  51.9 \\
\hline
       &  6GeV &   272.&  0.9     &    1.6 &    0.04    & 91.0 \\
       &       &    47.&  0.4     &   13.2 &    0.06    & 91.0 \\
\hline\hline
\end{tabular}
\end{center}
\end{table}

\begin{table}[hbtp]
\footnotesize
\caption{Expected number of interactions of $\ov{\nu}_\mu$ NBB.
Assumed parameters are the same as in \Tbref{bmtbl:nuflux}.}
\tblab{bmtbl:nubflux}
\begin{center}
\begin{tabular}{l|l|r|r|r|r|r}
\hline\hline
$L$       &   $E_{{\rm peak}}^{}$    &
$\nu_\mu$ CC &
$\ov{\nu}_\mu$ CC &
$\nu_e$ CC &
$\ov{\nu}_e$ CC &
$N_{\rm NC}$ \\
\hline
300km  &  3GeV &  160.0 & 2871.0    &   4.0   &   12.7   & 1184.7 \\
       &       &   66.0 & 2261.0    &   4.9  &   33.6   & 1184.7 \\
\hline
       &  6GeV &   141.0& 4076.0   &   6.7  &   13.3   & 1584.8 \\
       &       &    77.0& 3793.6   &   5.4  &   22.4   & 1584.8 \\
\hline\hline
700km  &  3GeV &   29.4&  527.0    &   0.7  &   2.4    & 217.6 \\
       &       &   13.9&  136.0    &   1.1  &   16.6    & 217.6 \\
\hline
       &  6GeV &   25.9& 748.6    &   1.2   &   2.5    & 291.1 \\
       &       &   11.5& 518.5    &   1.2   &  10.8     & 291.1 \\
\hline\hline
1,200km &  3GeV &    0.0&   179.0   &    0.2 &    0.8    &  74.0 \\
       &       &    3.5&    28.0   &    0.5 &    7.0    &  74.0 \\
\hline
       &  6GeV &    8.8&   254.7  &   0.4  &     0.8    &  99.1 \\
       &       &    3.4&    88.1  &   0.5  &    7.5     &  99.1 \\
\hline\hline
2,100km &  3GeV &    3.2& 58.6     &    0.1 &    0.3    &  24.2 \\
       &       &    1.9& 47.6     &    0.2 &    0.8    &  24.2 \\
\hline
       &  6GeV &    2.8&  83.1    &    0.1&    0.3    & 32.3 \\
       &       &    1.5&  12.9    &    0.2&    4.1    & 32.3 \\
\hline\hline
\end{tabular}
\end{center}
\end{table}

The results of the simulations are summarized in 
\Tbref{bmtbl:nuflux} and \Tbref{bmtbl:nubflux}.
In \Tbref{bmtbl:nuflux}, 
we show the expected number of CC and NC events for the 
3 GeV and 6 GeV $\nu_\mu$ NBB's for 100 kton$\cdot$year (10$^{21}$POT) at
four typical distances, 300 km, 700km, 1,200 km, and 2,100km from HIPA.
The upper numbers in each row and column show the numbers of 
events without oscillations, while the lower numbers are calculated by using
the three neutrino model for the following parameters :
\bea
&&\left(
~~\delta m^2_{_{\rm ATM}},~~~
~~\delta m^2_{_{\rm SOL}},~~~
\sin^22\theta_{_{\rm ATM}}, 
\sin^22\theta_{_{\rm SOL}},
\sin^22\theta_{_{\rm CHOOZ}},
\delta_{_{\rm MNS}}
\right)
\nn \\
&=&\left(
3.5\times10^{-3}, 
1.0\times10^{-4},
~~1.0,~~~~~~
~~0.8,~~~~~~~~
0.06,~~~~~~~~~~
0^\circ~~~
\right)
\eea
with the neutrino mass hierarchy I and for a constant matter density of
$\rho=3$ g/cm$^3$.
Because all the three neutrinos have identical NC interactions, the two
numbers are identical in $N_{\rm NC}$ column.
All the upper numbers simply follow the $1/L^2$ rule of the flux at
a distance $L$.

In Table 5, we show the corresponding numbers for the $\ov \nu_\mu$ NBB's.
The number of expected events are about a factor of 3 smaller than the 
corresponding ones in Table 4 because of the smaller CC and NC interactions
of $\ov \nu_\mu$ off nucleus target.

Details of all the NBB's generated for this study are available from 
\cite{TKHP}.

\subsection{Parameterization of the high-energy NBB}

In the numerical studies of the next section, we make the following 
parameterization of the $\nu_\mu^{}$ NBB with a single peak at 
$E_\nu^{}=E_{\rm peak}^{}$: 
\bea
MN_{A}\Phi(E_\nu) \sigma^{\rm CC}_\mu(E_\nu)
& =& 
f(E_{\rm peak})\left( \frac{E_\nu}{E_{\rm peak}}\right)^{b(E_{\rm peak})-1}
\left(
\dfrac{\sigma_\mu^{\rm CC}(E_\nu)}{\sigma_e^{\rm CC}(E_\nu)}
\right)\,,
\eqlab{def_flux}
\eea
where $E_{\nu}^{}$ is running from 0 to $E_{\rm peak}^{}$,
and $f(E_{\rm peak})$ and $b(E_{\rm peak})$ are parameterized as
\bseq
\bea
f(E_{\rm peak})&=&3.3E_{\rm peak}^2-76.8E_{\rm peak}+520\,,  \\
b(E_{\rm peak})&=&20.3E_{\rm peak}^{-1.4}+2.8\,,
\eea
\eseq
where $E_{\rm peak}^{}$ is measured as unit of GeV.
$M=100$ kton stands for the mass of the detector,
$N_A=6.017\times 10^{23}$ is the Avogadro number,
$\Phi(E_\nu)$ is the flux (in units of /GeV/cm$^2$/10$^{21}$POT)
at $L=2,100$ km.
The
$\sigma_\mu^{\rm CC}$ and $\sigma_e^{\rm CC}$
are, respectively, the $\nu_\mu$ and $\nu_e$ CC cross sections
per nucleon off water target \cite{cross},
and their ratio is approximately given by
\bea
\sigma_\mu^{\rm CC}(E_\nu)/\sigma_e^{\rm CC}(E_\nu) 
&\simeq& \l\{ 
\begin{array}{l}
1.0-0.056E_\nu^{-0.48} ~~~~ (0.7 \lsim E_\nu ) \\ 
0.83+0.16E_\nu ~~~~ (0.3 \lsim E_\nu  \lsim 0.7) \\
0.879(E_\nu-m_\mu)/(0.3-m_\mu) ~~~~(m_\mu \lsim E_\nu \lsim 0.3) 
\end{array}
\right.
\eqlab{def_cross}
\eea
where $m_\mu = 0.11$ GeV represents the muon mass.
This parameterization allows us to study the effects of changing
the peak energy of the NBB continuously.
We show in \Fgref{beamfit} our parameterizations of the NBB
neutrino spectra by thick solid curves for several peak energies.
For comparison, the corresponding original NBB
spectra are shown by histograms.
The parameterization reproduces well the main
part of the $\nu_\mu$ NBB's.
Because it does not account for the secondary high-energy peak from
$K^+\to \mu^+\nu_\mu$ decays (see \Fgref{bmfig:numuspec}), we check that
our main conclusions are not affected by those details (especially
the background from $\nu_\tau$ CC events).

We have not made parameterizations for the secondary
($\bar \nu_\mu, \nu_e, \bar \nu_e$) beams.
In the following analysis we use the MC generated secondary
beams at discrete energies ($E_{\rm peak}$) \cite{TKHP}
and
make interpolation for the needed $E_{\rm peak}$ values.
Fluxes at different distances and for different species are obtained
easily by multiplying the (2100 km/$L$)$^2$ flux factor and the ratio of 
the cross section at a given $E_\nu^{}$.
\begin{figure}
\begin{center}
{\scalebox{0.8}{\includegraphics{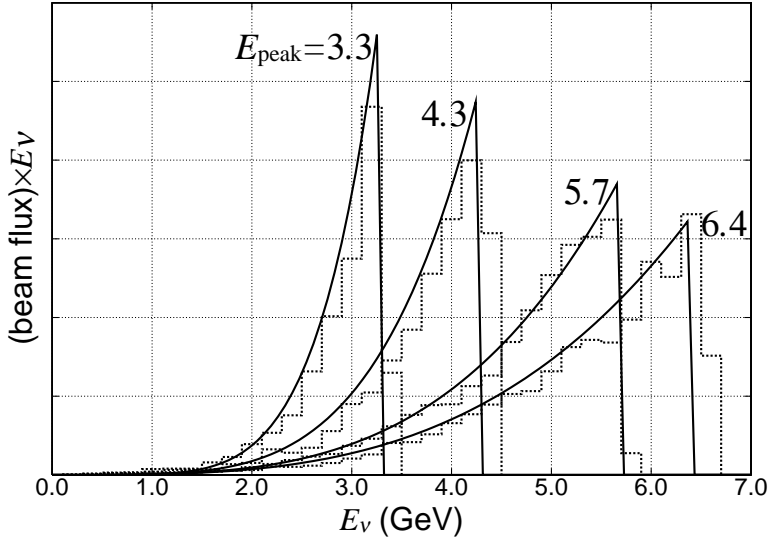}}}
\caption{The NBB's from HIPA (histograms), 
and our parameterization (solid lines).
The horizontal axis is the neutrino energy $(E_\nu^{})$
and the vertical axis is the beam flux times $E_\nu^{}$.
}
\Fglab{beamfit}
\end{center}
\end{figure}

\section{Results}
\clean
In this section we present results of our numerical studies on physics 
potential of VLBL experiments by using the NBB's from HIPA.

First, we present our basic strategy of the analysis
and explain our simplified
treatments of signals and backgrounds, and those of statistical and 
systematic errors.
In the next subsection, we give our reference predictions for the results
that may be obtained from the LBL experiment with HIPA and SK
$(L=295$km$)$.
In the latter two subsections, we give results of $L=2,100$ km
and $L=1,200$ km, respectively.

\subsection{Signals, backgrounds and systematic errors}

In order to explore the physics potential of a VLBL experiment with HIPA
at several base-line lengths, we make the following simple treatments in
estimating the signals and the backgrounds of a future experiment.
For a detector we envisage :
\begin{itemize}
 \item
 A 100 kton-level water-$\check {\rm C}$erenkov detector which has a capability
of distinguish $e^\pm$ CC events from $\mu^\pm$ CC events, but does not 
distinguish their charges.
 \item
 We do not require capability of the detector to reconstruct the neutrino 
energy. 
\end{itemize}
Although water-$\check{\rm C}$erenkov detectors
have the capability of measuring
the three-momentum of the produced $\mu^\pm$ and $e^\pm$ as well as a part
of hadronic activities, we do not make use of those information in our 
simplified analysis.
Instead we use only the total numbers of the produced $\mu^\pm$ and $e^\pm$ 
events from a NBB with a given peak energy.
For each base-line length $L$, we study the impacts of splitting the assumed
total experiment exposure of 1000 kton$\cdot$year (with $10^{21}$ POT/year)
into equi-partitioned runs of NBB's at several peak energies.
We find that the use of two different-energy NBB's improves the physics
resolving power of the experiment significantly, but we have 
not found further improvements by splitting the experiment into more than
two NBB's.
We therefore choose two appropriate NBB's at each $L$, whose peak energies are
chosen to make physics outputs (such as sensitivity to the neutrino mass
hierarchy, $\sin^22\theta_{_{\rm CHOOZ}}$ and $\delta_{_{\rm MNS}}$ angles)
significant.
Optimum choice of NBB's should depend on the model parameters
\bea
\left(~
\delta m^2_{_{\rm ATM}}\,,~
\delta m^2_{_{\rm SOL}}\,,~
\sin^22\theta_{_{\rm ATM}}\,,~ 
\sin^22\theta_{_{\rm SOL}}\,,~
\sin^22\theta_{_{\rm CHOOZ}}\,,~
\delta_{_{\rm MNS}}~
\right)\,,
\eea
especially
on $\delta m^2_{_{\rm ATM}}$ and $\sin^22\theta_{_{\rm ATM}}$, which will
be measured more accurately 
by K2K \cite{K2K}, MINOS \cite{MINOS} and by HIPA-to-SK \cite{H2SK}
in the future.
All our major findings will not be affected by
such details as long as appropriate NBB's are chosen according
to the data available at the time of the VLBL experiment.

The signals of our analysis are the numbers of $\nu_\mu$ CC events
and those of $\nu_e$ CC events from the $\nu_\mu$ beam.
They are calculated as
\bea
N(\mu, E_{\rm peak},L)=MN_A\int_0^{E_{\rm peak}} dE_\nu \Phi(E_{\rm peak}) 
\sigma^{\rm CC}_\mu
P_{\nu_\mu \to \nu_\mu} \eqlab{event_m}\,,\\
N(e, E_{\rm peak},L)=MN_A\int_0^{E_{\rm peak}} dE_\nu \Phi(E_{\rm peak}) 
\sigma^{\rm CC}_e
P_{\nu_\mu \to \nu_e}\,,
\eqlab{event_e}
\eea 
where the flux at a distance $L$ is calculated from the parameterization at
$L=2,100$ km \eqref{def_flux}
by multiplying the scale factor (2,100 km/$L$)$^2$.
The cross sections are obtained by assuming a pure water target.
At low $E_\nu$, the ratio of $\nu_e$ and $\nu_\mu$ CC cross sections is 
significant by different from unity, see \eqref{def_cross}.
Because of the vanishing of the NBB flux at low energies, our results are 
insensitive to the lower edge of the $E_\nu$ integration region.
The probabilities $P_{\nu_\mu \to \nu_\mu}$ and $P_{\nu_\mu \to \nu_e}$
are calculated for the following model parameters ;
\bseq
\eqlab{modelpara}
\bea
\sin^22\theta_{_{\rm ATM}}&=&1.0\,,~~~~~
\delta m^2_{_{\rm ATM}}=3.5\times10^{-3} ~{\rm eV}^2\,, \\
\sin^22\theta_{_{\rm SOL}}&=&0.8\,, ~~~~~
\delta m^2_{_{\rm SOL}}=10\times10^{-5} ~{\rm eV}^2\,, \\
\sin^22\theta_{_{\rm CHOOZ}}&=&0.00\,,
~0.02\,, ~0.04\,, ~0.06\,, ~0.08\,, ~0.10\,, \\
\delta_{_{\rm MNS}}&=&0^\circ\,, 90^\circ\,, 180^\circ\,, 270^\circ\,,
\eea
\eseq
for the neutrino mass hierarchy I
\bea
\delta m^2_{13}=\delta m^2_{_{\rm ATM}} > 0\,, 
~~~ 
\delta m^2_{12}=\delta m^2_{_{\rm SOL}} > 0\,,
~~~~~
({\rm hierarchy ~I~})\,,
\eea
and for a constant matter density
\bea
\rho=3 ~{\rm g/cm}^3,
\eqlab{rho}
\eea
throughout the base-line.
We show in \Fgref{Pro_fl} the oscillation probabilities calculated 
for the above parameters, (at $\sin^22\theta_{_{\rm CHOOZ}}=0.1$
and $ \delta_{_{\rm MNS}}=270^\circ$ ) at three base-line lengths,
$L=295$ km (SK), $L=1,200$ km and $L=2,100$ km.
The NBB flux ($\times E_\nu$) chosen for our analysis are overlayed
in each figure.
We use the result at SK ($L=295$ km) expected for the low-energy
NBB with $\langle p_\pi\rangle = 2$ GeV as a reference.

\begin{figure}
\begin{center}
{\scalebox{0.65}{\includegraphics{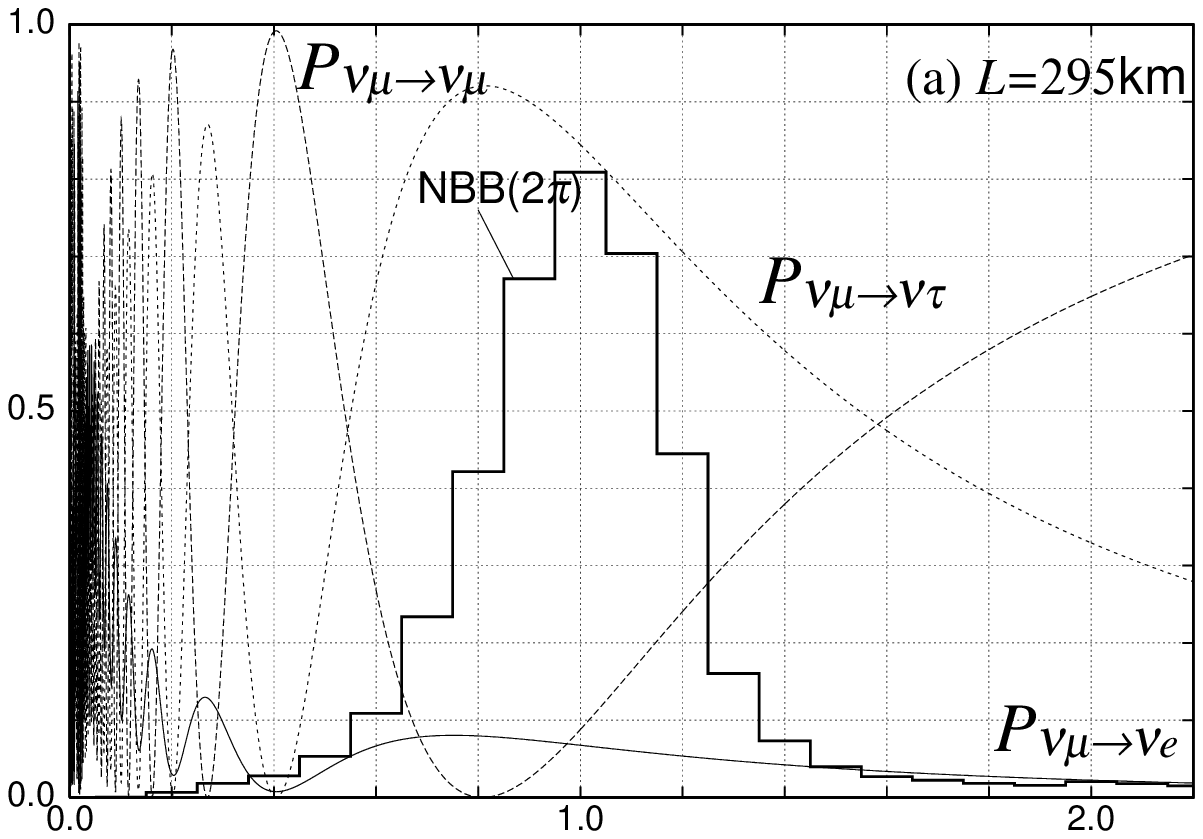}}}
{\scalebox{0.65}{\includegraphics{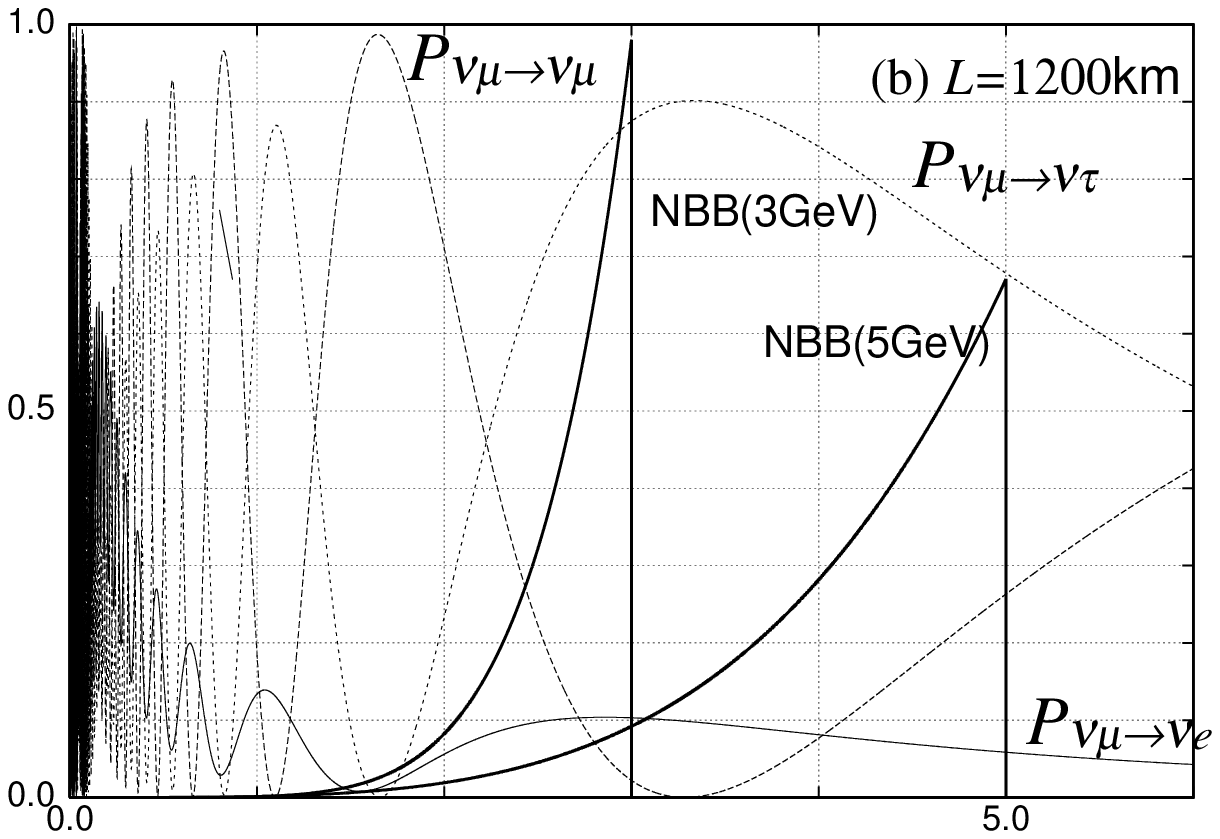}}}
{\scalebox{0.65}{\includegraphics{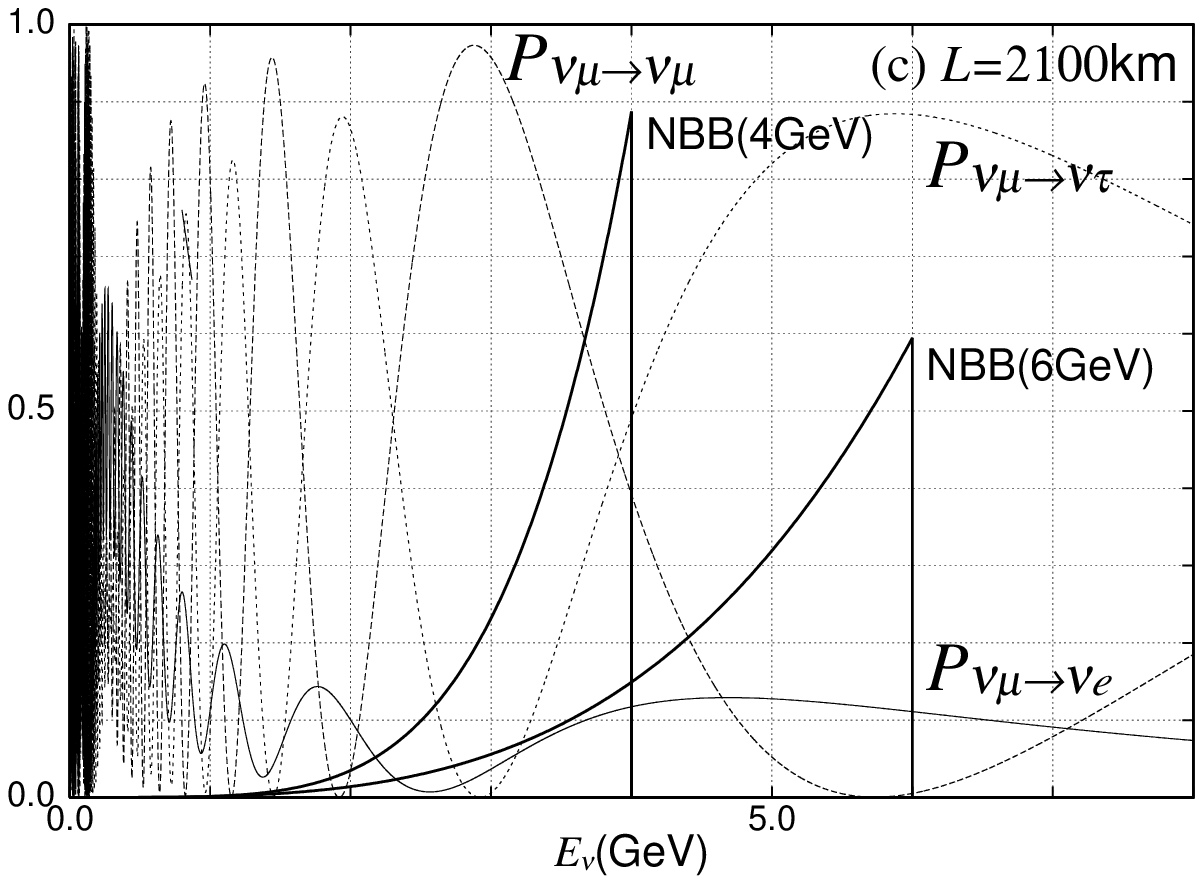}}} 
\end{center}
\caption{The neutrino oscillation probabilities at (a) $L=295$ km, 
(b)$L=1,200$ km and (c) $L=2,100$ km calculated for the parameter values ; 
$
\sin^22\theta_{_{\rm ATM}}=1.0,~
\delta m^2_{_{\rm ATM}}=3.5\times10^{-3} {\rm eV}^2, ~
\sin^22\theta_{_{\rm SOL}}=0.8,~
\delta m^2_{_{\rm SOL}}=10\times10^{-5} {\rm eV}^2, ~
\sin^22\theta_{_{\rm CHOOZ}}=0.1,~
\delta_{_{\rm MNS}}=270^\circ
$
and
$\rho =3$ g/cm$^3$ with the hierarchy I.
Overlayed are the NBB's (flux times energy) used in our
analysis.}
\Fglab{Pro_fl}
\end{figure}
  
The following background contributions to the `$\mu$' and the `$e$' events are
accounted for :
\bea
N(\mu,E_{\rm peak},L)_{\rm BG} &=& N(\mu; \ov \nu_\mu {\rm CC}) 
+N(\mu; \nu_\tau {\rm CC}:\tau \to \mu), 
\eqlab{bg_m}
\\
N(e,E_{\rm peak},L)_{\rm BG} &=& N(e; \nu_e {\rm CC}) 
   +N(e; \nu_\tau {\rm CC}:\tau \to e) +N(e;  {\rm NC}) \nn \\
&& ~~ +N(e; \nu_\tau {\rm CC}: \tau \to {\rm hadrons} ).
\eqlab{bg_e}
\eea
Here $\ov \nu_\mu$ CC and $\nu_e$ CC
contributions are
calculated
by interpolating the numerical integrations
\bea
N(l;  \stackrel{(-)}{\nu_l} {\rm CC} )=
MN_A\int dE_\nu \Phi_{\stackrel{(-)}{\nu_l}} (E_{\rm peak})  
\sigma_{\stackrel{(-)}{\nu_l}}^{\rm CC}
P_{\stackrel{(-)}{\nu_l} \to \stackrel{(-)}{\nu_l}} 
\eea
for the discrete set of the MC simulations \cite{TKHP}.
Here $\Phi_{\stackrel{(-)}{\nu_\mu}} (E_{\rm peak})$
and  $\Phi_{\nu_e} (E_{\rm peak})$ stand for,
respectively, the MC generated secondary $\ov \nu_\mu^{}$
and $\nu_e^{}$ flux of the primarily $\nu_\mu^{}$ beam.
The survival probabilities are calculated for the same set of the model
parameters,
\eqref{modelpara}.
We find that contributions from oscillations from the background beams, such
as
$\ov \nu_e \to \ov \nu_\mu$, are negligibly small and hence they are
not counted.
The contributions from $\tau$-lepton pure-leptonic decays are estimated as
\bea
N(l;\nu_\tau{\rm CC}: \tau \to l)=
MN_A\int dE_\nu  \Phi (E_{\rm peak})
P_{\nu_\mu \to \nu_\tau}  \sigma_{\nu_\tau}^{\rm CC}
Br(\tau \to l \ov \nu_l \nu_\tau)\,,
\eea
where we adopt $Br(\tau \to \mu \ov \nu_\mu \nu_\tau)=0.1737$ and
$Br(\tau \to e \ov \nu_e \nu_\tau)=0.1783$ \cite{PDG}.
The 10 $\%$ errors in these branching fractions are accounted for 
as systematic errors.
Because of the $\tau$-lepton threshold, $\tau$-backgrounds are significant
only at high
energies, NBB's with $E_{\rm peak} \gsim$ 4 GeV.
Because they receive contribution from the small high-energy secondary peak
due to Kaon decays, we use the interpolation of the results obtained for
discrete set of MC generated fluxes. 

The `$e$' events receive contributions from the NC events where
produced
$\pi^0$'s mimic electron shower in the water-$\check {\rm C}$erenkov detector.
By using the estimations from the SK experiments, we adopt
\bea
N(e;{\rm NC})=MN_A\int dE_\nu \Phi (E_{\rm peak}) \sigma_{\nu_\mu}^{\rm NC}
P_{e/{\rm NC}}
\eea
with
\bea
P_{e/{\rm NC}}=0.0055\pm 0.00055\,.
\eqlab{ncbg}
\eea
The error in the above $e/{\rm NC}$ mis-identification probability is
accounted for as a systematic error.
The last term in \eqref{bg_e} accounts for the probability that 
the $\nu_\tau$ CC events with hadronic $\tau$-decays are 
counted as $e$-like events.
In the absence of detailed study of such backgrounds, we use the same
misidentification probability for the NC events \eqref{ncbg} and obtain
\bea
N(e;\nu_\tau{\rm CC}: \tau &\to & {\rm hadrons})= \nn \\
&&MN_A\int dE_\nu  \Phi (E_{\rm peak})
P_{\nu_\mu \to \nu_\tau}  \sigma_{\nu_\tau}^{\rm CC}
Br(\tau \to \nu_\tau {\rm hadrons})P_{e/{\rm NC}}\,.
\eea
If the small misidentification probability of \eqref{ncbg} holds even for
$\tau \to$ hadron events, their background is only at the 2 $\%$ level of
the $\tau \to e$ background, $N(e; \nu_\tau {\rm CC} : \tau \to e)$,
and hence can safely be neglected.

In addition, we account for the following two effects as the major
part of the systematic uncertainty in the VLBL experiments.
One is the uncertainty in the total flux of the neutrino beam, for
which we adopt the estimate,
\bea
f_{_{\rm FLUX}}=1\pm 0.03\,, 
\eqlab{flux_unc}
\eea 
common for all the high-energy NBB's.
We allocate an independent flux uncertainty of 3$\%$ for the low-energy
NBB used for the SK experiment ($L=295$ km) since it uses different
optics.
Finally, we allocate 3.3 $\%$ uncertainty in the matter density along 
the base-line.
In our simplified analysis, we use
\bea
\rho=3.0 \pm 0.1 ~{\rm g/cm^3}
\eqlab{matter_unc}
\eea
as a representative density and the uncertainty.

\begin{table}[tb]
(a) $L=2,100$ km 
\begin{center}
\begin{tabular}{|c||c||r|r|r|r||r|r|}
\cline{3-8}
\multicolumn{2}{c|}{} & $N_{\rm signal}$ &
\multicolumn{3}{|c||}{ $N_{\rm BG}$ } & $N_{\rm tot}$& 
$\frac{N_{\rm signal}}{N_{\rm tot}}$ \\
  \cline{4-6}
\multicolumn{2}{c|}{} & &
     beams & $\nu_\tau$ CC & NC &    & \\
  \cline{3-8}
  \hline
NBB ($E_{\rm peak}=6$ GeV)& $N(\mu)$ &202.5 & 2.2 &15.2 & $-$ & 219.9 &.92\\
  \cline{2-8}
500kton$\cdot$year &   $N(e)$ &126.5      & 7.3 &15.9 &3.3& 153.0 &.83\\
  \hline
  \hline
NBB ($E_{\rm peak}=4$ GeV) & $N(\mu)$ &612.5 & 2.2 & 3.5 & $-$  & 618.2 &.99\\
  \cline{2-8}
500kton$\cdot$year &  $N(e)$& 66.4 & 8.5 & 3.7 &2.6& 81.2 &.82\\
  \hline
 \end{tabular}
\end{center}

(b) $L=1,200$ km 
\begin{center}
\begin{tabular}{|c||c||r|r|r|r||r|r|}
\cline{3-8}
\multicolumn{2}{c|}{} & $N_{\rm signal}$ &
\multicolumn{3}{|c||}{ $N_{\rm BG}$ } & $N_{\rm tot}$& 
$\frac{N_{\rm signal}}{N_{\rm tot}}$ \\
  \cline{4-6}
\multicolumn{2}{c|}{} & &
     beams & $\nu_\tau$ CC & NC &   & \\
  \cline{3-8}
  \hline
NBB ($E_{\rm peak}=5$ GeV)& $N(\mu)$ & 490.1 & 2.2 & 8.7 & $-$ & 501.0&.98\\
  \cline{2-8}
500kton$\cdot$year &   $N(e)$ & 239.1  & 10.5 & 9.0 & 9.4 & 268.0&.89\\
  \hline
  \hline
NBB ($E_{\rm peak}=3$ GeV) & $N(\mu)$& 413.8 & 2.3 & 0.0 & $-$  & 416.1&.99 \\
  \cline{2-8}
500kton$\cdot$year &  $N(e)$& 186.1 & 4.8 & 0.0 & 5.6 & 196.5&.95\\
  \hline
 \end{tabular}
\end{center}

(c) $L=295$ km 
\begin{center}
\begin{tabular}{|c||c||r|r|r|r||r|r|}
\cline{3-8}
\multicolumn{2}{c|}{} & $N_{\rm signal}$ &
\multicolumn{3}{|c||}{ $N_{\rm BG}$ } &$N_{\rm tot}$ & 
$\frac{N_{\rm signal}}{N_{\rm tot}}$ \\
  \cline{4-6}
\multicolumn{2}{c|}{} & &
     beams & $\nu_\tau$ CC & NC &    & \\
  \cline{3-8}
  \hline
NBB ($\langle p_\pi\rangle=2$ GeV) &  $N(\mu)$ & 464.9 &7.8 & 0.0 & $-$ & 472.7&.98\\
  \cline{2-8}
100kton$\cdot$year &  $N(e)$  &161.3 & 22.0 &0.0 & 7.4  & 190.7&.85 \\

  \hline
\end{tabular}
\end{center}
\caption{Expected signals and backgrounds for $\mu$-like and $e$-like
events. The results are shown for the parameters of \eqref{modelpara_B} at
$\delta_{_{\rm MNS}}=270^\circ$,
(a) 500kton$\cdot$year at $L=2,100$ km, 
(b) 500kton$\cdot$year at $L=1,200$ km,
(c) 100kton$\cdot$year at $L=295$ km.   }
\tblab{signal_bg}
\end{table}

Typical numbers of expected signals and backgrounds are tabulated 
in Table 6 for the parameter set of \eqref{modelpara} at
$\sin^22\theta_{_{\rm CHOOZ}}=0.1$ and
$\delta_{_{\rm MNS}}=270^\circ$.
The numerical values are given for the following sets of 
experimental conditions ;
\bea
\begin{tabular}{lll}
(a) $L=2,100$ km   & 500 kton$\cdot$year &
    NBB($E_{\rm peak}=4$ GeV),
    NBB($E_{\rm peak}=6$ GeV)
\\ [2mm]
(b) $L=1,200$ km   & 500 kton$\cdot$year &
    NBB($E_{\rm peak}=3$ GeV),
    NBB($E_{\rm peak}=5$ GeV)
\\ [2mm]
(c) $L=295$ km   & 100 kton$\cdot$year &
    NBB($\langle p_\pi \rangle=2$ GeV)
\end{tabular} 
\nn
\eea
We note here that 100 kton$\cdot$year at $L=295$ km is what SK can
gather in approximately 5 years with $10^{21}$ POT par year.
500 kton$\cdot$year at longer distances can be accumulated in 5 years for
a 100 kton detector with the same intensity beam.

A few remarks are in order.
A 100 kton-level detector at $L=2,100$ km or 1,200 km can detect comparable
numbers of $\mu$-like and $e$-like events as SK (22.5 kton) at $L=295$ km.
The backgrounds due to secondary beams (mostly from the $\nu_e$ beam ) are
significant in all cases for the $e$-like events.
The NC background for the $e$-like events remain
small at high energies if the estimate \eqref{ncbg} obtained from the
K2K experiment at SK remain valid.
We may expect gradual increase in the misidentification probability 
$P_{e/{\rm NC}}$ at high energies as the mean multiplicity of $\pi^0$
and charged particles grow.
Finally the $\tau$-decay background can be significant only for NBB's with
$E_{\rm peak} \gsim 5$ GeV.
If a detector is capable of distinguishing $\tau$-events from the   
$\nu_e$ and $\nu_\mu$ CC events, the overall fit quality improves slightly 
in the three-neutrino model, because the constraint 
$P_{\nu_\mu\to \nu_e}+P_{\nu_\mu\to \nu_\mu}+P_{\nu_\mu\to \nu_\tau}=1$
implies that the information obtained from the $P_{\nu_\mu\to\nu_e}$
measurement always diminishes by contaminations from the $\tau$-events.
As remarked in section 3.4, the $\nu_\tau$ CC event rates have
been corrected for the high-energy secondary peak contributions by
using the original MC generated NBB spectrum \cite{TKHP}.

\subsection{Results for $L=295$ km}

\begin{figure}[thbp]
 \begin{center}
 \scalebox{.85}{\includegraphics{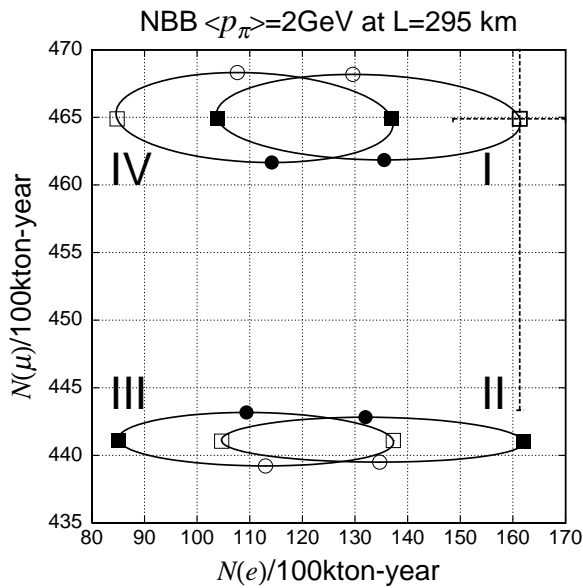}}
 \end{center} 
\caption{
CP phase dependence of $N(e)$ and $N(\mu)$
at SK($L=295$ km) for 100kton$\cdot$year with
the NBB($\langle p_\pi\rangle = 2$ GeV).
$\delta_{_{\rm MNS}}^{} = 0^\circ$ (solid-circle),
$90^\circ$ (solid-square),
$180^\circ$ (open-circle),
and $270^\circ$ (open-square).
 The results are shown for
$\sin^22\theta_{_{\rm ATM}} = 1.0$,
$\delta m^2_{_{\rm ATM}} = \numt{3.5}{-3} \mbox{eV$^2$}$,
$\sin^22\theta_{_{\rm SOL}} = 0.8$,
$\delta m^2_{_{\rm SOL}} = \numt{10}{-5} \mbox{eV$^2$}$,
$\sin^22\theta_{_{\rm CHOOZ}} = 0.1$,
and $\rho$ = 3 g/cm$^3$.
The predictions for the four types of the neutrino mass hierarchies
(\Fgref{cases}) are depicted as I, II, III and IV.
}
\Fglab{Cir_SK}
\end{figure}

In \Fgref{Cir_SK} we show the number of $\nu_\mu$ CC events, $N(\mu)$,
and that of $\nu_e$ CC events, $N(e)$, expected at SK ($L=295$ km)
for 100 kton$\cdot$year.
The NBB with $\langle p_\pi \rangle = 2$ GeV, 
NBB($\langle p_\pi \rangle = 2$ GeV), with $10^{21}$POT per year has been
assumed for simplicity.
The three-neutrino model parameters of \eqref{modelpara} at 
$\sin^22\theta_{_{\rm CHOOZ}}=0.1$ are assumed, 
and we take $\rho=3$g/cm$^3$, \eqref{rho}.
The predictions then depend only on the CP phase parameter 
$\delta_{_{\rm MNS}}$, and when we vary $\delta_{_{\rm MNS}}$ from $0^\circ$
to $360^\circ$, we have a circle on the plane of $N(\mu)$ vs $N(e)$.
The four representative cases, $\delta_{_{\rm MNS}}=0^\circ, 90^\circ, 
180^\circ$ and $270^\circ$, are marked by solid-circle, solid-square, 
open-circle, and open-square, respectively.
The four possible mass hierarchy cases of \Fgref{cases} are depicted 
as I, II, III and IV.

Only the expected event-numbers from the $\nu_\mu$ beam, \eqref{event_m}
and \eqref{event_e}, are counted in \Fgref{Cir_SK}, so that we can read off
the ultimate sensitivity of the experiment from the figure.
Statistical errors for such an experiment are shown for the 
$\delta_{_{\rm MNS}}=270^\circ$ point in the mass hierarchy I circle.
We can learn from the figure that if we know all the parameters of the
three-neutrino model except for the mass hierarchy and if we know 
the neutrino-beam flux exactly, then there is a possibility of
distinguishing the mass hierarchy I from III.
In practice, the LBL experiment 
between HIPA and SK can constrain mainly $\delta m^2_{_{\rm ATM}}$ and
$\sin^22\theta_{_{\rm ATM}}$ from $N(\mu)$, and $\sin^22\theta_{_{\rm CHOOZ}}$
from $N(e)$ \cite{H2SK}.
\Fgref{Cir_SK} shows that those measurements should suffer from uncertainties 
in the remaining parameters of the three-neutrino model, the neutrino mass 
hierarchy cases and $\delta_{_{\rm MNS}}$, as are explicitly shown,
as well as on
the solar-neutrino oscillation parameters, $\delta m^2_{_{\rm SOL}}$ and
$\sin^22\theta_{_{\rm SOL}}$.
Next generation of the solar-neutrino observation experiments \cite{next} 
and KamLAND experiment \cite{KAMLAND} may further constrain the latter two 
parameters, but the mass-hierarchy (between I and III) and 
$\delta_{_{\rm MNS}}$ should be determined by the next generation of
accelerator-based LBL experiments.

It is hence necessary that all the results from the LBL experiment between
HIPA and SK \cite{H2SK} should be expressed as constraints on the 
three primary parameters,
$\delta m^2_{_{\rm ATM}}$,
$\sin^22\theta_{_{\rm ATM}}$, and $\sin^22\theta_{_{\rm CHOOZ}}$,
which depend slightly on
the three remaining
parameters of the three-neutrino model, $\delta m^2_{_{\rm SOL}}$,
$\sin^22\theta_{_{\rm SOL}}$ and $\delta_{_{\rm MNS}}$, as well as 
on the mass-hierarchy cases.
In the following subsections, we show that the data obtained from the LBL
experiment between HIPA and SK ($L = 295$ km) are useful in determining 
the neutrino mass hierarchy, and in some cases even $\delta_{_{\rm MNS}}$,
when they are combined with the data from a VLBL experiments
($L = 2,100$ km or 1,200 km) with higher-energy neutrino beams from HIPA.

Before moving on to studying physics potential of VLBL experiments,
it is worth noting that the predictions for the mass hierarchy IV in 
\Fgref{Cir_SK} represents the prediction for the $\ov \nu_\mu \to 
\ov \nu_\mu, \ov \nu_e$ oscillation probabilities in the hierarchy I, 
according to the theorem \eqref{anti-P},
once the scales are connected by the factor
$\sigma(\ov\nu_l N)/\sigma(\nu_lN)$.
By comparing the $\delta_{_{\rm MNS}}$ dependences of the circle I 
and circle IV, we can clearly see that $P_{\nu_\mu \to \nu_e}$ and
$P_{\ov \nu_\mu \to \ov \nu_e}$ interchange approximately by exchanging 
$\delta_{_{\rm MNS}}=90^\circ$ and $270^\circ$.
The comparison of $\nu_\mu$ and $\ov \nu_\mu$ oscillation experiments at 
around $L = 295$ km hence has a potential of discovering CP-violation in
the lepton sector.
However, determination of the $\delta_{_{\rm MNS}}$ angle
by using the $\ov \nu_\mu$ beam from HIPA needs
a much bigger detector than SK \cite{H2SK,AHO in prep}.   

\subsection{Results for $L=2,100$ km}

\Fgref{Cir_B} shows the numbers of $\nu_\mu$ CC events and that of $\nu_e$ CC
events expected at the base-line length of $L = 2,100$ km from HIPA with
500 kton$\cdot$year.
The expected signal event numbers are shown for (a) the NBB with 
$E_{\rm peak}=4$ GeV and for (b) the NBB with $E_{\rm peak}=6$ GeV.
The parameters of the three neutrino model 
and the matter density are taken exactly
the same as in \Fgref{Cir_SK} :
\bseq
\eqlab{modelpara_B}
\bea
\sin^22\theta_{_{\rm ATM}}&=&1.0\,, ~~~~~
\delta m^2_{_{\rm ATM}}=3.5\times10^{-3} ~{\rm eV}^2\,, 
\eqlab{modelpara_B_atm}
\\
\sin^22\theta_{_{\rm SOL}}&=&0.8\,, ~~~~~
\delta m^2_{_{\rm SOL}}=10\times10^{-5} ~{\rm eV}^2\,, 
\eqlab{modelpara_B_sol}\\
\sin^22\theta_{_{\rm CHOOZ}}&=&0.1\,, 
\eqlab{modelpara_B_chooz}\\
\delta_{_{\rm MNS}}&=&0^\circ~-~ 360^\circ\,, 
\eqlab{modelpara_B_delta}\\
\rho&=&3 ~{\rm g/cm^3}\,. 
\eqlab{modelpara_B_matter}
\eea
\eseq
The predictions for the four neutrino mass hierarchy cases (\Fgref{cases})
are shown by separate circles when the CP-phase angle $\delta_{_{\rm MNS}}$
is allowed to vary freely.
The predictions for the four representative phase values are shown by
solid-circle ($\delta_{_{\rm MNS}}=0^\circ$), solid-square ($90^\circ$),
open-circle ($180^\circ$), and open-square ($270^\circ$).

\begin{figure}[thbp]
 \begin{center}
 \scalebox{.85}{\includegraphics{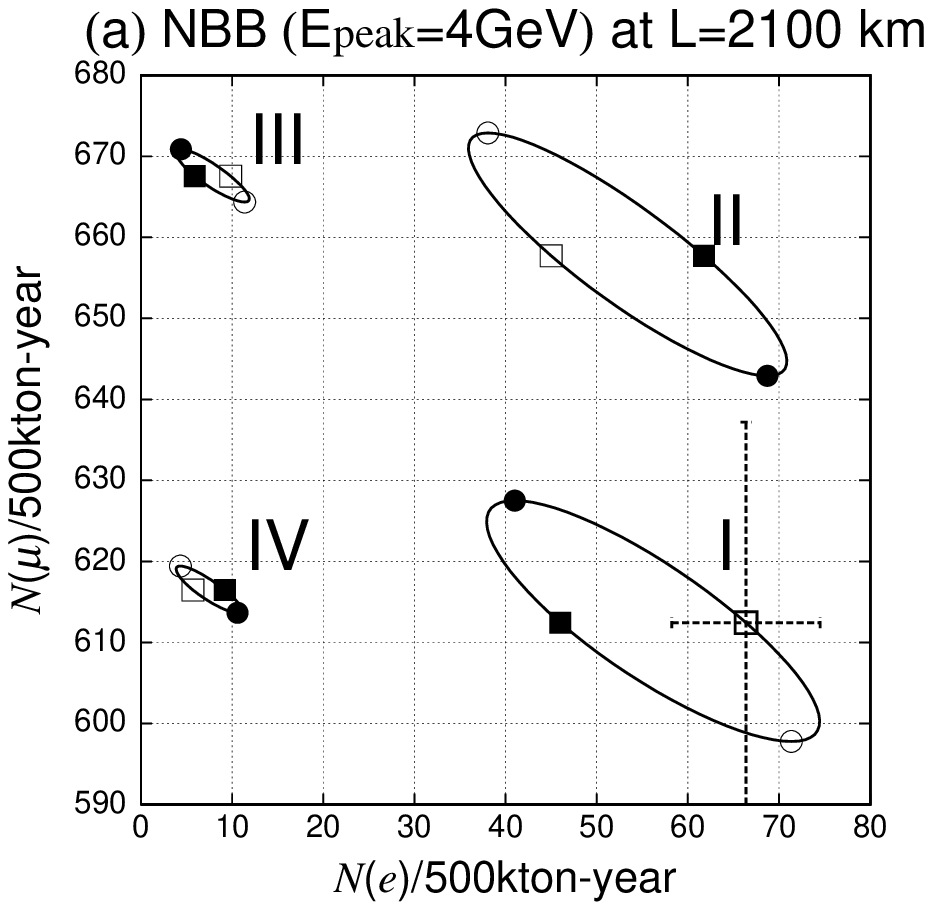}}
 \scalebox{.85}{\includegraphics{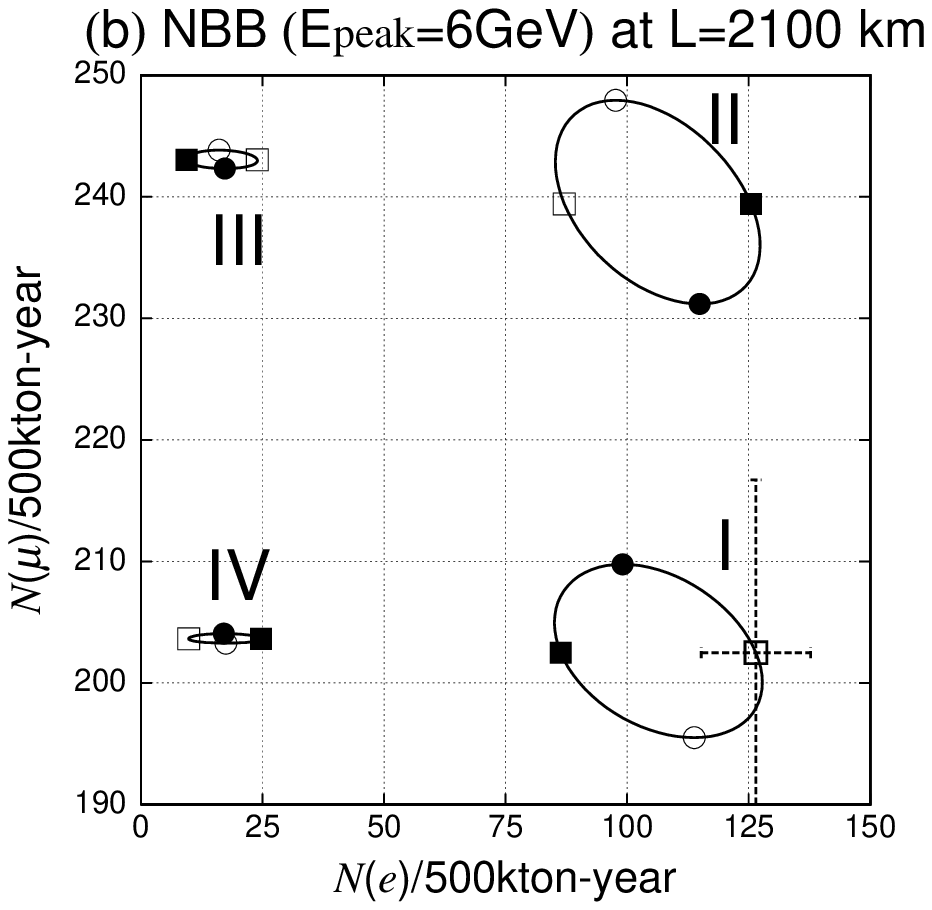}}
 \end{center} 
\caption{%
CP phase dependence of $N(e)$ and $N(\mu)$
at $L = 2100$ km for 500 kton$\cdot$year with
(a) NBB($E_{\rm peak}^{}=4$ GeV)  
and (b) NBB($E_{\rm peak}^{}=6$ GeV).
$\delta_{_{\rm MNS}}^{} = 0^\circ$ (solid-circle),
$90^\circ$ (solid-square),
$180^\circ$ (open-circle),
and $270^\circ$ (open-square).
The results are shown for
$\delta m^2_{_{\rm ATM}} = \numt{3.5}{-3} \mbox{eV$^2$}$,
$\sin^22\theta_{_{\rm ATM}} = 1.0$,
$\delta m^2_{_{\rm SOL}} = \numt{10}{-5} \mbox{eV$^2$}$,
$\sin^22\theta_{_{\rm SOL}} = 0.8$,
$\sin^22\theta_{_{\rm CHOOZ}} = 0.1$,
and $\rho=3$ g/cm$^3$.
The predictions for the four types of the neutrino mass hierarchies
(\Fgref{cases}) are depicted as I, II, III and IV.
}
\Fglab{Cir_B}
\end{figure}

\begin{figure}[htbp]
 \begin{center}
 \scalebox{.61}{\includegraphics{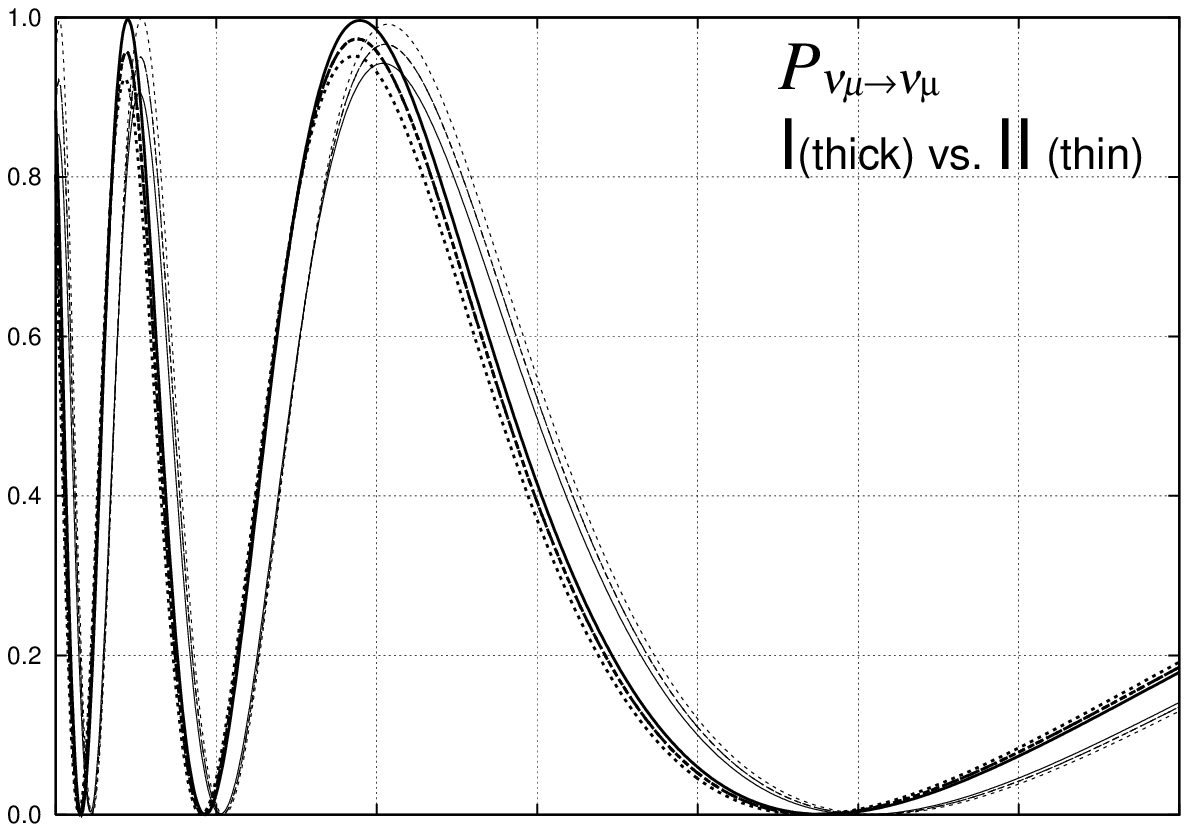}}
 \scalebox{.61}{\includegraphics{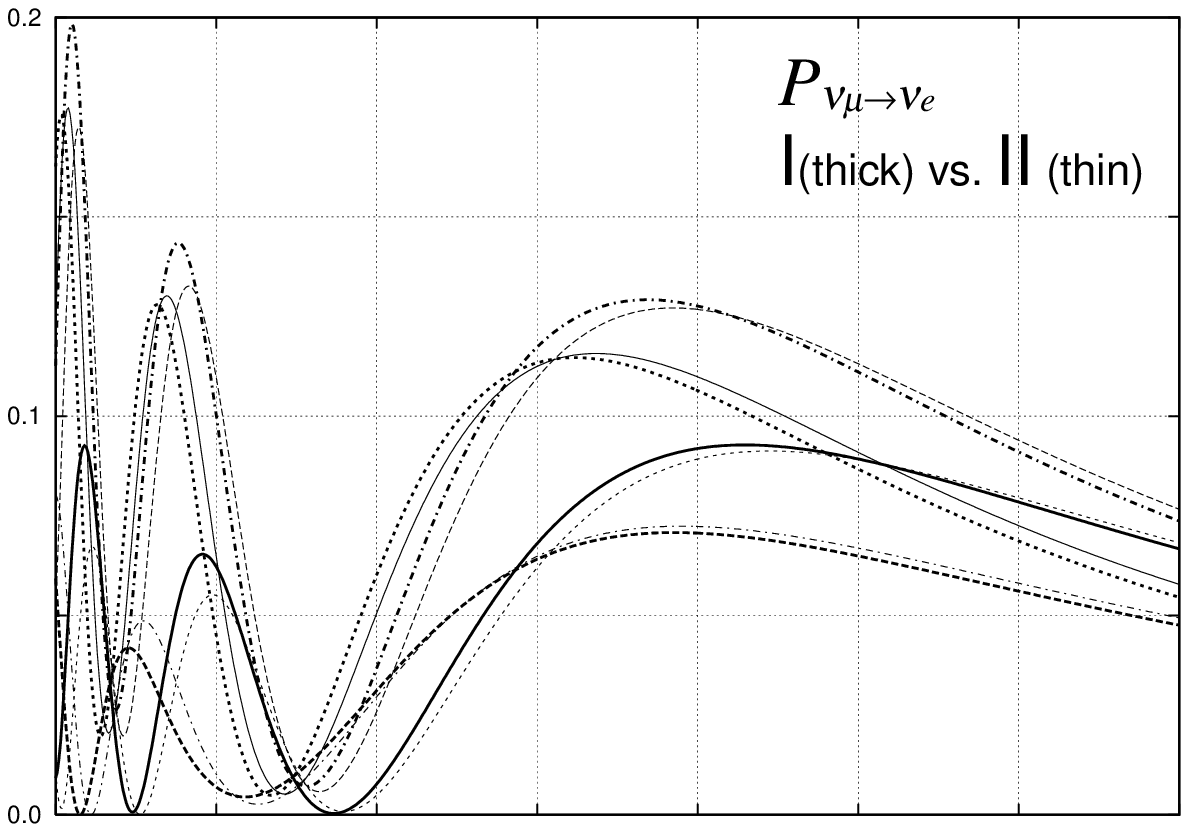}}
\vspace{5ex}

 \scalebox{.61}{\includegraphics{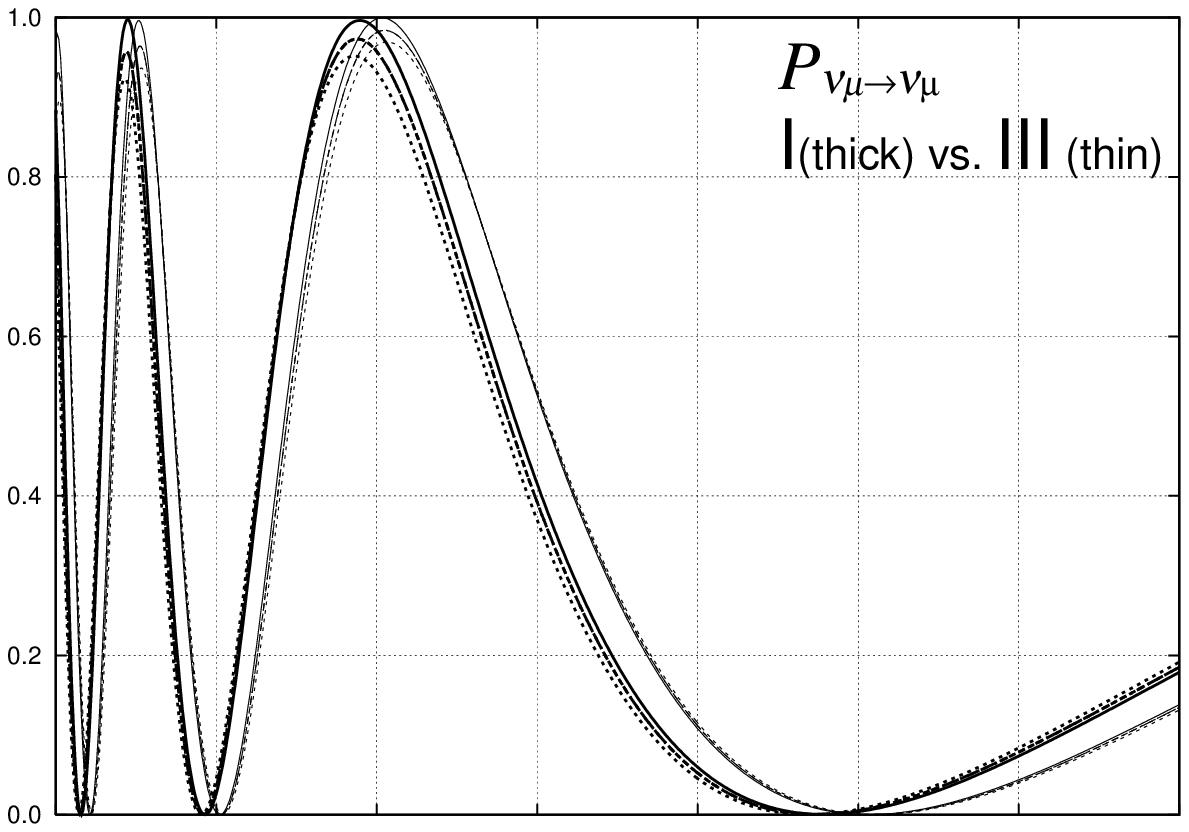}}
 \scalebox{.61}{\includegraphics{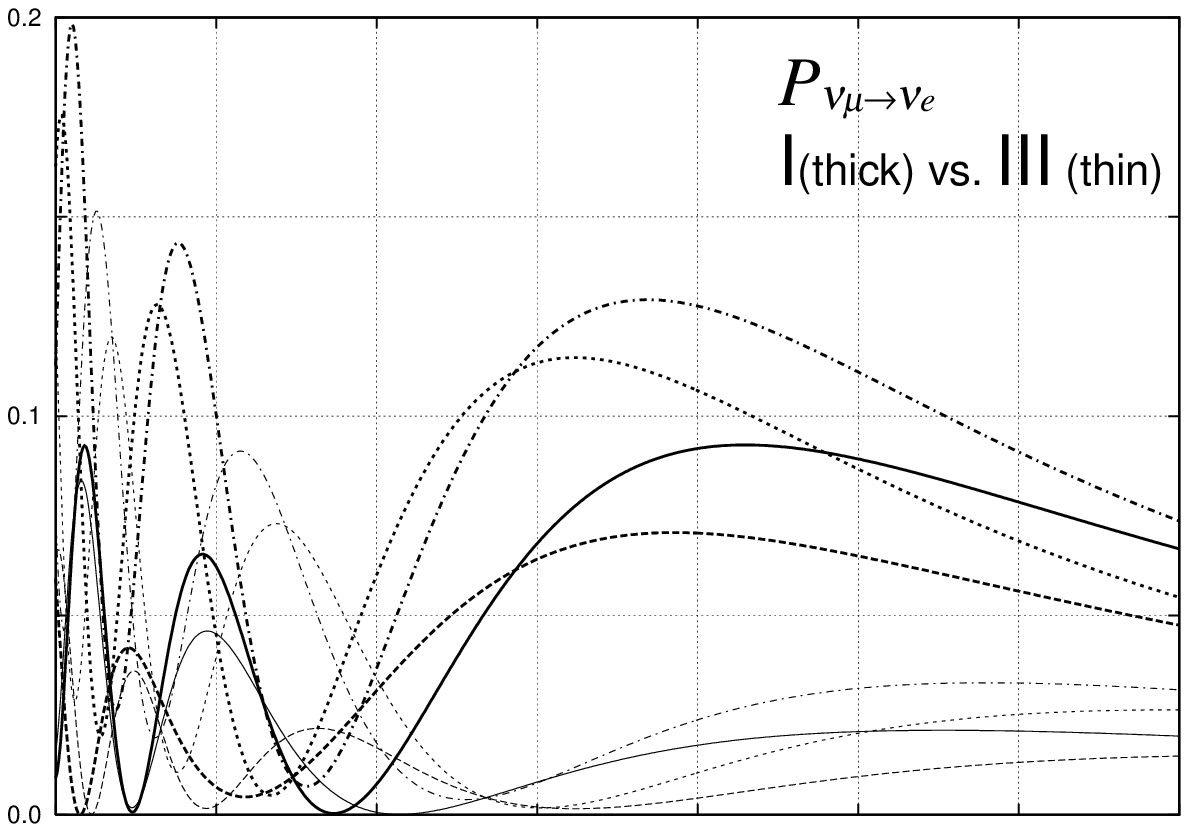}}
\vspace{5ex}

 \scalebox{.61}{\includegraphics{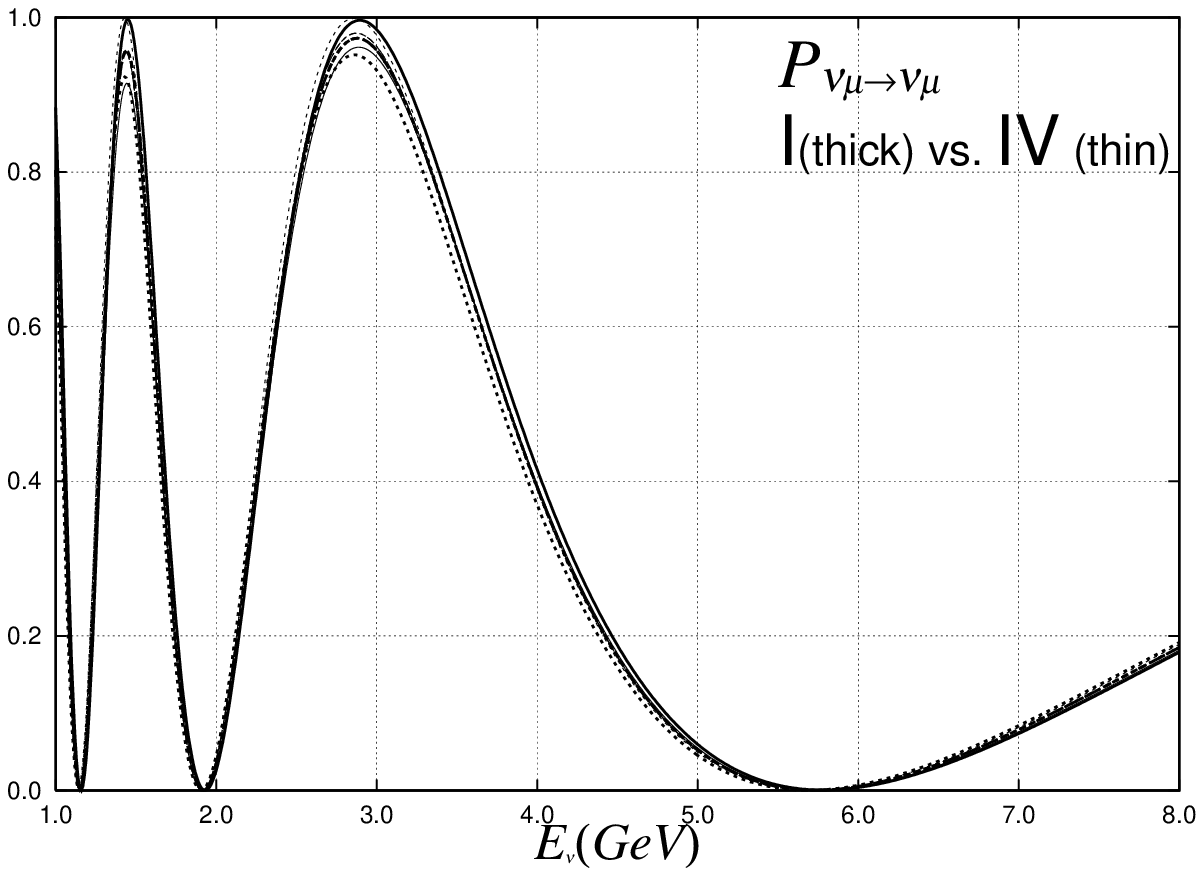}}
 \scalebox{.61}{\includegraphics{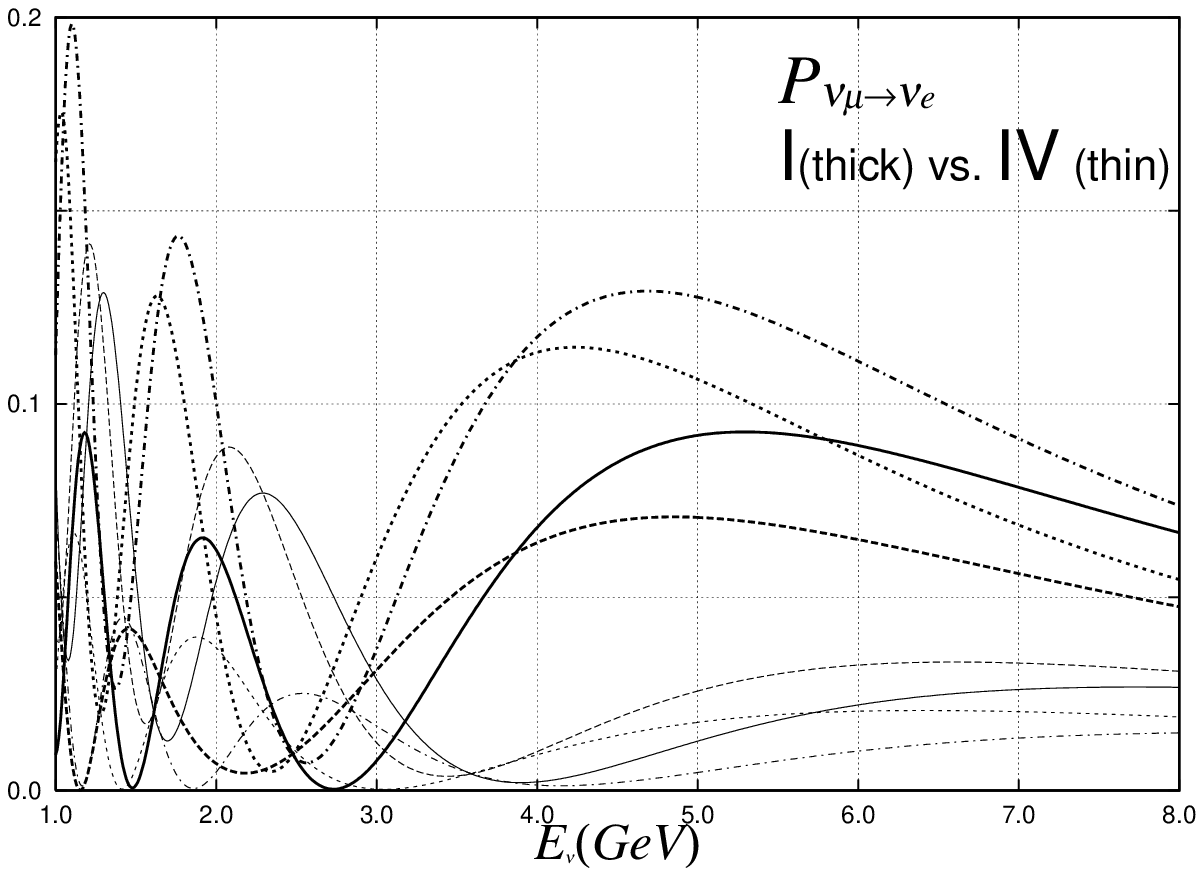}}
 \end{center} 
 \caption{ $P_{\nu_\mu \to \nu_\mu}$ and
$P_{\nu_\mu \to \nu_e}$
vs. $E_\nu$ (GeV) at $L = 2,100$ km.
Solid lines are for $\delta_{_{\rm MNS}} = 0^\circ$, 
long-dashed lines are for $\delta_{_{\rm MNS}} = 90^\circ$,
short-dashed lines are for $\delta_{_{\rm MNS}} = 180^\circ$, and
dot-dashed line are for $\delta_{_{\rm MNS}} = 270^\circ$.
}
\Fglab{pro_B}
\end{figure}

When comparing with the $L=295$ km case (\Fgref{Cir_SK}), it is most striking
to find that the predictions for the $\nu_e$ CC events, $N(e)$, 
differs by a factor of
5 or even larger in magnitude between the neutrino mass hierarchy I and III.
This is because of the enhancement of the matter effect at high energies as 
depicted in \Fgref{amplitude}.
This striking sensitivity of the probability $P_{\nu_\mu \to \nu_e}$
on the mass hierarchy cases is the bases of the capability of distinguishing
the cases in VLBL experiments
by using the HIPA beam.
On the other hand, we will find that the 5 $\%$ level differences in
$N(\mu)$ between the mass hierarchy cases are not useful for this purpose
because $P_{\nu_\mu \to \nu_\mu}$ depends strongly on the parameters
$\delta m^2_{_{\rm ATM}}$ and
$\sin^22\theta_{_{\rm ATM}}$, and also
because $N(\mu)$ suffers from the neutrino-beam flux uncertainty 
\eqref{flux_unc}.

We have examined NBB's with various peak energies, and find that the NBB
with $E_{\rm peak}=6$ GeV (see \Fgref{Pro_fl}) makes $N(e)$ largest
while keeping $N(\mu)$ small for the atmospheric-neutrino oscillation
parameter of \eqref{modelpara_B_atm}.
 The ration $N(e)/N(\mu)$ can be as large as $1/2$
if $\sin^22\theta_{_{\rm CHOOZ}}=0.1$
(\eqref{modelpara_B_chooz})
for the hierarchy case I.
The NBB with $E_{\rm peak}=4$ GeV is then chosen because it has the
$\delta_{_{\rm MNS}}$ dependence of $N(e)$ which is `orthogonal' to that
of the $E_{\rm peak}=6$ GeV case.
By comparing the hierarchy I circles of \Fgref{Cir_B}(a) and (b), we find that 
for the NBB with $E_{\rm peak}$ = 4 GeV,
$N(e)$ is
largest at around $\delta_{_{\rm MNS}}=180^\circ $
(open-circle)
and smallest at around $\delta_{_{\rm MNS}}=0^\circ $
(solid-circle),
whereas  
for the NBB with $E_{\rm peak}=6$ GeV,
$N(e)$ 
is largest at around $\delta_{_{\rm MNS}}=270^\circ$
(open-square)
and smallest at around $\delta_{_{\rm MNS}}=90^\circ$
(solid-square).

In order to understand the dependences of $N(\mu), N(e)$ on $E_{\rm peak}$
(the peak energy of the NBB)
and on $\delta_{_{\rm MNS}}$, and in particular to probe 
the reason for this peculiar
$\delta_{_{\rm MNS}}$-dependence of the expected number of $\nu_e$ CC events on
$E_{\rm peak}$, we show in \Fgref{pro_B} the oscillation probabilities
$P_{\nu_\mu \to \nu_\mu}$ and $P_{\nu_\mu \to \nu_e}$ plotted against the
neutrino
energy, $E_\nu$.
The three plots in the left-hand-side are for $P_{\nu_\mu \to \nu_\mu}$ and
the three in the right-hand-side are for $P_{\nu_\mu \to \nu_e}$.
In all the plots, the predictions of the three-neutrino model with the
parameters
of \eqref{modelpara_B} are shown by thick lines for the neutrino mass
hierarchy I.
In the upper, middle and bottom figures, the predictions for the same
parameters are
shown by thin lines for the hierarchy II, III, and IV, respectively.
In all the plots the predictions for $\delta_{_{\rm MNS}}=0^\circ, 90^\circ,
180^\circ,
270^\circ$ are shown by solid, long-dashed, short-dashed, and dot-dashed
lines,
respectively.
It is clear from these figures that the transition probability $P_{\nu_\mu
\to \nu_e}$
is more sensitive than 
the reduction probability $P_{\nu_\mu \to \nu_\mu}$ on the
neutrino mass hierarchy and $\delta_{_{\rm MNS}}$.
For the chosen parameters of \eqref{modelpara_B}, $P_{\nu_\mu \to \nu_\mu}$
hits
zero at around $E_\nu=6$ GeV at $L=2,100$ km, where the transition
probability
$P_{\nu_\mu \to \nu_e}$ becomes largest in the interval
4 GeV $< E_\nu < 6$ GeV.
Because the NBB with $E_{\rm peak}=6$ GeV is largest at the peak and
has significant tail down to around $E_\nu$ = 4 GeV, see
\Fgref{Pro_fl}(c) for
the NBB shape, we find largest $N(e)$ and smallest $N(\mu)$ for this NBB.
At the same time we clearly see
that in the interval 4 GeV $< E_\nu <$
6 GeV, $P_{\nu_\mu \to \nu_e}$ is largest at $\delta_{_{\rm MNS}}
=270^\circ$ (dot-dashed line), and smallest at 
$\delta_{_{\rm MNS}}=90^\circ$ (long-dashed line).
Just below $E_\nu=4$ GeV the ordering between the $\delta_{_{\rm MNS}}=
270^\circ$ (dot-dashed) curve and the $\delta_{_{\rm MNS}}=180^\circ$
(short-dashed) curves and that between the $\delta_{_{\rm MNS}}=0^\circ$
(solid) curve and the $\delta_{_{\rm MNS}}=90^\circ$ (long-dashed) curve
is reversed.
Consequently the NBB with $E_{\rm peak}=4$ GeV gives largest $N(e)$ at
around $\delta_{_{\rm MNS}}=180^\circ$ and
smallest $N(e)$ at around $\delta_{_{\rm MNS}}=0^\circ$, among the four 
representative $\delta_{_{\rm MNS}}$ angles.
The overall normalization of $N(e)$ is smaller, and that of $N(\mu)$
is larger for NBB ($E_{\rm peak} = 4$ GeV) than the predictions of
NBB ($E_{\rm peak}=6$ GeV).

In \Fgref{Cir_B}, we show the statistical errors of the $N(\mu)$ and
$N(e)$ measurements at 500 kton$\cdot$year on the 
$\delta_{_{\rm MNS}}=270^\circ$ point for the hierarchy case I.
The size of the error bars suggest that a 100 kton-level detector is needed
to explore the model parameters in VLBL experiments at $L\simeq$ 2,100 km.
It also tells us that such detector has the potential of discriminating 
the neutrino mass hierarchies and constraining the $\delta_{_{\rm MNS}}$
angle in a certain region of the three-neutrino model parameter space.
 A more careful error analysis that accounts for backgrounds
and systematic errors is given in the next subsection.

We are now ready to study the physics capability of such VLBL experiments in
some detail.
In \Fgref{main_B}, we show the expected numbers of signal events, $N(\mu)$
and $N(e)$, for the same NBB's and for the same volume of 500 
kton$\cdot$year as in \Fgref{Cir_B}.
The circles of \Fgref{Cir_B} are now made of 
36 $\delta_{_{\rm MNS}}$ points,  $\delta_{_{\rm MNS}}=n\times 10^\circ$
for $n=1$ to 36, for each set of the model parameters.
The common parameters for all the points are
\bseq
\eqlab{modelpara_main}
\bea
\sin^22\theta_{_{\rm ATM}}&=&1.0\,,~~
\delta m^2_{_{\rm ATM}}=3.5\times10^{-3} ~{\rm eV}^2\,, 
\eqlab{modelpara_atm}\\
\sin^22\theta_{_{\rm CHOOZ}}&=&
0.00\,,~0.02\,,~0.04\,,~0.06\,,~0.08\,,~0.10\,, 
\eqlab{modelpara_chooz} \\
\delta_{_{\rm MNS}}&=&n\times 10^\circ
~~(n=1~{\rm to}~36)\,, 
\eqlab{modelpara_delta}\\
\rho&=&3 ~{\rm g/cm^3}\,.
\eea
\eseq
The predictions for the six $\sin^22\theta_{_{\rm CHOOZ}}$
cases \eqref{modelpara_chooz}
can be recognized as five circles and
one point with decreasing $N(e)$ values,
as $\sin^22\theta_{_{\rm CHOOZ}}$ decreases from 0.1 to 0.0.
The remaining two parameters are constrained by the solar-neutrino oscillation
experiments, and we chose the following representative parameter sets for
the three possible solutions to the solar-neutrino deficit anomaly~:
\bseq
\eqlab{modelpara_sol}
\bea
{\rm LMA}~~~&:&~~~
\sin^22\theta_{_{\rm SOL}}=0.8\,, 
~~~~~~~~~~\delta m^2_{_{\rm SOL}}= \left\{
\begin{array}{l}
15 \times 10^{-5} ~{\rm eV}^2\,, \\
5 \times 10^{-5} ~{\rm eV}^2\,, 
\end{array}
\right.
\eqlab{modelpara_sol_lma}\\[-1mm]
{\rm SMA}~~~&:&~~~
\sin^22\theta_{_{\rm SOL}}=7\times 10^{-3} \,, 
~~~\delta m^2_{_{\rm SOL}}= 5 \times 10^{-6} ~{\rm eV}^2\,, 
\eqlab{modelpara_sol_sma}\\
{\rm VO~}~~~&:&~~~
\sin^22\theta_{_{\rm SOL}}=0.7\,, 
~~~~~~~~~~
\delta m^2_{_{\rm SOL}}=7\times10^{-11} ~{\rm eV}^2\,.
\eqlab{modelpara_sol_vo}
\eea
\eseq
The predictions of the four types of the neutrino mass hierarchy, I, II,
III, IV are indicated
explicitly.

\begin{figure}[th]
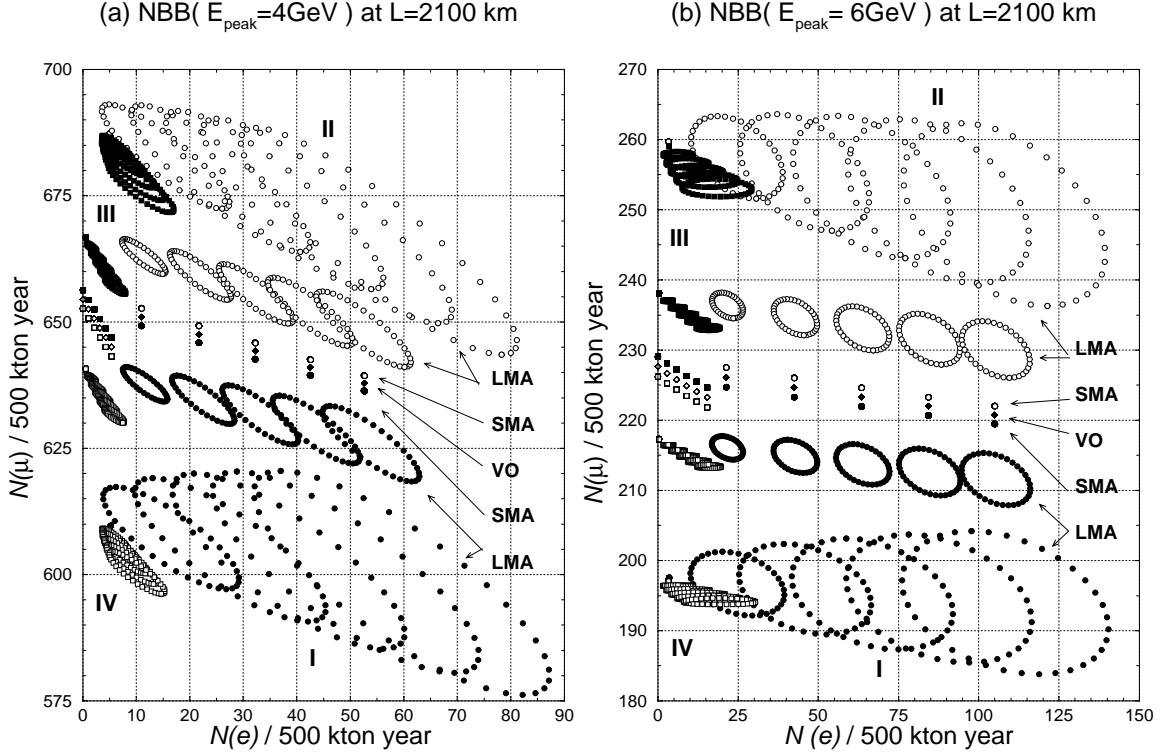

 \begin{center}
{\scalebox{0.45}{\includegraphics{./FigS/main_4gev.eps}}} 
{\scalebox{0.45}{\includegraphics{./FigS/main_6gev.eps}}} 
\end{center}
\caption{%
The neutrino parameter dependences of the expected numbers
of $\nu_e^{}$ CC and $\nu_\mu^{}$ CC events,
$N(e)$ and $N(\mu)$, respectively,
for the NBB with $E_{\rm peak}^{}=4$ GeV (a) and 6 GeV (b)
with 500 kton$\cdot$year at $L=2,100$ km.
The predictions are shown for the LMA, SMA, and VO scenarios
of the solar neutrino oscillations and for the four neutrino
mass hierarchies I to IV.
All the numbers are calculated for
$\delta m^2_{_{\rm ATM}}=\numt{3.5}{-3}$eV$^2$
and $\sin^22\theta_{_{\rm ATM}}^{}=1.0$
at $\sin^22\theta_{_{\rm CHOOZ}}^{}= 0.02 \times m$ ($m=0$ to 5)
and $\delta_{_{\rm MNS}}=10^\circ\times n$ ($n=1$ to 36).
The five larger circles in each hierarchy show the
$\delta_{_{\rm MNS}}^{}$ predictions of the LMA with
$\delta m^2_{_{\rm SOL}}=\numt{15}{-5}$eV$^2$ and
the five smaller circles are for the LMA with
$\delta m^2_{_{\rm SOL}}=\numt{5}{-5}$eV$^2$,
both for $m=1$ to 5. 
$\delta_{_{\rm MNS}}^{}$-dependence is not recognized
for the predictions of SMA and VO, and also at
$\sin^22\theta_{_{\rm CHOOZ}}=0$ ($m=0$).
The VO predictions are the same for the hierarchy I and II,
and III and IV.
}
\Fglab{main_B}
\end{figure}

The solid-circle points in \Fgref{main_B} show the predictions of all
the models with neutrino mass hierarchy I.
They reside in the corner at large $N(e)$ and small $N(\mu)$.
The five grand circles with smallest $N(\mu)$ give the predictions of the
LMA solution
for $\delta m^2_{_{\rm SOL }}= 15\times 10^{-5}$ eV$^2$, and those with
larger
$N(\mu)$ are for the LMA with $\delta m^2_{_{\rm SOL }}= 5\times 10^{-5}$
eV$^2$.
It is clearly seen that the $\delta_{_{\rm MNS}}$ dependence (the size of
the grand circles)
is larger for larger $\delta m^2_{_{\rm SOL }}$ and for larger
$\sin^22\theta_{_{\rm CHOOZ}}$ as expected.
It is worth pointing out that the LMA scenario with
$\delta m^2_{_{\rm SOL }} = 15\times 10^{-5}$ eV$^2$
predicts non-zero $N(e), N(e)\simeq 5$/500 kton$\cdot$year
even when  $\sin^22\theta_{_{\rm CHOOZ}}=0$.
This is because the `higher' oscillation modes of
the three-neutrino model grow
as $\delta m^2_{_{\rm SOL }}/\delta m^2_{_{\rm ATM }}$ rises.
The predictions of the SMA parameters appear just above the upper LMA grand
circles,
where the $\delta_{_{\rm MNS}}$ dependence (the size of the grand circle)
diminishes
to zero for each  $\sin^22\theta_{_{\rm CHOOZ}}$ value of 
\eqref{modelpara_chooz}.
Just five solid-circle points appear in \Fgref{main_B} (a) and (b) for
the SMA parameters
because the $\delta_{_{\rm MNS}}$-dependence diminishes as
$\delta m^2_{_{\rm SOL }}/\delta m^2_{_{\rm ATM }}$ and
$\sin^22\theta_{_{\rm SOL}}$
are both significantly smaller than unity.
This is expected from the Hamiltonian \eqref{hamiltonian1}
that governs the neutrino oscillation in matter,
because the $\delta_{_{\rm MNS}}$-dependence of observables diminished
whenever the second terms proportional to
$\delta m_{12}^2/\delta m_{13}^2$ are much smaller than the first term.
The predictions of the VO parameters, which are given by the solid-diamonds,
are shown just above those of the SMA parameters.
Here the $\delta_{_{\rm MNS}}$ dependence vanishes because of the extreme
smallness
of $\delta m^2_{_{\rm SOL }}/\delta m^2_{_{\rm ATM }}$.
With the same token, we cannot distinguish the VO predictions of the
neutrino mass
hierarchy I and II.
Because the magnitude of  $\delta m^2_{_{\rm SOL }}$ is so small, the sign of
$\delta m^2_{12}=\pm\delta m^2_{_{\rm SOL }}$  does not have observable
consequences in terrestrial LBL experiments.

The predictions of the four scenarios of the solar neutrino oscillations 
(VO, SMA, and two $\delta m^2_{_{\rm SOL}}$ cases of LMA in 
\eqref{modelpara_sol})
with the neutrino mass hierarchy II
($\delta m^2_{12}=-\delta m^2_{_{\rm SOL }}, \delta m^2_{13}=\delta
m^2_{_{\rm ATM}}$) are shown by open-circle points, which are located 
in the corner of large $N(e)$ large $N(\mu)$.
As explained above, the predictions of the VO scenario do not differ from
those for the hierarchy I.
The predictions of the SMA scenario appear slightly above
those of VO.
The two LMA scenarios predict larger $N(\mu)$ and give visible
$\delta_{_{\rm MNS}}$ dependences, which lead to five grand circles 
each for the five discrete values of assumed $\sin^22\theta_{_{\rm CHOOZ}}$.
Again the $\delta_{_{\rm MNS}}$-dependence
(the size of the grand circles) is larger
for the large $\sin^22\theta_{_{\rm CHOOZ}}$ and  
large $\delta m^2_{_{\rm SOL }}$.
Summarizing the predictions of the four scenarios in the neutrino mass
hierarchies I and II, both of which have $\delta m^2_{13}=
\delta m^2_{_{\rm ATM}}>0$, $N(e)$ grows linearly with increasing 
$\sin^22\theta_{_{\rm CHOOZ}}$ in all four scenarios and for both hierarchies.
The predictions of the hierarchy II differ from those of the hierarchy I
only by slightly larger $N(\mu)$ in each scenarios.
The difference in $N(\mu)$ is largest in the LMA scenarios with
$\delta m^2_{_{\rm SOL }}= 15\times 10^{-5}$ eV$^2$, for which the hierarchy II
predicts about 30 $\%$ (10 $\%$) larger $N(\mu)$ than the predictions of the
hierarchy I for the NBB with $E_{\rm peak}=6$ GeV (4 GeV).

All the predictions of the hierarchy cases III and IV have very small $N(e)$, 
as we explained in section 2.4 for the hierarchy III.
This is because the two mass hierarchies have
$\delta m^2_{13}=-\delta m^2_{_{\rm ATM}}$ common.
The predictions of the hierarchy III are shown by solid-squares
and those of the hierarchy IV are shown by open-squares.
As in the case of hierarchy I vs II, the predictions of the VO scenario do
not show visible dependence on the sign of $\delta m^2_{12}=
\pm\delta m^2_{_{\rm SOL}}$, and its common predictions for the hierarchy III 
and IV are shown by open-diamonds.
Even though compressed to the very small $N(e)$ region, the largest $N(e)$
point is for $\sin^22\theta_{_{\rm CHOOZ}}=0.1$.
The predictions of the SMA scenario appear just above (below) the VO points
for the hierarchy III (IV).
In the LMA scenario, the predictions for $N(\mu)$ increases as
$\delta m^2_{_{\rm SOL }}$ grows for the hierarchy III, 
while $N(\mu)$ decreases 
with $\delta m^2_{_{\rm SOL }}$ 
for the hierarchy IV.
By noting that the $\nu_\mu \to \nu_e$ oscillation probability of the
hierarchy IV is
approximately that of the $\bar \nu_\mu \to \bar \nu_e$ oscillation
probability of the hierarchy I, see \eqref{anti-P},
and by noting that the $\bar \nu_e$ CC cross section is a factor
of three smaller than the $\nu_e$ CC cross section, we can conclude
from the figures that the $\bar \nu_\mu \to \bar \nu_e$ oscillation
experiments by
using the $\bar \nu_\mu$ beam from HIPA is not effective if the
neutrino mass hierarchy
is indeed type I \cite{SN1987A}.

\subsection{A case study of semi-quantitative analysis}
Now that the intrinsic sensitivity of the very simple observables
$N(\mu, E_{\rm peak})$ and $N(e, E_{\rm peak})$ on the
three-neutrino model parameters, 
$\sin^22\theta_{_{\rm CHOOZ}}$, $\delta_{_{\rm MNS}}$,
$\delta m^2_{_{\rm SOL}}$ and the neutrino mass hierarchy cases are
shown clearly in \Fgref{main_B}, we would like to examine the capability
of such VLBL experiments in determining the model parameters.
The following 4 questions are of our concern: \\
\begin{tabular}{ll}
1. & Can we distinguish the neutrino mass hierarchy cases ? \\
2. & Can we distinguish the solar-neutrino oscillation scenarios
($\delta m^2_{_{\rm SOL}}$ and $\sin^22\theta_{_{\rm SOL}}$) ? \\
3. & Can we measure the two unknown parameters of the model,
$\sin^22\theta_{_{\rm CHOOZ}}$ and $\delta_{_{\rm MNS}}$ ? \\
4. & How much can we improve the measurements of $\delta m^2_{_{\rm ATM}}$
and $\sin^22\theta_{_{\rm ATM}}$ ? \\
\end{tabular}
In this subsection we examine these questions for a VLBL experiment
at $L=2,100$ km in combination with the LBL experiment of $L=295$ km
between HIPA and SK.
\vspace{8mm}

For definiteness, we assume that the VLBL experiment at $L=2,100$ km
accumulates 500 kton$\cdot$year each with the NBB ($E_{\rm peak}=6$ GeV)
and NBB ($E_{\rm peak}=4$ GeV).
As for the LBL experiment at $L=295$ km, we assume that 100 kton$\cdot$year
data is obtained for the NBB ($\langle p_\pi \rangle=2$ GeV).
Although this latter assumption does not agree with the present plan of
the HIPA-to-SK project \cite{KEKTC5} where the OAB may be
used in the first stage, we find that essentially the same final results
follow as long as the total amount of data at $L=$ 295 km is of the
order of 500 kton$\cdot$year.
The expected numbers of $\mu$-like and $e$-like events 
\begin{eqnarray}
N(\mu)   &=&
       N(\mu,E_{\rm peak},L) + N(\mu,E_{\rm peak},L)_{\rm BG}\,, \\
N(e)   &=&
       N(e,E_{\rm peak},L) + N(e,E_{\rm peak},L)_{\rm BG}\,,
\end{eqnarray}
are tabulated in \Tbref{signal_bg} for the parameters of \eqref{modelpara_B}
at $\delta_{_{\rm MNS}}=0^\circ$.
We see from the \Tbref{signal_bg} that roughly 
the same number of $\nu_\mu \to \nu_e$
signal events is expected for an experiment at $L=2,100$ km with
500 kton$\cdot$year and for an experiment at $L=295$ km with 100
kton$\cdot$year.
This agrees with the naive scaling low of $N \sim 1/L$ at same $L/E_\nu$.

Our program proceeds as follows.
For a given set of the three-neutrino model parameters, we calculate
predictions of $N(\mu)$ and $N(e)$, including both the signal and the
background, with a parameterized neutrino-beam flux and
a constant matter density of the earth $(\rho=3$g/cm$^3)$.
We assume that the detection efficiencies of the $\mu$-like and $e$-like
events are $100 \%$ for simplicity.
The statistical errors of each predictions are then simply the 
square roots of the observed numbers of events, $N(\mu)$ and $N(e)$.
In our analysis, we account for the following systematic errors:
\begin{center}
\begin{tabular}{cl}
$\bullet$ & Overall flux normalization error of $3\%$ \\
$\bullet$ & Uncertainty in  the matter density along the baseline of
$3.3\%$ \\
$\bullet$ & Relative uncertainty in the misidentification probability
$P_{e/{\rm NC}}$ of $10\%$  \\
$\bullet$ & Relative uncertainty in the $\tau$-backgrounds of $10\%$  \\
\end{tabular}
\end{center}
The $\chi^2$ function of the fit to the LBL experiments
may then be expressed as the sum
\bea
\chi^2=\chi^2(L=2,100{\rm ~km})+\chi^2(L=295 {\rm ~km})\,, 
\eea
where the first term is 
\begin{eqnarray}
&&\chi^2 (L=2,100 {\rm ~km})\nn\\
&&~~~
=\sum_{E_{\rm peak},~{\rm kton}\cdot{\rm year}}^{}
\left\{
\left(
\frac{
   f_{_{\rm FLUX}}{\cdot}N^{fit}(\mu)
 - N^{true}(\mu)
}{
\sigma(\mu)
}
\right)^2 
+ \left(\frac{f_{_{\rm FLUX}}{\cdot}N^{fit}(e)
- N^{true}(e)}{\sigma(e)}
\right)^2 \right\} \nn \\
&&~~~~~~~
+\left( \frac{f_{_{\rm FLUX}} - 1}{0.03} \right)^2
+\left( \frac{\rho - 3}{0.1} \right)^2\,,
\end{eqnarray}
\begin{eqnarray}
\sigma(\mu)
&=& \sqrt{N^{true}(\mu)+ \left(0.1 N^{true}(\mu; \nu_\tau{\rm CC}) \right)^2 
		    }\,, \\
\sigma(e)
  &=& \sqrt{ N^{true}(e)+ \left(0.1 N^{true}(e;{\rm NC}) \right)^2
+ \left(0.1 N^{true}(e; \nu_\tau{\rm CC}) \right)^2
		    }\,.
\end{eqnarray}
Here the $\chi^2$ is a function of the parameters of the three-neutrino model
(the two mass-squared differences, three angles and one phase),
the flux normalization factor $f_{_{\rm FLUX}}$ and
the matter density $\rho$.
We assign the overall flux normalization error of $3\%$ which is common for
all high-energy NBB($E_{\rm peak}$)'s, including the secondary beams.
Besides this common flux normalization error, the individual error of
$N(\mu)$ is the sum of the statistical error
and the systematic error coming from the 
uncertainty in the $\tau$-background. 
The error of $N(e)$ is a sum of 
the statistical error and the systematic errors coming from the uncertainty
in the $e/{\rm NC}$ misidentification probability and the $\tau$-background.
The $\chi^2$  function for the HIPA-to-SK experiment is obtained
similarly where the $3\%$ flux normalization error and the $3.3\%$ 
matter density error are assumed to be independent from those for
the $L=2,100$ km experiment. 

\begin{figure}[ht]
\begin{center}
{\scalebox{0.48}{\includegraphics{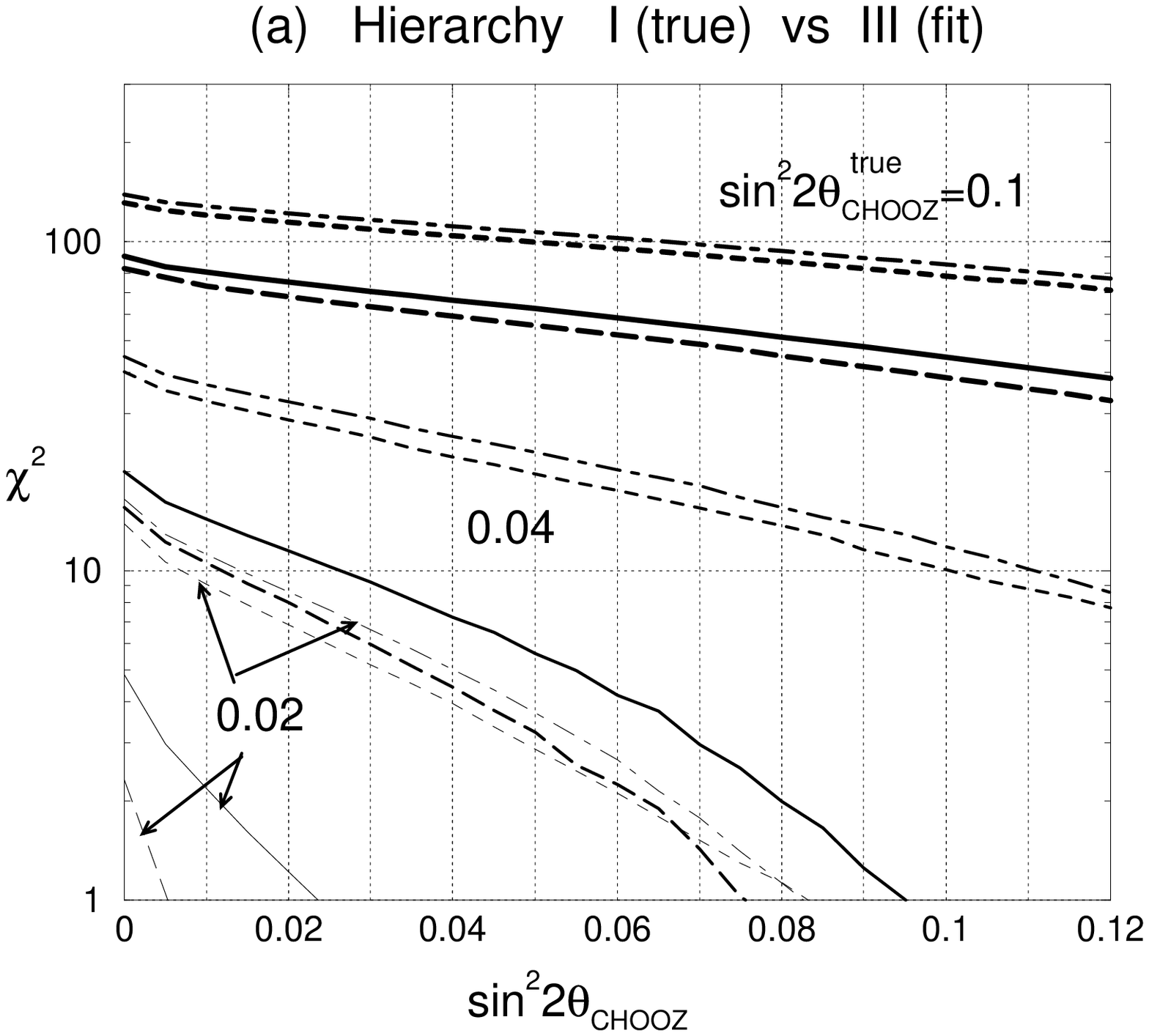}}} 
{\scalebox{0.48}{\includegraphics{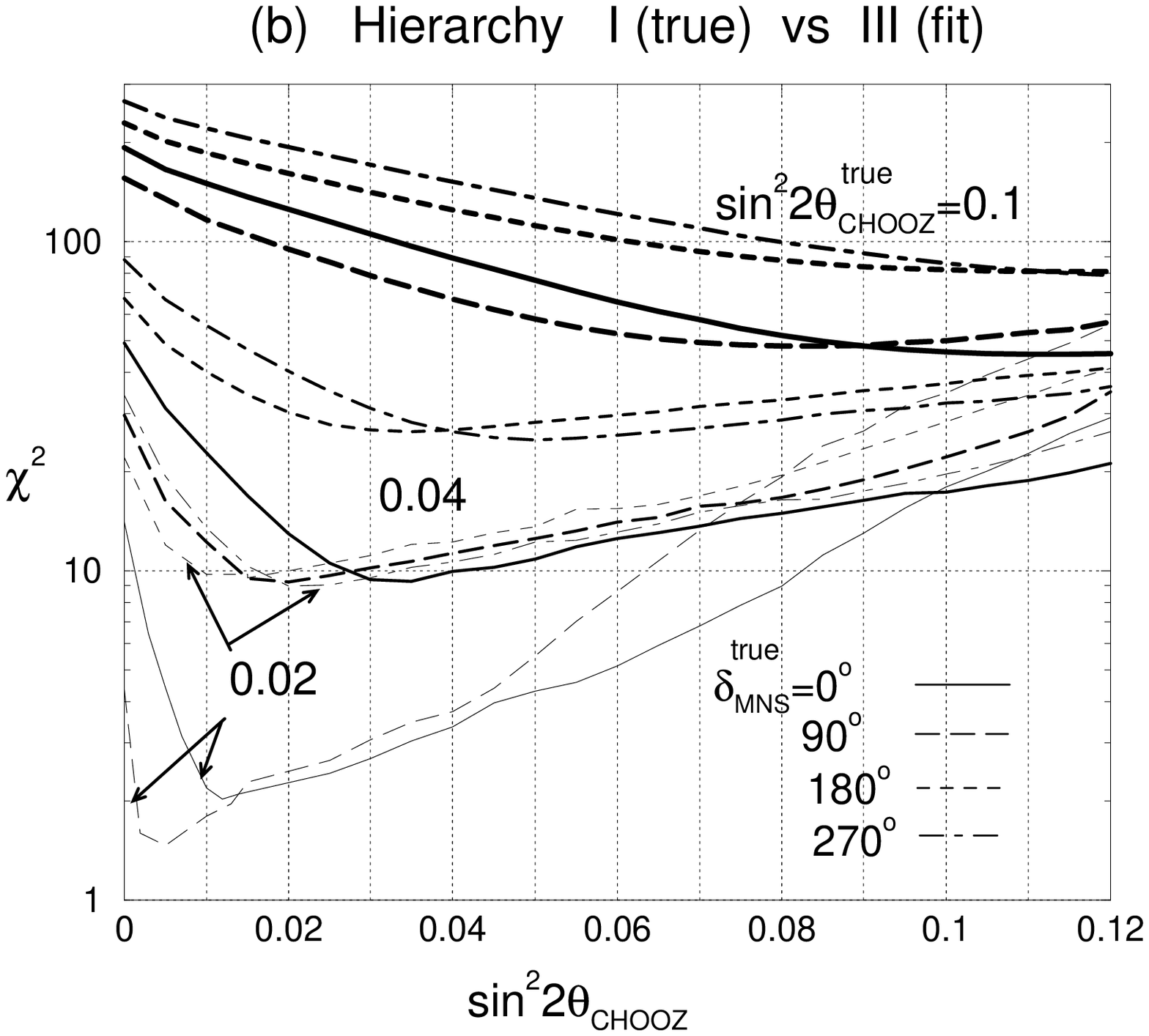}}} 
\end{center}
\caption{%
Minimum $\chi^2$ as functions of the fitting parameter 
$\sin^22\theta_{_{\rm CHOOZ}}$ by assuming the hierarchy III, 
when the mean values of $N(\mu, E_{\rm peak})$ and $N(e, E_{\rm peak})$ 
are calculated for the LMA points with hierarchy I (\eqref{IvsIII_true}). 
In total 12 cases of the input
data are labeled by the input (`true') values of 
$\sin^22\theta_{_{\rm CHOOZ}}^{true}$ 
= 0.02 (thin lines), 0.04 (medium-thick lines), 
0.1 (thick lines) and
$\delta_{_{\rm MNS}}^{true}$
= $0^\circ$ (solid lines), $90^\circ$ (long-dashed lines), 
$180^\circ$ (short-dashed lines), $270^\circ$ (dot-dashed lines). 
The minimum of the $\chi^2$ function is found by 
assuming the hierarchy III for a 
since $\sin^22\theta_{_{\rm CHOOZ}}$ value, by varying the parameters
$\delta m^2_{_{\rm SOL}}$ and $\sin^22\theta_{_{\rm SOL}}$  
within the LMA allowed region, and the remaining three parameters, 
$\delta m^2_{_{\rm ATM}}$, $\sin^22\theta_{_{\rm ATM}}$ and 
$\delta_{_{\rm MNS}}$ freely (\eqref{IvsIII_fit}). 
(a) Result with 500 kton$\cdot$year each for NBB($E_{\rm peak}=6$ GeV) and
NBB($E_{\rm peak}=4$ GeV) at $L=2,100$ km.
(b) In addition, 100 kton$\cdot$year data from
NBB($\langle p_\pi \rangle=2$ GeV)
at $L=295$ km are included in the fit.}
\Fglab{IvsIII_B}	     
\end{figure}

We first examine the capability of the VLBL experiment in distinguishing
the neutrino mass hierarchy cases I and III.
For this purpose, we calculate the expected numbers of signals and
backgrounds for a certain set of the three-neutrino model parameters
with the hierarchy I, and the examine if the generated data can be
interpreted under the assumption of the hierarchy III.

\subsubsection{Neutrino mass hierarchy}
We show in \Fgref{IvsIII_B} the minimum $\chi^2$ as functions of 
the parameter $\sin^22\theta_{_{\rm CHOOZ}}$ by assuming the hierarchy III, 
when the mean values of
$N(\mu, E_{\rm peak})$ and $N(e, E_{\rm peak})$ are calculated 
for the LMA points with the hierarchy I.
The results in \Fgref{IvsIII_B} (a) and (b) 
are given for the following sets
of experimental conditions (A) and (B), respectively:
\bseq
\bea
&&\hspace*{-10ex}
\begin{tabular}{ll}
(A)& 500 kton$\cdot$year each for NBB($E_{\rm peak}=6$ GeV) and
NBB($E_{\rm peak}=4$ GeV) \\
 & at $L=2,100$ km   \\
\end{tabular}
\eqlab{ex_condition_A}
\\
&&\hspace*{-10ex}
\begin{tabular}{ll}
(B)& In addition to (A), 100 kton$\cdot$year data from 
NBB($\langle p_\pi \rangle=2$ GeV) \\
& at $L=295$ km are included in the fit. 
\end{tabular}
\eqlab{ex_condition_B}
\eea
\eqlab{ex_condition}
\eseq

We take the 12 sets of  
$N(\mu, E_{\rm peak})$ and $N(e, E_{\rm peak})$, which are generated for 
the LMA parameters of,  
\eqref{modelpara} at $\sin^22\theta_{_{\rm CHOOZ}}=0.02,~0.04,~0.1$,
with the matter density $\rho=3$ g/cm$^3$
by assuming the hierarchy I.
Summing up, the expected numbers of signals and background events
are calculated for the following input (`true') parameters:
\bseq
\eqlab{IvsIII_true}
\bea
\sin^22\theta_{_{\rm ATM}}^{true}&=&1.0
\,, ~~~~~~~~~~~~~~~~~~~~~~~
\delta m^{2~true}_{_{\rm ATM}}=3.5\times10^{-3} ~{\rm eV}^2
\,, \eqlab{IvsIII_t_atm}
\\
\sin^22\theta_{_{\rm SOL}}^{true}&=&0.8
\,, ~~~~~~~~~~~~~~~~~~~~~~~
\delta m^{2~true}_{_{\rm SOL}}=10\times10^{-5} ~{\rm eV}^2
\,, \eqlab{IvsIII_t_sol}
\\
\sin^22\theta_{_{\rm CHOOZ}}^{true}&=&0.02\,,~0.04\,,~0.10
\,,~~~~~
\delta_{_{\rm MNS}}^{true}=
 0^\circ,~90^\circ,~180^\circ,~ 270^\circ\,, 
\eqlab{IvsIII_t_delta}\\
\delta m^2_{12}&=&\delta m^{2~true}_{_{\rm SOL}}\,,
~~~~~~~~~~~~~~~~
\delta m^2_{13}=\delta m^{2~true}_{_{\rm ATM}}\,, 
~~({\rm hierarchy~I})
\eqlab{IvsIII_t_case} \\
f_{_{\rm FLUX}}^{true}&=&1.0,
~~~~~~~~~~~~~~~~~~~~~~~
\rho^{true}=3{\mbox{{g/cm}}}^3\,.
\eea
\eseq
The 12 cases of the input data sets
are labeled by the input (`true') values of 
$\sin^22\theta_{_{\rm CHOOZ}}^{true}$ = 
0.02 (thin lines), 0.04 (medium-thick lines), 0.1 (thick lines) 
and 
$\delta_{_{\rm MNS}}^{true}$
= $0^\circ$ (solid lines), $90^\circ$ (long-dashed lines), 
$180^\circ$ (short-dashed lines), $270^\circ$ (dot-dashed lines).

The $\chi^2$ fit has been performed by assuming the hierarchy III.
The minimum of the $\chi^2$ function is found for a 
$\sin^22\theta_{_{\rm CHOOZ}}$ 
value in the range below 0.12, by varying the parameters
$\delta m^2_{_{\rm SOL}}$ and $\sin^22\theta_{_{\rm SOL}}$  
within the LMA allowed region, \eqref{MSW_L},
and the remaining three parameters, 
$\delta m^2_{_{\rm ATM}}$, $\sin^22\theta_{_{\rm ATM}}$ and 
$\delta_{_{\rm MNS}}$ freely. 
The uncertainties in the total fluxes of the neutrino beams, \eqref{flux_unc}, 
and the matter density, \eqref{matter_unc}, are taken into account.
Summing up, the fitting parameters used to obtain the $\chi^2$ functions of 
\Fgref{IvsIII_B} are :
\bseq
\eqlab{IvsIII_fit}
\bea
\sin^22\theta_{_{\rm ATM}}&:&{\rm free}\,,
~~~~~~~~~
\delta m^2_{_{\rm ATM}}~:~ {\rm free}\,,
\eqlab{IvsIII_f_atm}\\
\sin^22\theta_{_{\rm SOL}}&=&0.7 -0.9\,, 
~~~
\delta m^2_{_{\rm SOL}}=(3-15)\times10^{-5} ~{\rm eV}^2\,,
{\rm~~~(LMA)}
\eqlab{IvsIII_f_sol}\\
\sin^22\theta_{_{\rm CHOOZ}}~&:&~{\rm free}\,, 
~~~~~~~~
\delta_{_{\rm MNS}}~:~{\rm free}\,, \eqlab{IvsIII_f_delta}\\ 
\delta m^2_{12}&=&\delta m^2_{_{\rm SOL}}\,, 
~~~~~~
\delta m^2_{13}=-\delta m^2_{_{\rm ATM}}\,, ~~~({\rm hierarchy~III})\\
f_{_{\rm FLUX}}~&:&~{\rm free}\,,
~~~~~~~~
\rho ~:~{\rm free}\,.
\eea
\eseq

In \Fgref{IvsIII_B} (a),
we show the resulting $\chi^2_{min}$ from the VLBL experiments
at $L=2,100$ km.
The minimum $\chi^2$ for
$\sin^22\theta_{_{\rm CHOOZ}}^{true}$ = 0.1
is always larger than 30 
for $\sin^22\theta_{_{\rm CHOOZ}}< 0.12$.
We can hence distinguish the neutrino mass hierarchy I from III.
It is also possible to make the distinction 
at more than $3\sigma$ level for
the cases with $\delta_{_{\rm MNS}}^{true}=180^\circ$ and $270^\circ$
at $\sin^22\theta_{_{\rm CHOOZ}}^{true}=0.04$ whereas 
the distinction disappears for $\delta_{_{\rm MNS}}^{true}=0^\circ$ and
$90^\circ$.
This is because,
the hierarchy I predictions of $N(e)$'s for
$\delta_{_{\rm MNS}}^{true}=0^\circ$ and $90^{\circ}$
at $\sin^22\theta_{_{\rm CHOOZ}}^{true}=0.04$
can be reproduced by the hierarchy-III model
if we choose larger $\sin^22\theta_{_{\rm CHOOZ}}$ $(\gsim0.08)$;
see \Fgref{Cir_B} and \Fgref{main_B}.

In \Fgref{IvsIII_B}(b), the minimum $\chi^2$ values are 
shown when data from HIPA-to-SK experiment are added.
The remarkable difference between \Fgref{IvsIII_B}(a) and (b) is
found for the small $\sin^22\theta_{_{\rm CHOOZ}}^{true}$ cases when 
the fitting parameter $\sin^22\theta_{_{\rm CHOOZ}}$ is large.
This can be explained as follows.
In the HIPA-to-SK LBL experiment,
the predicted $N(e,E_{\rm peak})$ of the hierarchy III is not much 
smaller than that of the hierarchy I (see \Fgref{Cir_SK}).
In particular, $N(e,E_{\rm peak}^{})$ calculated for the
hierarchy I at $\sin^22\theta_{_{\rm CHOOZ}}^{}<0.04$ is
significantly smaller than that for the hierarchy III
at $\sin^22\theta_{_{\rm CHOOZ}}>0.1$.
This leads to the enhancement of 
the minimum $\chi^2$ in \Fgref{IvsIII_B}(b)
at large $\sin^22\theta_{_{\rm CHOOZ}}^{}$.
We find that
the data obtained from the HIPA-to-SK experiments
are useful,
which allow us to determine 
the neutrino mass hierarchy (between I and III) for all four values of 
$\delta_{_{\rm MNS}}^{true}$ at $3\sigma$ level 
when $\sin^22\theta_{_{\rm CHOOZ}}^{true} \gsim 0.04$, 
and at one-sigma level when
$\sin^22\theta_{_{\rm CHOOZ}}^{true}\gsim 0.02$. 

Let us now study the possibility of distinguishing the
solar-neutrino oscillation scenarios by using the VLBL experiments
with HIPA.
Because the predictions of the SMA and VO scenarios differ very little
in \Fgref{main_B},
we compare the predictions of the SMA and LMA scenarios.

\begin{figure}[th]
\begin{center}
{\scalebox{0.48}{\includegraphics{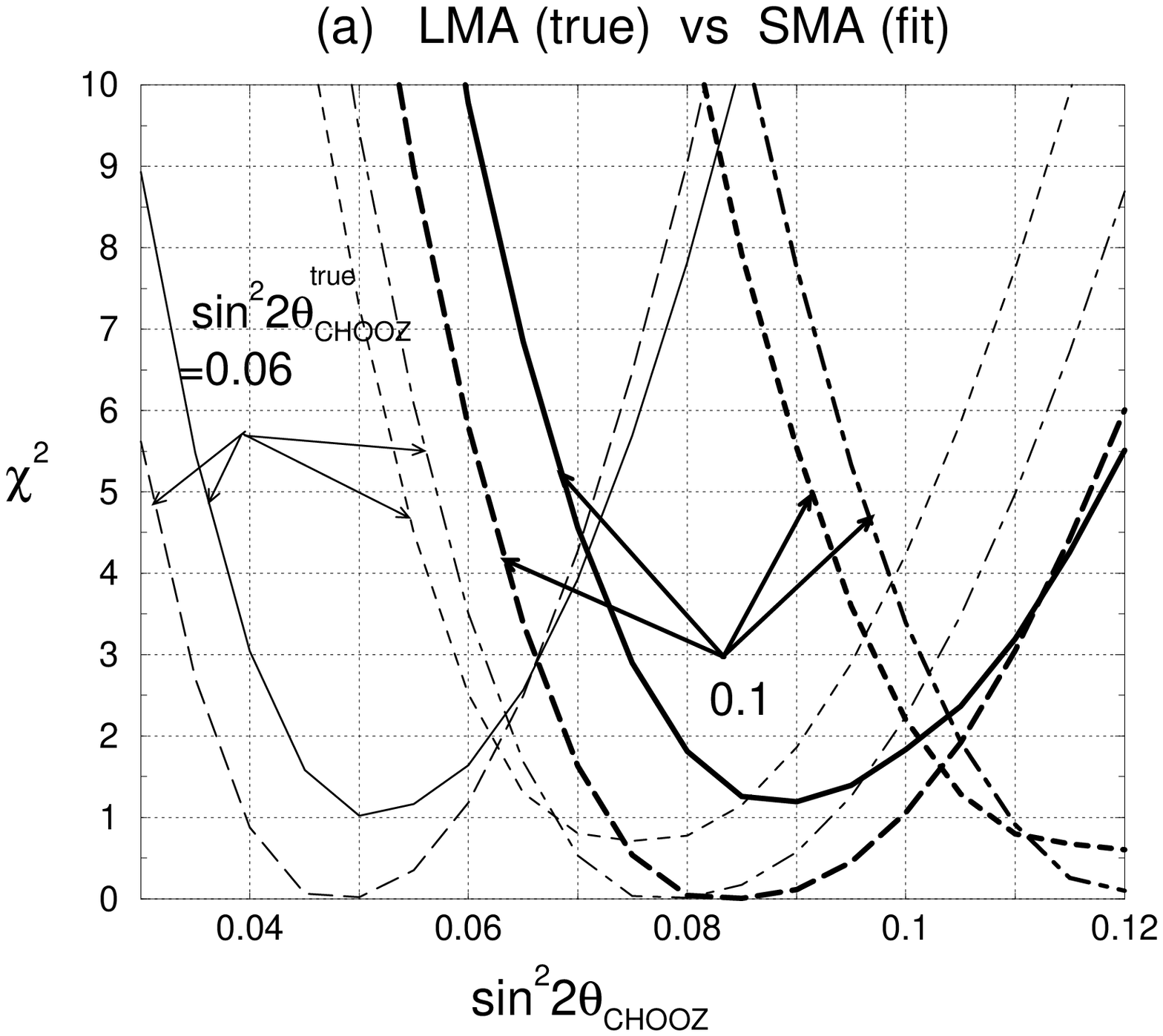}}} 
{\scalebox{0.48}{\includegraphics{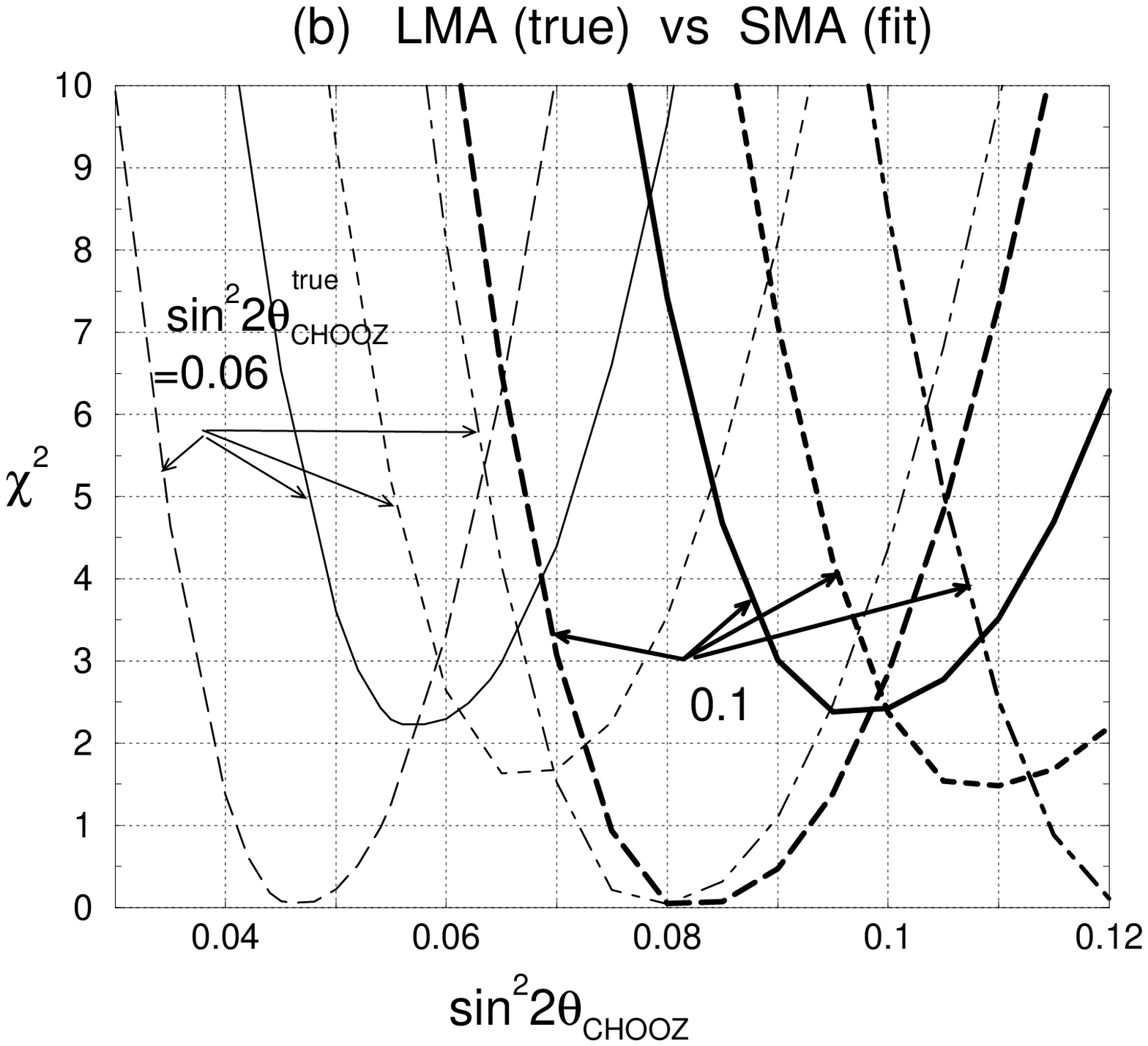}}} 
\end{center}	     
\caption{
Minimum $\chi^2$ as functions of the  
parameter $\sin^22\theta_{_{\rm CHOOZ}}$.
The mean values of the input data,
$N(\mu, E_{\rm peak})$ and $N(e, E_{\rm peak})$,
are calculated for the 8 LMA points:
$\sin^22\theta_{_{\rm CHOOZ}}^{true}$ = 
0.06 (thin lines), 0.1 (thick lines) and
$\delta_{_{\rm MNS}}^{true}$
= $0^\circ$ (solid lines), $90^\circ$ (long-dashed lines), 
$180^\circ$ (short-dashed lines), $270^\circ$ (dot-dashed lines).
The $\chi^2$ fit has been performed by assuming the SMA scenario for
$\delta m^2_{_{\rm SOL}}$ and $\sin^22\theta_{_{\rm SOL}}$
and varying the remaining four parameters, 
$\delta m^2_{_{\rm ATM}}$, $\sin^22\theta_{_{\rm ATM}}$, 
$\delta_{_{\rm MNS}}$ and $\sin^22\theta_{_{\rm CHOOZ}}^{}$.
(a) With 500 kton$\cdot$year data each with NBB($E_{\rm peak}=6$ GeV) and
NBB($E_{\rm peak}=4$ GeV) at $L=2,100$ km.
(b) In addition, 100 kton$\cdot$year data from
NBB($\langle p_\pi \rangle=2$ GeV)
at $L=295$ km are included.}
\Fglab{LMAvsSMA_B}
\end{figure}

\subsubsection{LMA v.s. SMA/LOW/VO}
We show in \Fgref{LMAvsSMA_B} the minimum $\chi^2$ as functions of 
the parameter $\sin^22\theta_{_{\rm CHOOZ}}$ by assuming the SMA
scenario, when the mean values of 
$N(\mu, E_{\rm peak})$ and $N(e, E_{\rm peak})$ are calculated for the LMA 
point with hierarchy I (\eqref{IvsIII_true}). 
The results of \Fgref{LMAvsSMA_B} (a)
are obtained from the $L=2,100$ km VLBL experiments only,
(b) are obtained after adding the HIPA-to-SK data; see
(A) and (B) in \eqref{ex_condition}. 
 
We use the same input parameters as in \eqref{IvsIII_true},
but only two cases of $\sin^22\theta_{_{\rm CHOOZ}}^{true}$\,,
0.06 and 0.1, are examined.
The 8 cases of the input data are again 
labeled by the input (true) values of 
$\sin^22\theta_{_{\rm CHOOZ}}^{true}$ = 0.06 (thin lines),
0.1 (thick lines)
and
$\delta_{_{\rm MNS}}^{true}$
=
$0^\circ$ (solid lines),
$90^\circ$ (long-dashed lines), 
$180^\circ$ (short-dashed lines),
$270^\circ$ (dot-dashed lines).

The $\chi^2$ fit has been performed by assuming the SMA solution, 
\eqref{modelpara_sol_sma},
with the hierarchy I.
The other fitting parameters in the three-neutrino model
are taken for the same as in \Fgref{IvsIII_B}. 
Summing up, the parameters used in the fit are
\bseq
\eqlab{LMAvsSMA_fit}
\bea
\sin^22\theta_{_{\rm ATM}}&:&{\rm free}\,, 
~~~~~~~~
\delta m^2_{_{\rm ATM}}~:~ {\rm free}\,,\\
\sin^22\theta_{_{\rm SOL}}&=&7 \times 10^{-3}\,, 
~~
\delta m^2_{_{\rm SOL}}=5\times 10^{-6} ~{\rm eV}^2\,,
~~~{\rm (SMA)}\\
\sin^22\theta_{_{\rm CHOOZ}}&:&{\rm free}\,, 
~~~~~~~~~
\delta_{_{\rm MNS}}~:~ {\rm free}\,,\\
\delta m^2_{12}&=&\delta m^2_{_{\rm SOL}}\,, 
~~~~~~
\delta m^2_{13}=\delta m^2_{_{\rm ATM}}\,, ~~~({\rm hierarchy~I})\\
f_{_{\rm FLUX}}~&:&~{\rm free}\,,
~~~~~~~~
\rho ~:~{\rm free}\,.
\eea
\eseq

In \Fgref{LMAvsSMA_B}(a),
the minimum $\chi^2$ becomes almost zero when
$\delta_{_{\rm MNS}}^{true}=90^\circ$ or $270^\circ$,
whereas
the results with the $\delta_{_{\rm MNS}}^{true}=0^\circ$ and $180^\circ$
give non-zero values of the minimum $\chi^2$.
This is because the two beams, NBB
($E_{\rm peak} = 4$ GeV) and NBB($E_{\rm peak} = 6$ GeV),
give `orthogonal' $\delta_{_{\rm MNS}}$ dependences
of $N(e,E_{\rm peak})$  in \Fgref{main_B}.
For $\delta_{_{\rm MNS}}^{true}=90^\circ$ and $270^\circ$,
the predictions of the LMA scenario for
$N(e,4~{\rm GeV})$ and $N(e,6~{\rm GeV})$ can be 
reproduced in the SMA scenarios where the fitting parameter
$\sin^22\theta_{_{\rm CHOOZ}}$ is approximately chosen,
but such reproduction does not work perfectly
for the $\delta_{_{\rm MNS}}^{true}=0^\circ$ and $180^\circ$.
Although the minimum $\chi^2$ value is raised in \Fgref{LMAvsSMA_B}(b)
after combining the HIPA-to-SK data,
it seems difficult to distinguish the LMA scenario from the SMA scenario.

The $\chi^2$ fit by assuming the LOW or VO scenarios give the same results
as those obtained by assuming the SMA scenario.
This is understood from the Hamiltonian \eqref{hamiltonian1}
that governs the neutrino oscillation in the vacuum.
The effects of solar-neutrino oscillation modes
are proportional to the term with
$\delta m_{12}^2/m_{13}^2$.
 For the LOW and VO scenarios it is the extreme smallness of
$\delta m_{_{\rm SOL}}^2/m_{_{\rm ATM}}^2$
that makes the effects unobservable.
 In case of the SMA scenario, the effects
are also tiny because the mixing matrix elements
multiplying the $\delta m_{_{\rm SOL}}^2/m_{_{\rm ATM}}^2$
factor are also very small.

\subsubsection{$\sin^22\theta_{_{\rm CHOOZ}}^{}$ and $\delta_{_{\rm MNS}}^{}$}
\begin{figure}[tb]
\begin{center}
{\scalebox{0.67}{\includegraphics{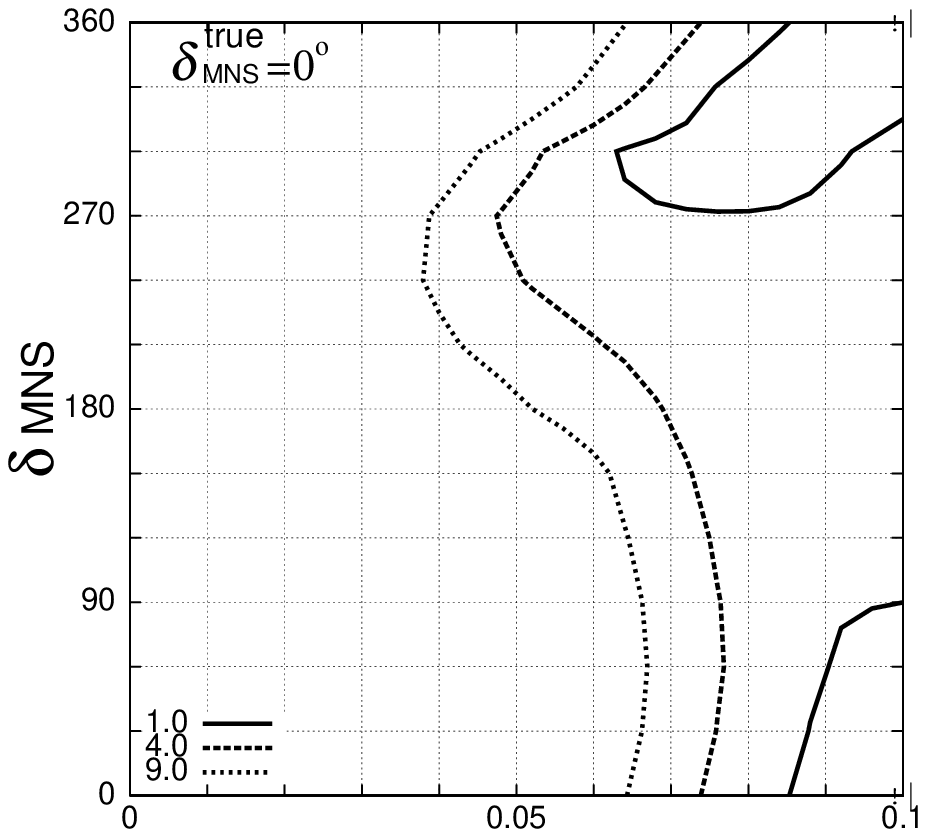}}} 
{\scalebox{0.67}{\includegraphics{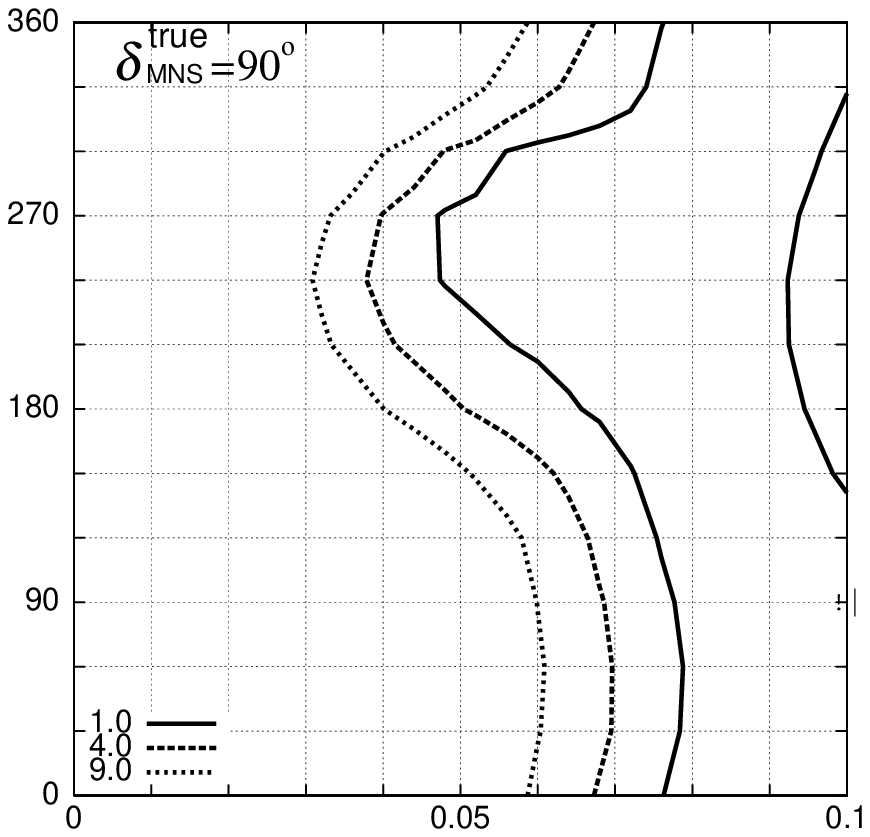}}} 
{\scalebox{0.67}{\includegraphics{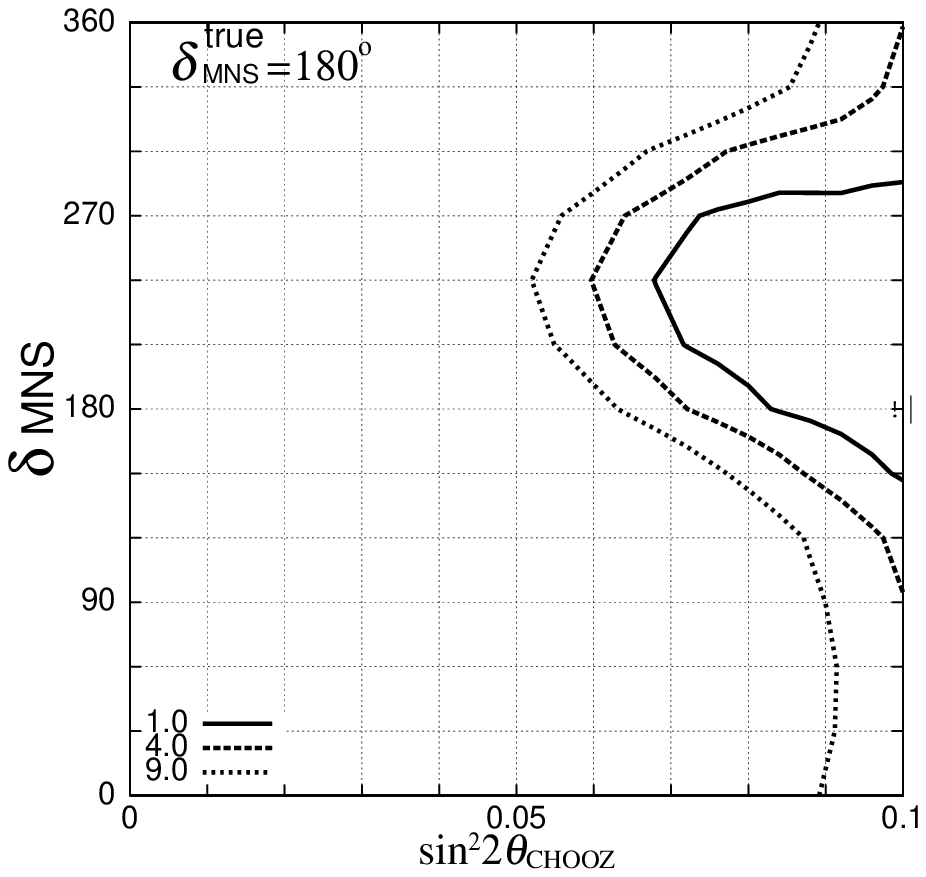}}} 
{\scalebox{0.67}{\includegraphics{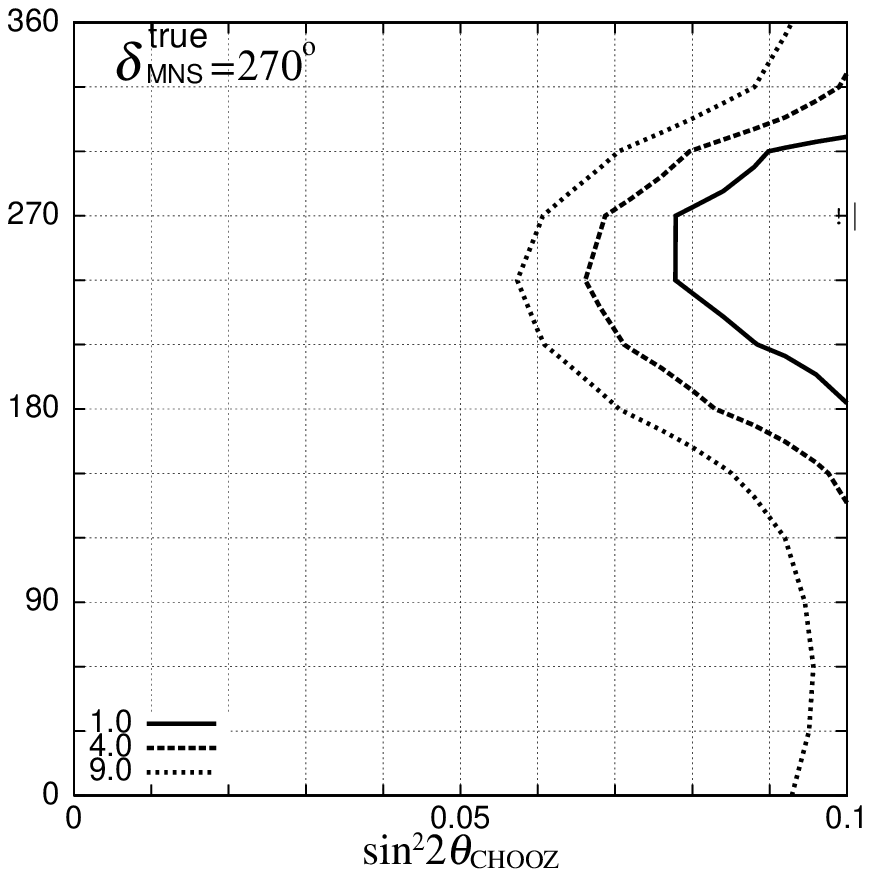}}} 
\end{center}	     
\caption{
Regions allowed by the VLBL experiment at 
$L=$2,100 km are shown
in the plain of
$\sin^2 2\theta_{_{\rm CHOOZ}}^{}$
and $\delta_{_{\rm MNS}}^{}$.
The experimental conditions are given in \eqref{ex_condition_A}.
The input data are calculated for the LMA parameters,
\eqref{IvsIII_true}, at $\sin^22\theta_{_{\rm CHOOZ}}^{true}=0.1$
and for four values of $\delta_{_{\rm MNS}}^{true}$\,,
0$^\circ$, 90$^\circ$, 180$^\circ$, and 270$^\circ$.
In each figure, the input parameter point
$(\sin^22\theta_{_{\rm CHOOZ}}^{true}\,,
\delta_{_{\rm MNS}}^{true})$ is shown by a solid-circle,
and the regions where $\chi^2_{min}<$1, 4, and 9 are
depicted by solid, dashed, and dotted boundaries, respectively.
}
\Fglab{H2B_wo_SK}
\end{figure}
\begin{figure}[tb]
\begin{center}
{\scalebox{0.670}{\includegraphics{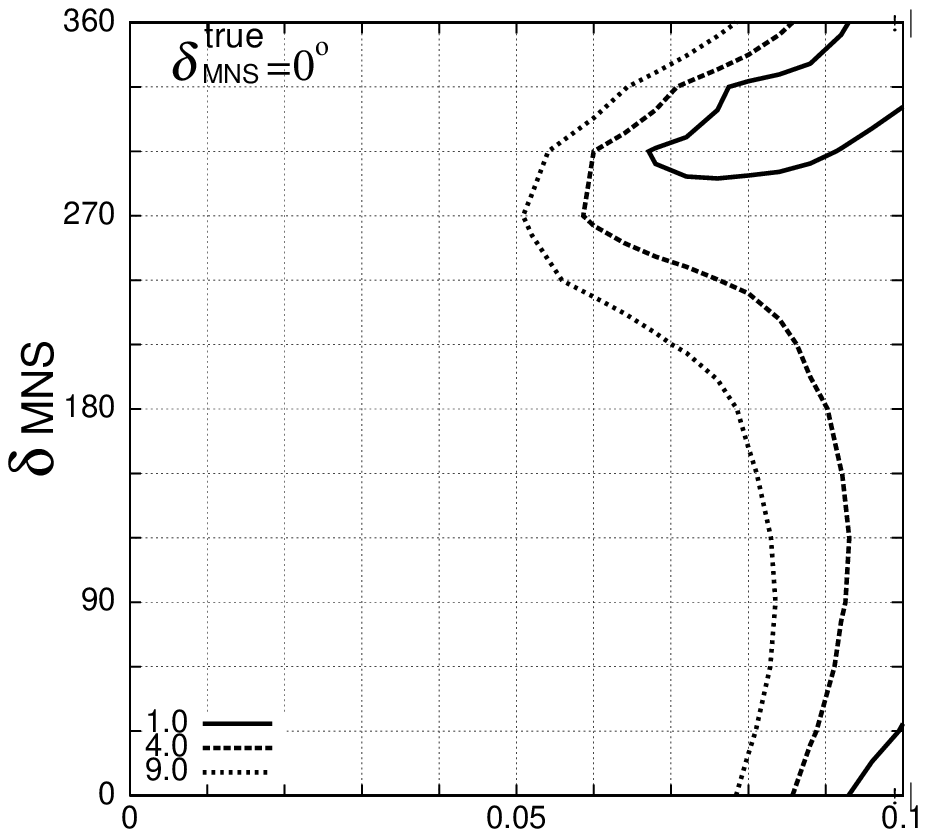}}} 
{\scalebox{0.670}{\includegraphics{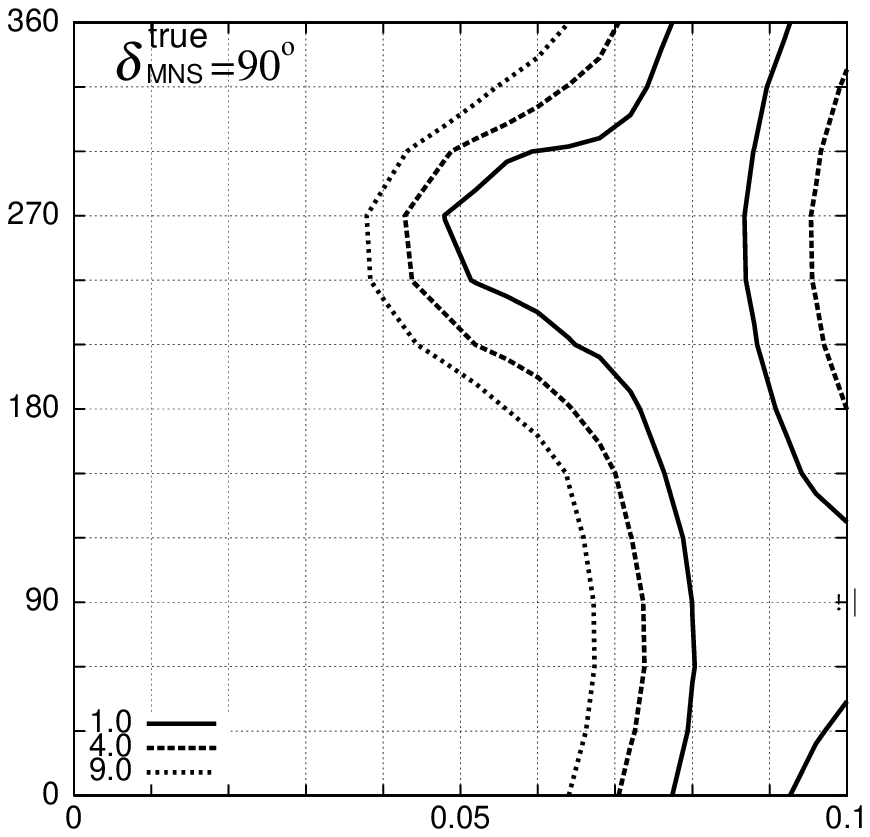}}} 
{\scalebox{0.670}{\includegraphics{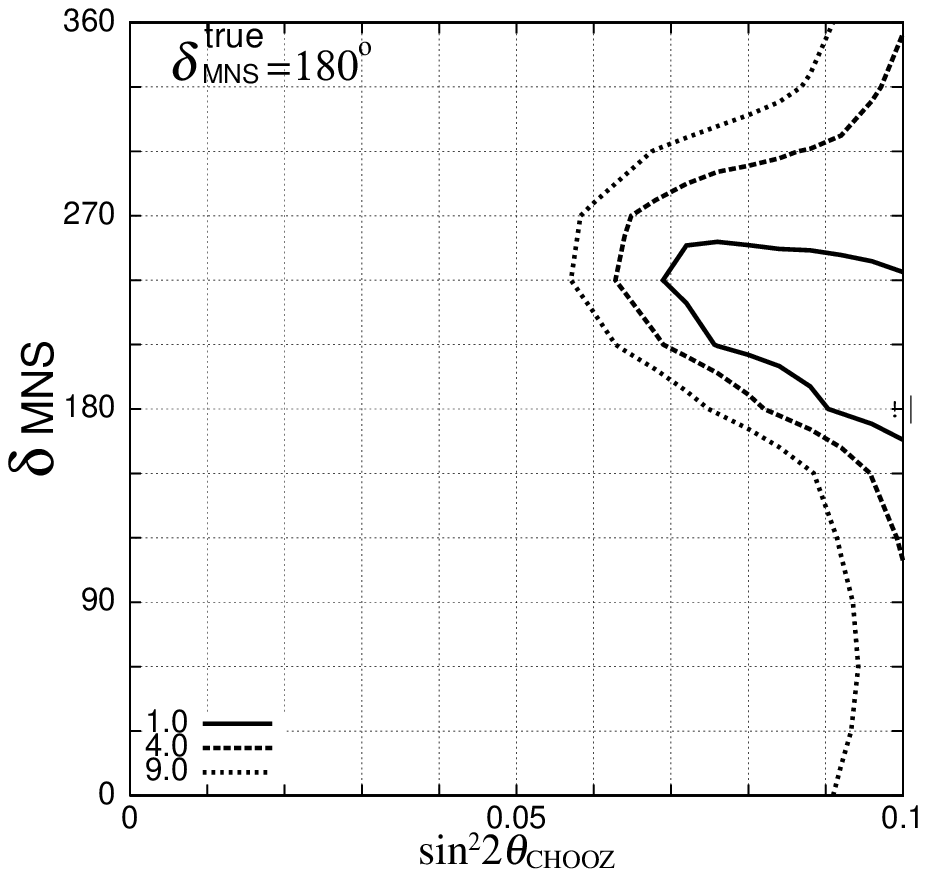}}} 
{\scalebox{0.670}{\includegraphics{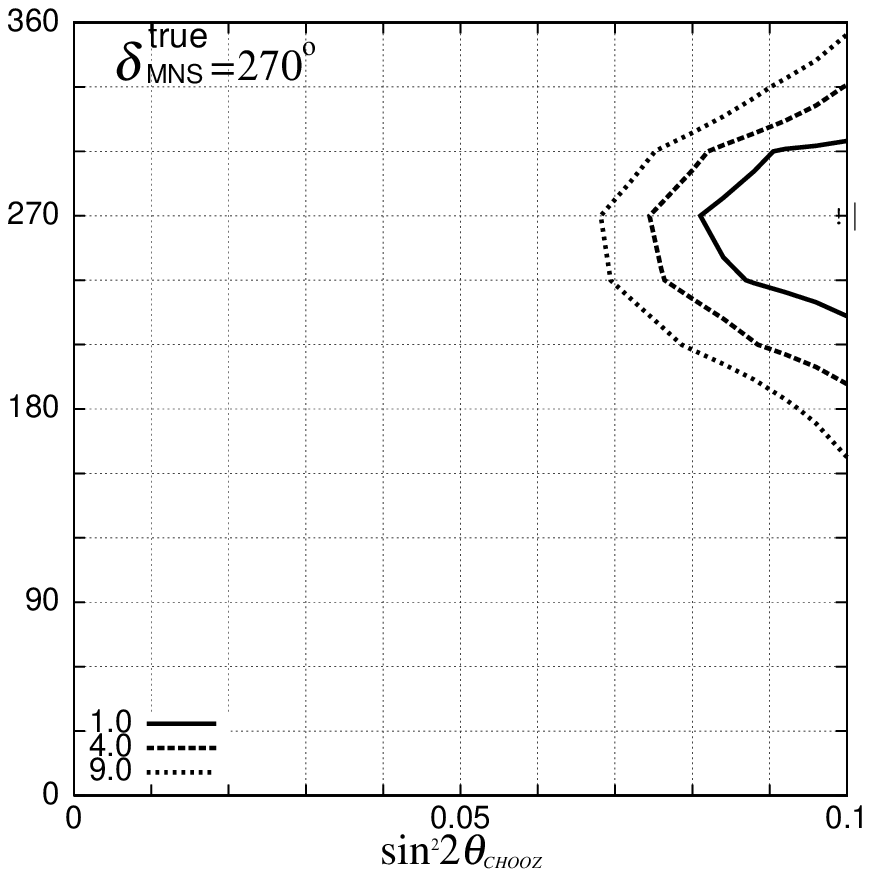}}} 
\end{center}	     
\caption{
Regions allowed by the VLBL experiment at $L=$2,100 km
and the LBL experiment at $L=$295 km,
with the experimental condition of \eqref{ex_condition_B}.
The symbols are the same as those in \Fgref{H2B_wo_SK}.
}
\Fglab{H2B_w_SK}
\end{figure}
 \Fgref{H2B_wo_SK} shows the regions 
in the $\sin^2 2 \theta_{_{\rm CHOOZ}}$ v.s.
$\delta_{_{\rm MNS}}$ plane
which are allowed by the VLBL
experiment at $L=$2,100 km with 500 kton$\cdot$year
each for NBB($E_{\rm peak}=6$GeV) and
NBB($E_{\rm peak}=4$GeV);
\eqref{ex_condition_A}.
The input data are calculated for the LMA parameters,
\eqref{IvsIII_true}, at $\sin^22\theta_{_{\rm CHOOZ}}^{true}=0.1$
and for four values of $\delta_{_{\rm MNS}}^{true}$\,,
0$^\circ$, 90$^\circ$, 180$^\circ$, and 270$^\circ$.
The $\chi^2$ fit has been performed by assuming that
$\delta m^2_{_{\rm SOL}}$ and $\sin^22\theta_{_{\rm SOL}}$
are in the LMA region \eqref{MSW_L} while the rest of
the parameters are freely varied;
\bseq
\eqlab{IvsI_fit}
\bea
\sin^22\theta_{_{\rm ATM}}&:&{\rm free}\,,
~~~~~~~~~
\delta m^2_{_{\rm ATM}}~:~ {\rm free}\,,
\eqlab{IvsI_f_atm}\\
\sin^22\theta_{_{\rm SOL}}&=&0.7 -0.9\,, 
~~~
\delta m^2_{_{\rm SOL}}=(3-15)\times10^{-5} ~{\rm eV}^2\,,
{\rm~~~(LMA)}
\eqlab{IvsI_f_sol}\\
\sin^22\theta_{_{\rm CHOOZ}}~&:&~{\rm free}\,, 
~~~~~~~~
\delta_{_{\rm MNS}}~:~{\rm free}\,, \eqlab{IvsI_f_delta}\\ 
\delta m^2_{12}&=&\delta m^2_{_{\rm SOL}}\,, 
~~~~~~
\delta m^2_{13}=\delta m^2_{_{\rm ATM}}\,, ~~~({\rm hierarchy~I})\\
f_{_{\rm FLUX}}~&:&~{\rm free}\,,
~~~~~~~~
\rho ~:~{\rm free}\,.
\eea
\eseq
In each figure, the input parameter point
$(\sin^22\theta_{_{\rm CHOOZ}}^{true}\,,~
\delta_{_{\rm MNS}}^{true})$ is shown by a solid-circle,
and the regions where $\chi^2_{min}<$1, 4, and 9 are
depicted by solid, dashed, and dotted boundaries, respectively.

A few comments are in order.
From the top-right figure for $\delta_{_{\rm MNS}}^{true}=90^\circ$,
we learn that $\delta_{_{\rm MNS}}^{}$ is not constrained by this
experiment. 
The reason can be qualitatively understood by studying the LMA
predictions shown in \Fgref{Cir_B} and \Fgref{main_B}.
Note, however, that the LMA cases shown in \Fgref{main_B}
are for $\delta m_{_{\rm SOL}}^2=\numt{5}{-5}$eV$^2$
and $\numt{15}{-5}$eV$^2$, while the input data in our example
are obtained for $\delta m_{_{\rm SOL}}^2=\numt{10}{-5}$eV$^2$.
An appropriate interpolation between the two cases is hence needed.
From \Fgref{Cir_B}, we find that $\delta_{_{\rm MNS}}^{}=90^\circ$
(solid-square) points for the hierarchy I lie in the lower $N(e)$
corner of the grand circle made of the predictions of all
$\delta_{_{\rm MNS}}$.
On the other hand, we can tell from \Fgref{main_B} that the same
values of low $N(e)$ can be obtained for the other $\delta_{_{\rm MNS}}$
values typically by reducing the $\sin^22\theta_{_{\rm CHOOZ}}^{}$
parameter.
In case of $\delta_{_{\rm MNS}}^{true}=270^\circ$, both
$\delta_{_{\rm MNS}}$ and $\sin^22\theta_{_{\rm CHOOZ}}^{}$
are constrained in the bottom-right figure,
because $\delta_{_{\rm MNS}}^{}=270^\circ$ gives the largest $N(e)$
along the relevant grand circle in \Fgref{Cir_B}(b).
When $\delta_{_{\rm MNS}}$ are chosen differently,
we need to increase $\sin^22\theta_{_{\rm CHOOZ}}^{}$ to make
$N(e)$ large, but it then gets into conflict with the CHOOZ bound
\eqref{chooz}.

\begin{figure}[tb]
\begin{center}
{\scalebox{0.670}{\includegraphics{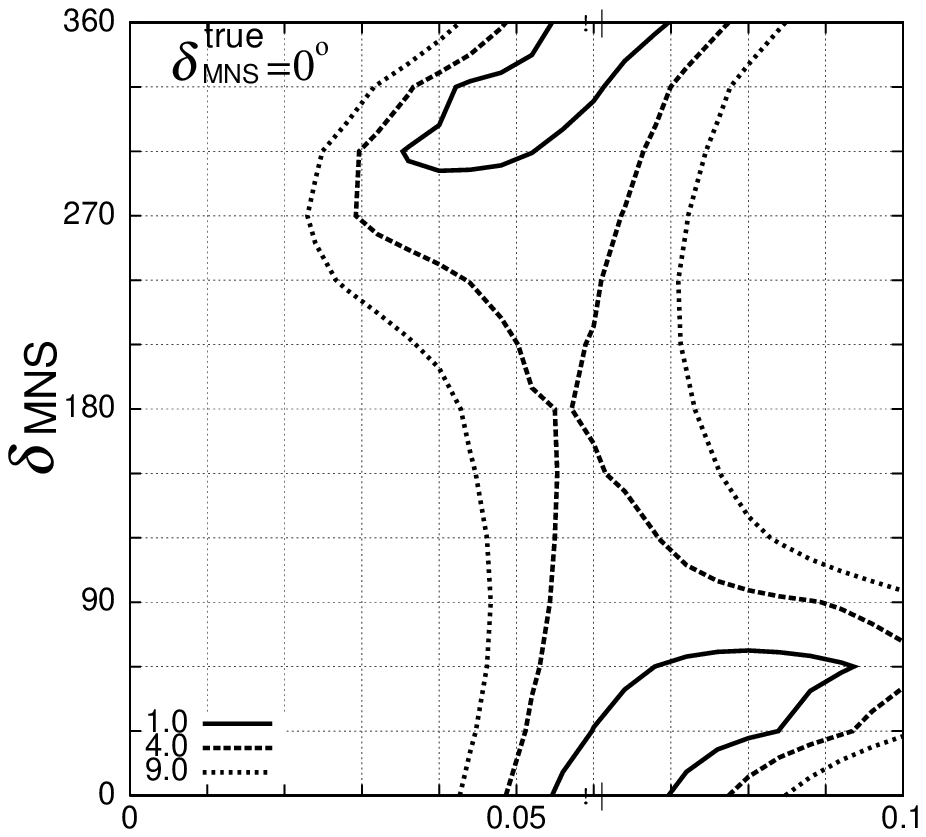}}} 
{\scalebox{0.670}{\includegraphics{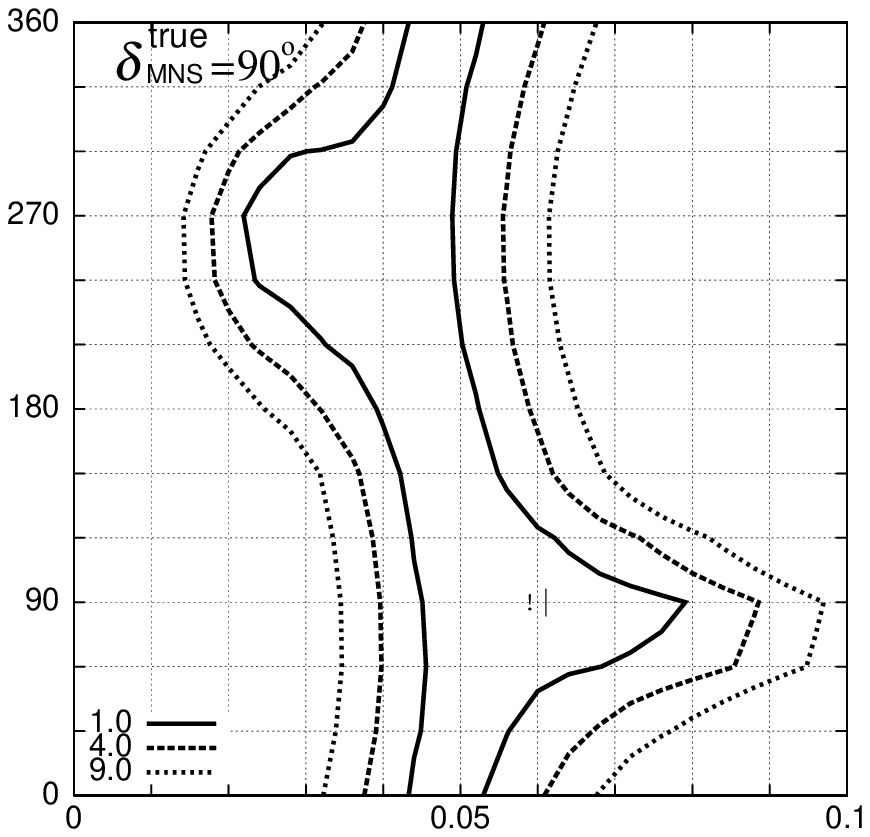}}} 
{\scalebox{0.670}{\includegraphics{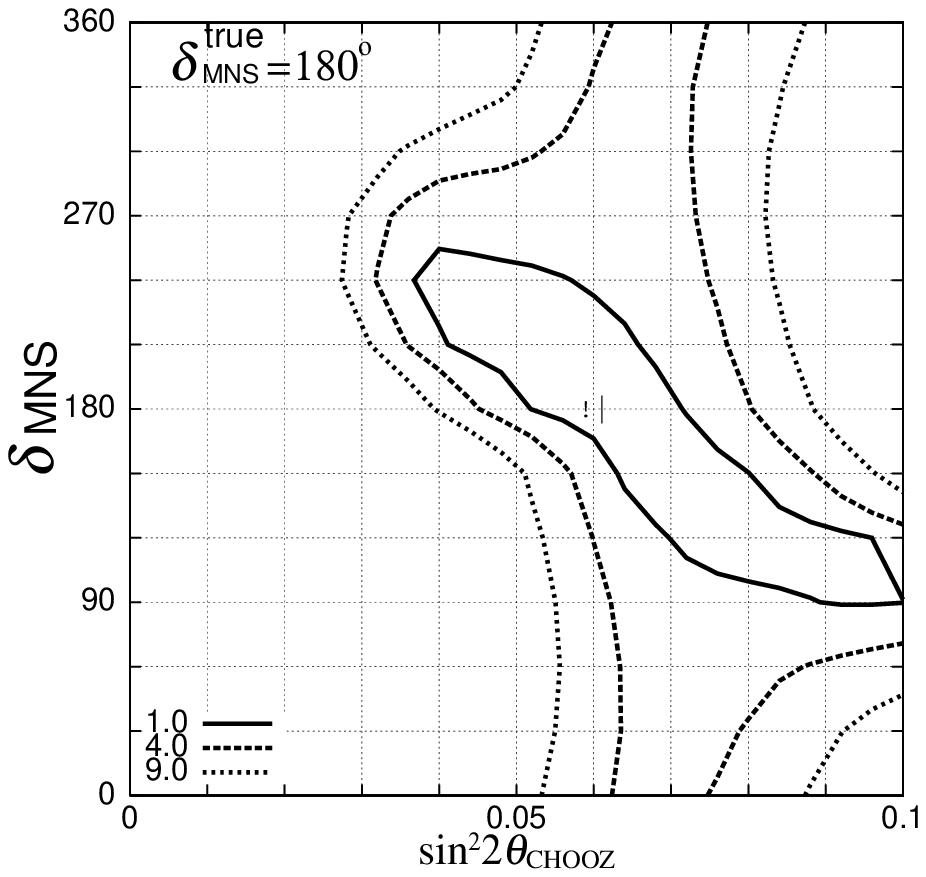}}} 
{\scalebox{0.670}{\includegraphics{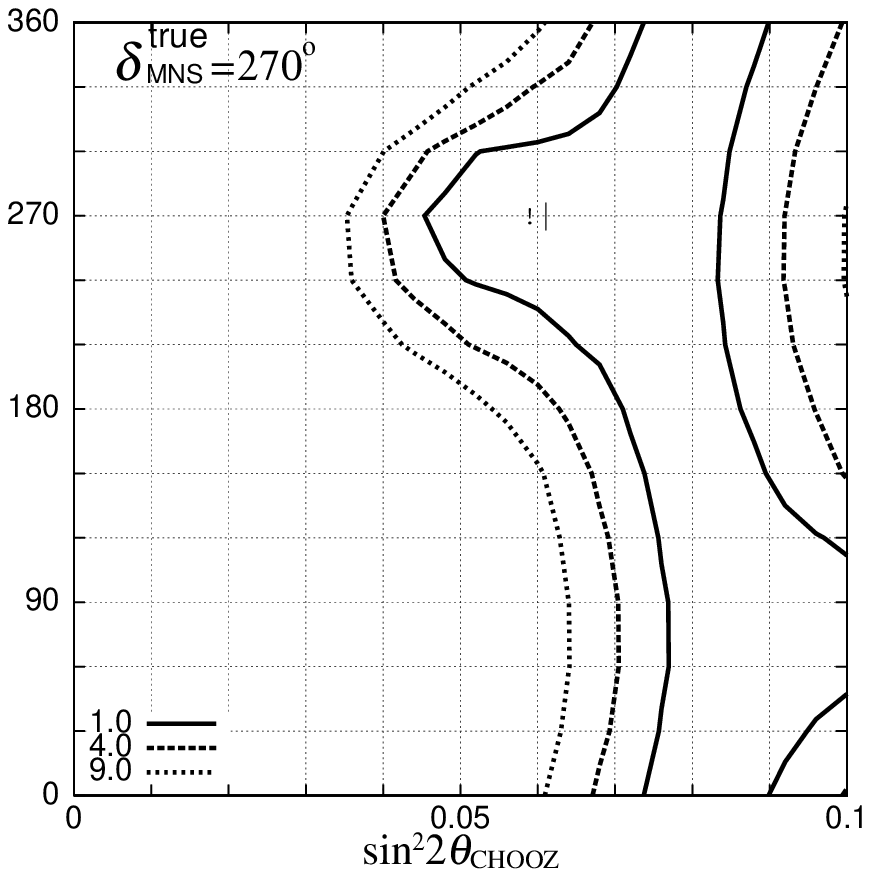}}} 
\end{center}	     
\caption{
The same as \Fgref{H2B_w_SK}, but when the input data are calculated
for $\sin^22\theta_{_{\rm CHOOZ}}^{true}=0.06$.
}
\Fglab{H2B_w_SK2}
\end{figure}
\begin{table}[t]
\begin{center}
\begin{tabular}{|c|c||c|c|c|c|}
\hline
\multicolumn{1}{|c}{} &  \multicolumn{1}{c||}{}
  & $\delta_{_{\rm MNS}}^{true}=0^\circ$ & $90^\circ$ & $180^\circ$ & $270^\circ$  \\
  \hline
  \hline
$\delta m^2_{_{\rm ATM}}$
& (A)  
& 3.50$^{+0.08}_{-0.10}$
& 3.50$^{+0.10}_{-0.15}$
& 3.50$^{+0.11}_{-0.16}$
& 3.50$^{+0.11}_{-0.15}$
\\    
  \cline{2-6}
$(\times 10^{-3})$ (eV$^2$)
& (B) 
& 3.50$^{+0.07}_{-0.06}$
& 3.50$^{+0.09}_{-0.11}$
& 3.50$^{+0.10}_{-0.10}$
& 3.50$^{+0.10}_{-0.11}$
\\    
  \hline
  \hline
$ \delta \l({\rm sin}^22\theta_{_{\rm ATM}} \r) $
& (A)
& $>0.976$
& $>0.976$ 
& $>0.975$ 
& $>0.976$ \\
  \cline{2-6}
& (B) 
& $>0.991$
& $>0.991$
& $>0.991$
& $>0.991$\\
  \hline
 \end{tabular}
\end{center}
\caption{
The mean and the one-sigma errors of
$\delta m^{2}_{_{\rm ATM}}$ and 
the one-sigma lower bound of 
$\sin^22\theta_{_{\rm ATM}}^{true}$
when the input data are calculated for the four
LMA points (\eqref{IvsIII_true} at
$\sin^22\theta_{_{\rm CHOOZ}}^{true}$ = 0.06) 
and the fits are performed by assuming
the LMA scenarios.
(A) Results with 500 kton$\cdot$year each for NBB($E_{\rm peak} = 6$ GeV) and
NBB($E_{\rm peak}=4$ GeV) at $L=2,100$ km.
(B) In addition, 100 kton$\cdot$year data from
NBB($\langle p_\pi \rangle=2$ GeV)
at $L = 295$ km are included in the fit.  
}
\tblab{atm_lma}
\end{table}

 The left figures of \Fgref{H2B_w_SK2} show that
$\delta_{_{\rm MNS}}^{}$ can be constrained at $1\sigma$
level when $\delta_{_{\rm MNS}}^{true}=0^\circ$
(top-left), or at $2\sigma$ level when $180^\circ$
(bottom-left).
 The reason for this behavior is more subtle.
 We find that it is essentially the ratio of $N(e,4$GeV$)/N(e,6$GeV$)$
which is difficult to reproduce with the other $\delta_{_{\rm MNS}}^{}$.
In \Fgref{Cir_B}, we can see among the four representative
$\delta_{_{\rm MNS}}^{}$ predictions that $N(e,4$GeV$)/N(e,6$GeV$)$
is smallest at $\delta_{_{\rm MNS}}=0^\circ$ (solid-circle)
and it is largest at $\delta_{_{\rm MNS}}=180^\circ$ (open-circle),
while this ratio is almost the same for
$\delta_{_{\rm MNS}}=90^\circ$ (solid-square) and
$\delta_{_{\rm MNS}}=270^\circ$ (open-square).
Because it is the energy-dependence of $N(e)$ that has some 
discriminating power for $\delta_{_{\rm MNS}}^{}$,
detectors with the capability of
measuring neutrino energy \cite{BAND} may have better
sensitivity for the $\delta_{_{\rm MNS}}^{}$ angle.

In \Fgref{H2B_w_SK} we show the same plots as
\Fgref{H2B_wo_SK} where in addition to the data from
the VLBL experiment at $L=$2,100 km,
we also include the 100 kton$\cdot$year data from
the $L=$295 km experiment;
see \eqref{ex_condition_B}.
Although the area of $\chi^2_{min}<$1, 4, and 9 regions
decrease significantly,
the qualitative features of our findings remain.
$\delta_{_{\rm MNS}}$ can be constrained when
$\delta_{_{\rm MNS}}^{true}=270^\circ$,
because it predicts largest $N(e)$.
$\delta_{_{\rm MNS}}^{}$ can be weakly constrained when
$\delta_{_{\rm MNS}}^{true}=0^\circ$ or $180^\circ$,
it cannot be constrained when 
$\delta_{_{\rm MNS}}^{true}=90^\circ$.
The constraints on $\sin^22\theta_{_{\rm CHOOZ}}^{}$
improves, but only slightly.

\Fgref{H2B_w_SK2} shows the same constraints as in
\Fgref{H2B_w_SK} but when the input data are calculated
for $\sin^22\theta_{_{\rm CHOOZ}}^{true}=0.06$.
Now that the input $\sin^22\theta_{_{\rm CHOOZ}}^{}$ is
chosen significantly below the CHOOZ bound \eqref{chooz},
the global picture of the constraint from
the VLBL+LBL experiments with HIPA is more clearly seen.
The constraints for the case with $\delta_{_{\rm MNS}}^{true}=90^\circ$
(top-right) and 270$^\circ$ (bottom-right) are now look more symmetric.
In case of $\delta_{_{\rm MNS}}^{true}=90^\circ$, the same
$N(e)$ can be reproduced for the other angles by choosing
appropriately small $\sin^22\theta_{_{\rm CHOOZ}}^{}$,
whereas for $\delta_{_{\rm MNS}}^{true}=270^\circ$,
the same $N(e)$ can be obtained for larger
$\sin^22\theta_{_{\rm CHOOZ}}^{}$.
The constraints for $\delta_{_{\rm MNS}}=0^\circ$ and $180^\circ$
cases also look more or less symmetric.
Locally, a small shift in $\delta_{_{\rm MNS}}^{}$ can be
compensated by a small shift in $\sin^22\theta_{_{\rm CHOOZ}}^{}$
to make $N(e)$ similar at all experiments.
This compensation does not work perfectly
because the ratio
$N(e,4$GeV$)/N(e,6$GeV$)$ hits the minimum
($\simeq 0.41$) at around $\delta_{_{\rm MNS}}^{}=5^\circ$,
and the maximum ($\simeq 0.63$) at around
$\delta_{_{\rm MNS}}^{}=165^\circ$.
These trends can be read off from \Fgref{Cir_B}.
Unfortunately,
the region allowed by $\chi^2<4$ extends over
to all $\delta_{_{\rm MNS}}$ in both cases.
From \Fgsref{H2B_wo_SK}-\Fgvref{H2B_w_SK2},
we can tell that $\sin^22\theta_{_{\rm CHOOZ}}^{}$
can be constrained by these experiments,
but the constraint can be improved significantly
if $\delta_{_{\rm MNS}}^{}$ can be
constrained  independently by other means.
If indeed the LMA scenarios is chosen by the nature,
it is important that we present the constraints on
$\sin^22\theta_{_{\rm CHOOZ}}^{}$ and $\delta_{_{\rm MNS}}^{}$
simultaneously.

\subsubsection{$\delta m^2_{_{\rm ATM}}$ and
 $\sin^22\theta_{_{\rm ATM}}^{}$}

Finally, we study the capability of the VLBL experiment in measuring
the atmospheric neutrino oscillation parameters,
$\delta m^2_{_{\rm ATM}}$ and $\sin^22\theta_{_{\rm ATM}}^{}$,
accurately.
\Tbref{atm_lma} shows the mean and one-sigma errors of
$\delta m^2_{_{\rm ATM}}$ and the one-sigma lower bound on
$\sin^22\theta_{_{\rm ATM}}^{true}$
when the data are calculated for 
$\delta m^2_{_{\rm ATM}}=\numt{3.5}{-3}$eV$^2$
and $\sin^22\theta_{_{\rm ATM}}^{}=1.0$
in the LMA scenario with the hierarchy I.
The input parameters are chosen for the LMA point of
\eqref{IvsIII_true}, but for
$\sin^22\theta_{_{\rm CHOOZ}}^{true} = 0.06$
and the four $\delta_{_{\rm MNS}}$ analysis. 
 The $\chi^2$ fit has been performed by assuming the hierarchy I
but allowing all the model parameters to vary freely
within the LMA constraint \eqref{MSW_L}.
 The fitting conditions are the same as in \eqref{IvsI_fit}.
The (A) rows gives the results with 500 kton$\cdot$year each for
NBB($E_{\rm peak}=6$ GeV) and
NBB($E_{\rm peak}=4$ GeV) at $L=2,100$ km.
The (B) rows gives the results when in addition,
100 kton$\cdot$year data from
NBB($\langle p_\pi \rangle=2$ GeV)
at $L=295$ km are included in the fit.  
The sensitivities to $\delta m^2_{_{\rm ATM}}$ and $\sin^22\theta_{_{\rm ATM}}$
are improved by using the data from the HIPA-to-SK experiments.
The expected sensitivity for $\delta m^2_{_{\rm ATM}}$ is
about $3 \sim 4.5\%$ and that for $\sin^22\theta_{_{\rm ATM}}$ is
about $2.4\%$.
After including the HIPA-to-SK data the sensitivities
improve to about $2 \sim 3 \%$ and $1\%$, respectively.

\begin{table}[t]
\begin{center}
\begin{tabular}{|c|c||r|r|r|r|}
\hline
\multicolumn{1}{|c}{} &  \multicolumn{1}{c||}{}
  & $\delta_{_{\rm MNS}}^{true}=0^\circ$ 
& $90^\circ$
& $180^\circ$
& $270^\circ$  \\
  \hline
  \hline
$\chi^2_{min}$ 
&  (A) 
& 1.0
& 0
& 0.7
& 0
\\
  \cline{2-6}
&  (B)
& 2.2
& 0
& 1.6
& 0
\\
 \hline
 \hline
$\sin^22\theta_{_{\rm CHOOZ}}$ 
&  (A)
& $0.050^{+0.013}_{-0.005}$
& $0.048^{+0.011}_{-0.008}$
& $0.074^{+0.014}_{-0.012}$
& $0.079^{+0.014}_{-0.011}$
\\
  \cline{2-6}
&   (B) 
& $0.059^{+0.006}_{-0.009}$
& $0.046^{+0.008}_{-0.005}$
& $0.068^{+0.010}_{-0.008}$
& $0.080^{+0.010}_{-0.008}$
\\
 \hline
 \hline
$\delta m^2_{_{\rm ATM}} $
& (A)
& $3.46^{+0.055}_{-0.07}$
& $3.42^{+0.06}_{-0.09}$
& $3.38^{+0.07}_{-0.085}$
& $3.43^{+0.06}_{-0.09}$
\\
  \cline{2-6}
$(\times 10^{-3})$ (eV$^2$)
&  (B) 
& $3.75^{+0.05}_{-0.05}$
& $3.42^{+0.05}_{-0.04}$
& $3.39^{+0.05}_{-0.045}$
& $3.43^{+0.055}_{-0.045}$
\\    
  \hline
  \hline
$ {\rm sin}^22\theta_{_{\rm ATM}} $
& (A)  
& $>0.981$
& $>0.975$
& $>0.974$
& $>0.979$ 
\\
  \cline{2-6}
& (B)
& $>0.992$
& $>0.992$ 
& $>0.993$ 
& $>0.992$ 
\\
  \hline
 \end{tabular}
\end{center}
\caption{
Constraints on the three-neutrino model parameters,
$\sin^22\theta_{_{\rm CHOOZ}}$, 
$\delta m^2_{_{\rm ATM}}$ and $\sin^22\theta_{_{\rm ATM}}$,
when SMA or VO scenario 
is assumed in the fit.
The input data are generated for
the same 4 points as those in \Tbref{atm_lma} 
(\eqref{IvsIII_true} at $\sin^22\theta_{_{\rm CHOOZ}}^{true}=0.06$ ).
(A) Results with 500 kton$\cdot$year each for NBB($E_{\rm peak}=6$ GeV) and
NBB($E_{\rm peak}=4$ GeV) at $L=2,100$ km.
(B) In addition, 100 kton$\cdot$year data from
NBB($\langle p_\pi \rangle=2$ GeV)
at $L=295$ km are included in the fit.  
}
\tblab{atm_sma}
\end{table}

\Tbref{atm_sma} shows
the results of the fit to the same sets of data but
when the SMA scenario is assumed in the fit;
see \eqref{LMAvsSMA_fit}.
As we have seen in subsection 4.4.2 and in \Fgref{LMAvsSMA_B},
the SMA (or LOW or VO) gives a reasonably good fit to the data
that are calculated by assuming the LMA scenario.
The minimum $\chi^2$ can become zero when
$\delta_{_{\rm MNS}}^{true}=90^\circ$, $270^\circ$,
whereas $\chi^2_{min}$ is as low as 1 or 2 when
$\delta_{_{\rm MNS}}^{true}=0^\circ$ or $180^\circ$.
On the other hand the fitted model parameters are significantly
different from their input (`true') values.
The magnitudes and the significances of these differences
between the input parameters and their fitted values
can have read off from \Tbref{atm_sma}.
The shifts and errors of $\sin^22\theta_{_{\rm CHOOZ}}^{}$
can be read off from \Fgref{LMAvsSMA_B}.
\Tbref{atm_sma} shows that the fitted $\delta m^2_{_{\rm ATM}}$
can be more than $3\%$ smaller than the input values.
It is therefore important that the SMA/LOW/VO scenarios are excluded
with good confidence level, \eg by KamLAND experiment \cite{KAMLAND},
in order to constrain the parameters of the three-neutrino model
by the LBL experiments.

\subsection{Results for $L=1,200$ km}
In order to examine the sensitivity of the physics
outputs of the VLBL experiments to the base-line length,
we report the whole analysis for $L=1,200$ km,
which is approximately the distance between HIPA
and Seoul.
\begin{figure}[htbp]
 \begin{center}
 \scalebox{.85}{\includegraphics{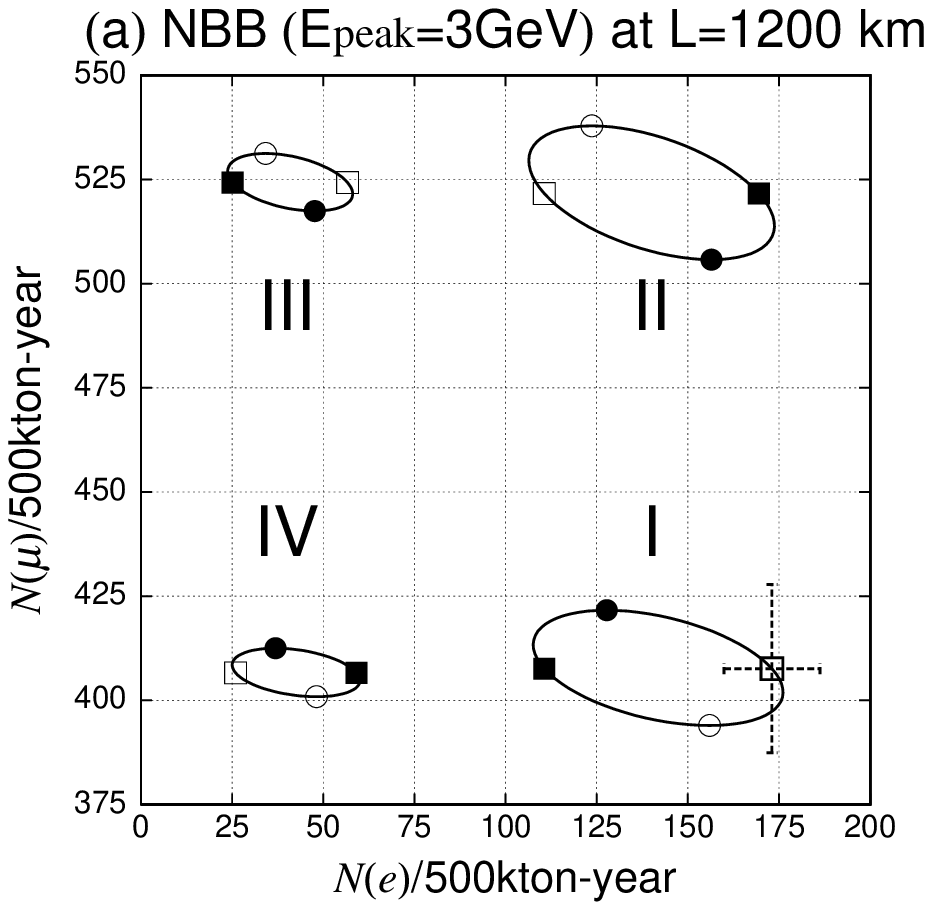}}
 \scalebox{.85}{\includegraphics{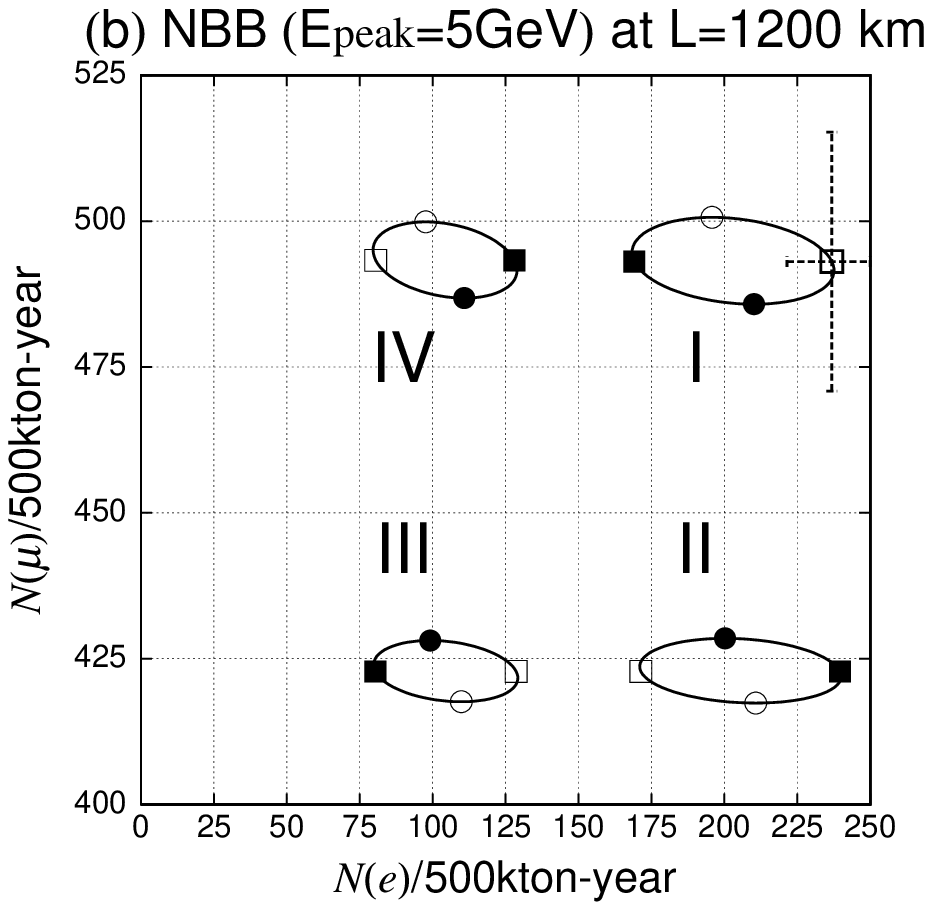}}
 \end{center} 
\caption{%
CP phase dependence of $N(e)$ and $N(\mu)$
at $L=1,200$ km  for 100kton$\cdot$year with
(a) NBB($E_{\rm peak}^{}=3$ GeV) 
and (b)NBB($E_{\rm peak}^{}=5$ GeV).
$\delta_{_{\rm MNS}}^{} = 0^\circ$ (solid-circle),
$90^\circ$ (solid-square),
$180^\circ$ (open-circle),
and $270^\circ$ (open-square).
The input parameters are
$\delta m^2_{_{\rm ATM}} = \numt{3.5}{-3} \mbox{eV$^2$}$,
$\sin^22\theta_{_{\rm ATM}} = 1.0$,
$\delta m^2_{_{\rm SOL}} = \numt{10}{-5} \mbox{eV$^2$}$,
$\sin^22\theta_{_{\rm SOL}} = 0.8$,
$\sin^22\theta_{_{\rm CHOOZ}} = 0.1$,
and $\rho$ = 3 g/cm$^3$.
The predictions for the four types of the neutrino mass hierarchies
(\Fgref{cases})
are depicted as I, II, III and IV.
}
\Fglab{Cir_S}
\end{figure}

We show in \Fgref{Cir_S} the expected signal event numbers, 
$N(e)$ and $N(\mu)$, at the base-line length of $L=1,200$ km from HIPA with
500kton$\cdot$year
for (a) the NBB with 
$E_{\rm peak}=3$ GeV and for (b) the NBB with $E_{\rm peak}=5$ GeV.
The predictions are calculated for exactly the same
three-neutrino model parameters and the matter density as in 
\Fgref{Cir_SK}
and 
\Fgref{Cir_B} ;
see \eqref{modelpara_B}.
The predictions for the neutrino mass hierarchy cases
I to IV, see \Fgref{cases} and \Tbref{cases},
are shown by separate circles when the $\delta_{_{\rm MNS}}$
is allowed to vary freely.
The four representative phase values are shown by
solid-circle ($\delta_{_{\rm MNS}}=0^\circ$), solid-square ($90^\circ$),
open-circle ($180^\circ$), and open-square ($270^\circ$).
The statistical errors of the $N(e)$ and
$N(\mu)$ measurements at 500 kton$\cdot$year are shown 
on the 
$\delta_{_{\rm MNS}}=270^\circ$ point for the hierarchy case I.

Because we have learned from the analysis at $L=2,100$ km
the number of $\nu_e^{}$ CC events, $N(e)$, is most sensitive
to the neutrino model parameters,
we choose the peak energies ($E_{\rm peak}^{}$) of the NBB
by requiring large $N(e)$ and suppressed $N(\mu)$.
A pair of NBB's is then chosen such that the
$\delta_{_{\rm MNS}}^{}$-dependence of the ratio of $N(e)$'s
is significant.
The chosen peak energies,
$E_\nu=3$ GeV and 5 GeV at $L=1,200$ km have the same $L/E_\nu$ with
$E_\nu \simeq 5$ GeV and 9 GeV at $L=2,100$ km, respectively. 
As shown in the middle figure in the left-hand-side in
\Fgref{pro_B}, 
the magnitude of
the transition probabilities $P_{\nu_\mu \to \nu_\mu}$ for
the mass hierarchy I 
is smaller than that for the hierarchy III in the interval
3 GeV $\lsim E_\nu \lsim$ 6 GeV, 
whereas this ordering is reversed for 6 GeV$\lsim E_\nu$.
Accordingly, in \Fgref{Cir_S}(b) for 
NBB($E_{\rm peak}=5$ GeV) at $L=1,200$ km,
the predicted $N(\mu)$ in the hierarchy I is larger than that 
in the hierarchy III.
It turns out that this reversing at the ordering of
$N(\mu)$ is not very effective in distinguishing the
neutrino hierarchy cases
because the trend can easily be accounted for by shifting the
atmospheric neutrino oscillation parameters,
$\delta m_{_{\rm ATM}}^2$ and $\sin^22\theta_{_{\rm ATM}}^{}$,
slightly.
The fact that the two chosen NBB's cover the opposite sides of
the $P_{\nu_\mu \to \nu_\mu}=0$ node leads to a significant
improvement in the measurement of $\delta m^2_{_{\rm ATM}}$
and $\sin^22\theta_{_{\rm ATM}}^{}$, see \Tbref{atm_lma_S}.

When we compare \Fgref{Cir_S} for $L=1,200$ km with
\Fgref{Cir_B} for $L=2,100$ km,
we notice that the reduction of $N(e)$ for the
hierarchy III at $L=1,200$ km is not as drastic as
the reduction at $L=2,100$ km,
but that the expected statistical errors of the signal is
smaller at $L=1,200$ km because of a factor of
two larger $N(e)$ for the same size of the detector. 
\begin{figure}[ht]
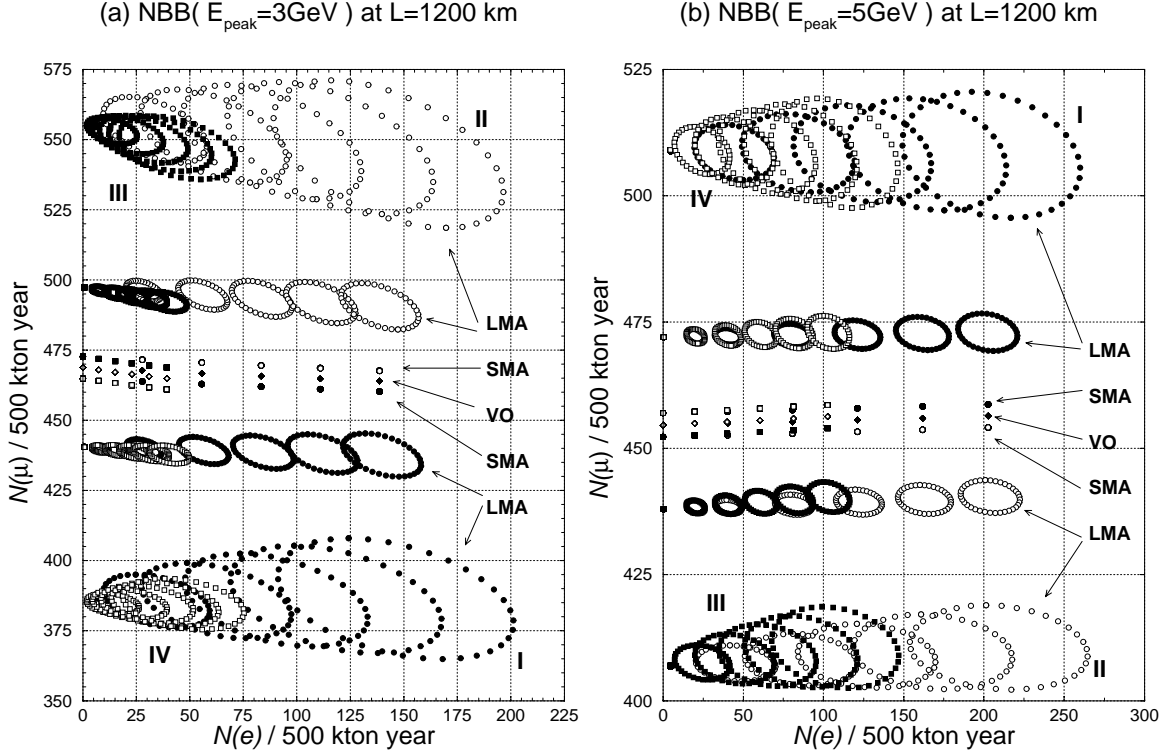

 \begin{center}
{\scalebox{0.45}{\includegraphics{./FigS/main_3gev.eps}}} 
{\scalebox{0.45}{\includegraphics{./FigS/main_5gev.eps}}} 
\end{center}
\caption{
The neutrino parameter dependences of the expected
numbers
of $\nu_e^{}$ CC events and $\nu_\mu^{}$ CC events, 
$N(e)$ and $N(\mu)$, respectively,
for the NBB with $E_{\rm peak}^{}=3$ GeV (a) and 5 GeV (b)
500 kton$\cdot$year at $L=1,200$ km.
All the symbols are the same as those in \Fgref{main_B}.
}
\Fglab{main_S}
\end{figure}

In \Fgref{main_S}, we show the expected numbers of signal events,
$N(\mu)$ and $N(e)$, for the same set of NBB's,
(a) NBB ($E_{\rm peak}=3$GeV) and
(b) NBB ($E_{\rm peak}=5$GeV), each
with 500 kton$\cdot$year at $L=1,200$ km.
The three-neutrino model parameters and the matter density
used for calculating those numbers are the same as those used to
generate \Fgref{main_B} for $L=2,100$ km;
see  \eqref{modelpara_main} and \eqref{modelpara_sol}.
All the symbols are the same as those adopted in \Fgref{main_B}.
The dependences of $N(e)$ on the input 
$\sin^22\theta_{_{\rm CHOOZ}}$
are the same as those found in \Fgref{main_B};
$N(e)$ decreases as input $\sin^22\theta_{_{\rm CHOOZ}}$
is decreased from 0.1 to 0.08, 0.06, 0.04, 0.02 and 0.
The solid-circle, open-circle, solid-square and open-square 
points show the predictions of the LMA and SMA scenarios
with neutrino mass hierarchy I, II, III and IV, respectively.
For each hierarchy,
the five larger grand circles give the predictions of the
LMA scenario with
$\delta m^2_{_{\rm SOL }}= 15\times 10^{-5}$ eV$^2$, and 
the smaller circles are
for the LMA with $\delta m^2_{_{\rm SOL }}= 5\times 10^{-5}$
eV$^2$.
The VO predictions of the neutrino mass
hierarchies I and II (III and IV) cannot be distinguished, 
and they are given by the solid (open)-diamonds. 
The difference in $N(\mu)$ is largest in the LMA scenarios with
$\delta m^2_{_{\rm SOL }}= 15\times 10^{-5}$ eV$^2$, 
for which the hierarchy III predicts about 40 $\%$
larger (20 $\%$ smaller) $N(\mu)$ than the predictions of the
hierarchy I for the NBB with $E_{\rm peak}=3$ GeV (5 GeV).
When we compare \Fgref{main_S} for $L=1,200$ km with the 
corresponding \Fgref{main_B} for $L=2,100$ km,
we notice that the prediction for $N(e)$ in the hierarchy III are
significantly larger for $L=1,200$ km than those for $L=2,100$ km.
If it were only $N(e)$ in \Fgref{main_S} that effectively
discriminates the neutrino mass hierarchy,
we should expect significant reduction of the hierarchy
discriminating capability of the VLBL experiments
at $L=1,200$ km.

\begin{figure}[t]
\begin{center}
{\scalebox{0.48}{\includegraphics{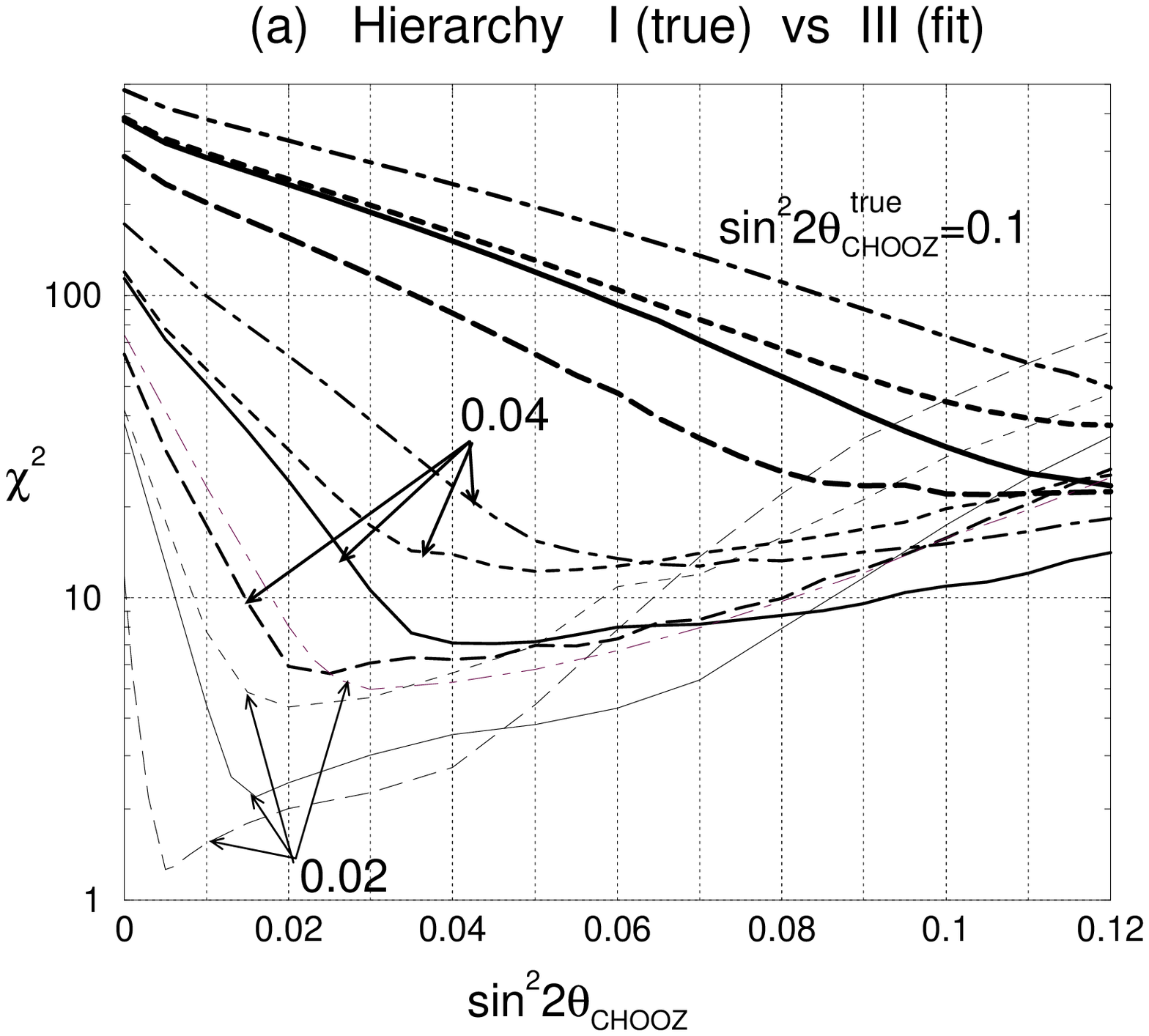}}} 
{\scalebox{0.48}{\includegraphics{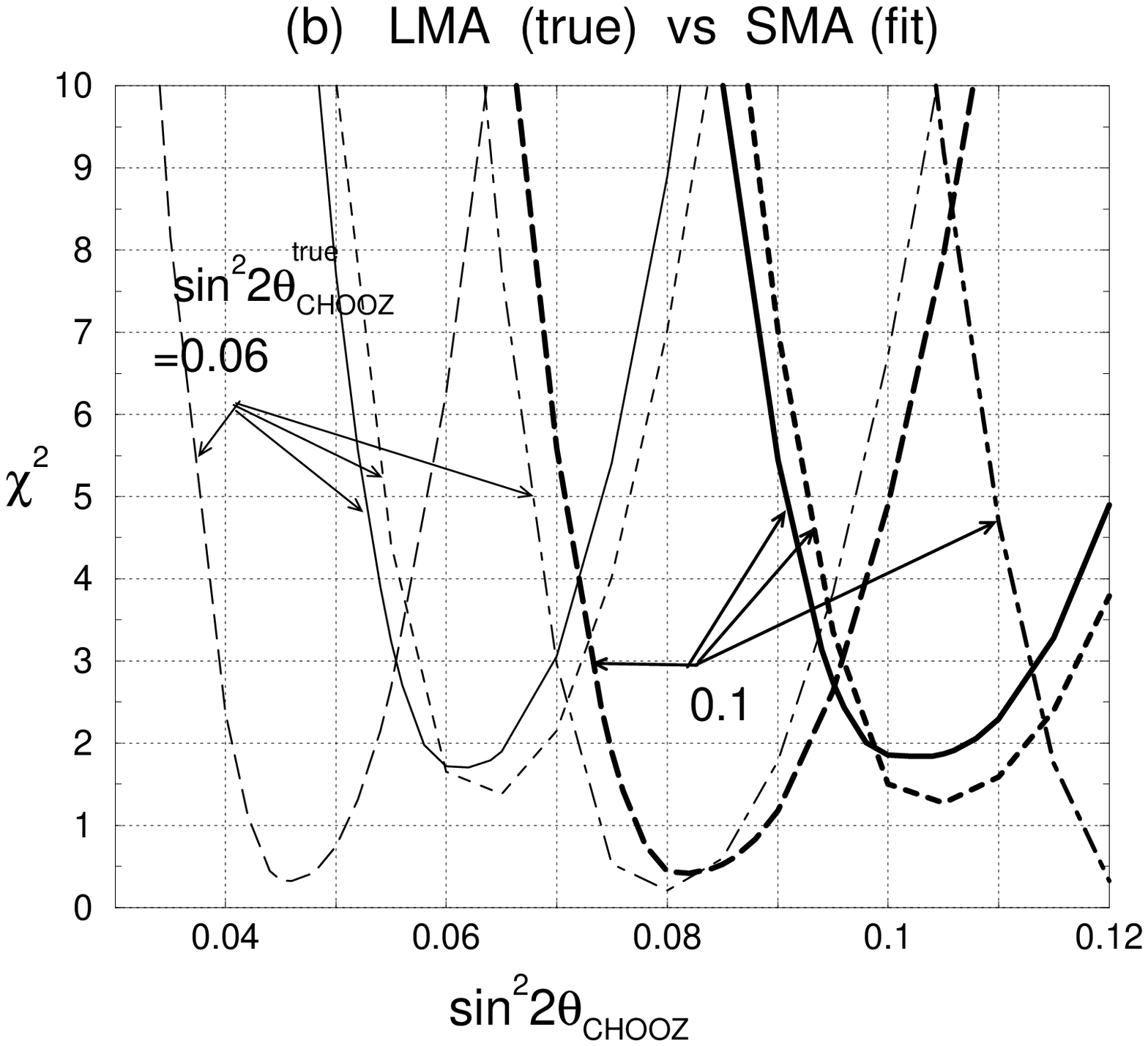}}} 
\end{center}	     
\caption{
The same as \Fgref{IvsIII_B}(b) and \Fgref{LMAvsSMA_B}(b),
but for $L=1,200$ km.
}
\Fglab{H2S_chi}
\end{figure}

In order to study these questions quantitatively,
we repeat the $\chi^2$ fit for the following sets of
experimental conditions ;
\bseq
\bea
&&\hspace*{-10ex}
\begin{tabular}{ll}
(A')&
500 kton$\cdot$year each for NBB($E_{\rm peak}=5$ GeV) and
NBB($E_{\rm peak}=3$ GeV) \\
 & at $L=1,200$ km
\end{tabular}
\eqlab{ex_condition_s_A}\\
&&\hspace*{-10ex}
\begin{tabular}{ll}
(B')
& In addition to (A'), 100 kton$\cdot$year data from 
NBB($\langle p_\pi \rangle=2$ GeV) \\
& at $L=295$ km are included in the fit. 
\end{tabular}
\eqlab{ex_condition_s_B}
\eea
\eqlab{ex_condition_s}
\eseq

\noindent
In \Fgref{H2S_chi} we show the minimum $\chi^2$ as functions of the
fit parameter $\sin^22\theta_{_{\rm CHOOZ}}$ 
for the data set (B').
The mean values of $N(\mu, E_{\rm peak})$ and $N(e, E_{\rm peak})$
at $L=1,200$ km and those at $L=295$ km are calculated
for the experimental condition of \eqref{ex_condition_s_B} 
by assuming the LMA scenario and the hierarchy I,
and by choosing 12 sets of the model parameters as in
\eqref{IvsIII_true}; $\sin^22\theta_{_{\rm CHOOZ}}^{true}=
0.02$, 0.04, and 0.10 and $\delta_{_{\rm MNS}}^{true}=
0^\circ$, 90$^\circ$, 180$^\circ$, and 270$^\circ$.
The $\chi^2$ fit has then been performed by assuming
the LMA scenario with the hierarchy III;
see \eqref{IvsIII_fit} for details.
The 12 lines in \Fgref{H2S_chi}(a) correspond to
$\sin^22\theta_{_{\rm CHOOZ}}^{true} = 0.02$ (thin lines), 
0.04 (medium-thick lines), 0.1 (thick lines) 
and $\delta_{_{\rm MNS}}^{true}
= 0^\circ$ (solid lines), $90^\circ$ (long-dashed lines), 
$180^\circ$ (short-dashed lines), $270^\circ$ (dot-dashed lines). 
We find that
the neutrino mass hierarchy (between I and III) 
can be determined at $3\sigma$ level, 
for all four values of $\delta_{_{\rm MNS}}^{true}$ and 
when $\sin^22\theta_{_{\rm CHOOZ}} =0.1$, and  
for $\delta_{_{\rm MNS}}^{true}=180^\circ$ and 270$^\circ$ 
when $\sin^22\theta_{_{\rm CHOOZ}}^{true} = 0.04$.
$\chi^2_{min}$ is greater than about 7 for
$\delta_{_{\rm MNS}}^{true}=0^\circ$ and 90$^\circ$
when 
$\sin^22\theta_{_{\rm CHOOZ}}^{true}=0.04$.
Even when 
$\sin^22\theta_{_{\rm CHOOZ}}^{true}=0.02$,
$\chi^2_{min}>4$ for
$\delta_{_{\rm MNS}}^{true}=180^\circ$ and 270$^\circ$.
These results at $L=1,200$ km are not much worse than
the results at $L=2,100$ km.
We find that this is because the combination of the VLBL experiment
and the $L=295$ km LBL experiment is still effective
in distinguishing the hierarchy cases,
even though the VLBL experiment at $L=1,200$ km
itself cannot distinguish the hierarchy cases
when $\sin^22\theta_{_{\rm CHOOZ}}^{true}\lsim 0.04$.

In \Fgref{H2S_chi} (b), 
the mean values of $N(\mu, E_{\rm peak})$ and $N(e, E_{\rm peak})$ 
are calculated for the 8 LMA points with hierarchy I;
$\sin^22\theta_{_{\rm CHOOZ}}^{}=$0.06 and 0.10
for the four $\delta_{_{\rm MNS}}$ points in \eqref{IvsIII_true}.
The $\chi^2$ fit has then been performed by assuming
the SMA scenario within the hierarchy I; see \eqref{LMAvsSMA_fit}.
The results of the fit are given for
$\sin^22\theta_{_{\rm CHOOZ}}^{true}$ = 0.06 (thin lines), 0.1 (thick lines) 
and $\delta_{_{\rm MNS}}^{true}$
= $0^\circ$ (solid lines), $90^\circ$ (long-dashed lines), 
$180^\circ$ (short-dashed lines), $270^\circ$ (dot-dashed lines). 
The qualitative features of the results shown in \Fgref{H2S_chi}(b)
are very similar to those of \Fgref{LMAvsSMA_B}(b) at $L=2,100$ km.
The VLBL experiment with HIPA do not seem to have the capability to
distinguish the scenarios of the solar-neutrino oscillation models.
This is mainly because the magnitude of oscillation phase
$\Delta_{12}^{}$ is smaller than 2.1 even for 
$\delta m^2_{_{\rm SOL}}=\numt{10}{-5}$eV$^2$ and
$L/E_{\nu}^{}=2,100$km/4GeV.
We should rely on the results of the solar-neutrino oscillation
measurements and the forthcoming reactor neutrino oscillation experiments
\cite{KAMLAND}.

\begin{figure}[tb]
\begin{center}
{\scalebox{0.7}{\includegraphics{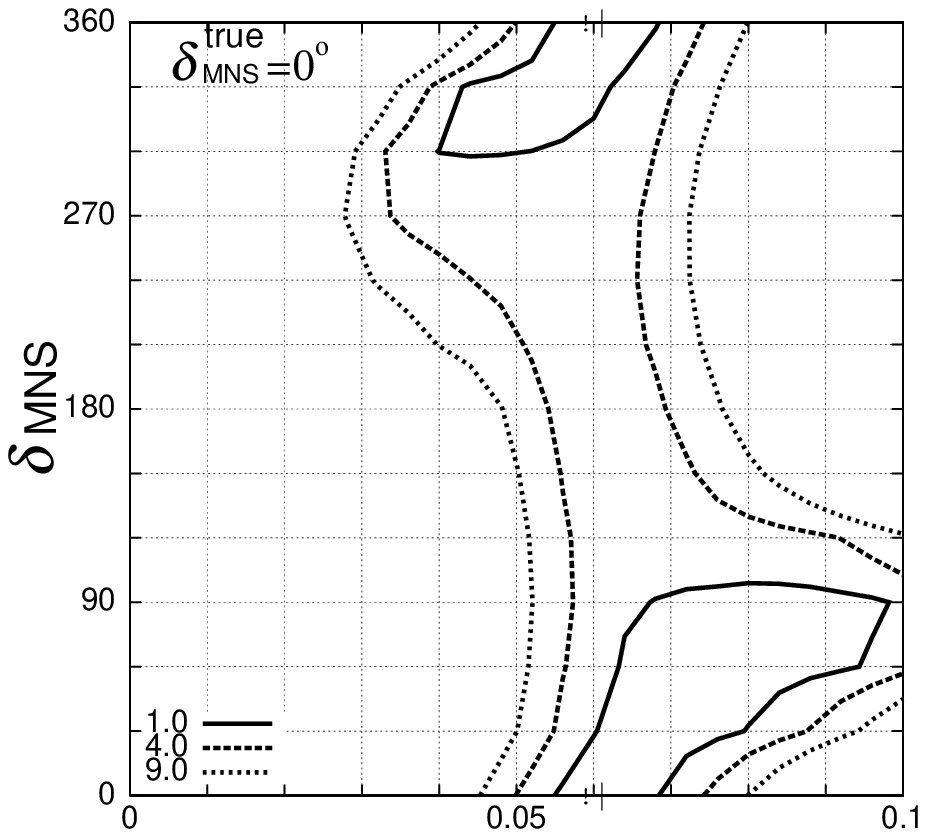}}} 
{\scalebox{0.7}{\includegraphics{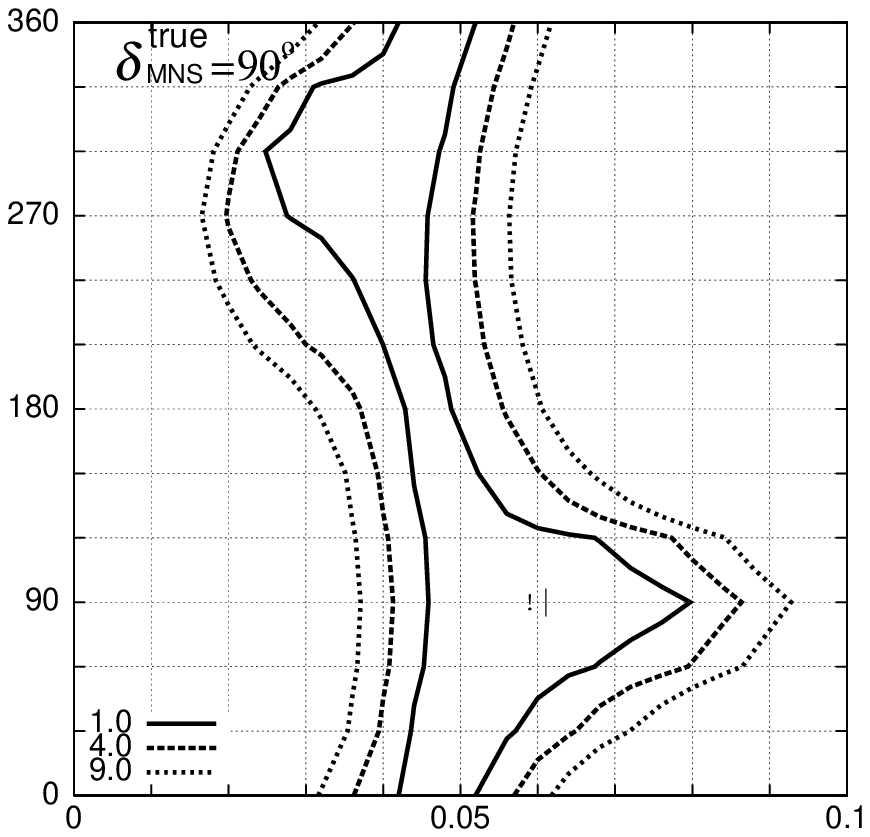}}} 
{\scalebox{0.7}{\includegraphics{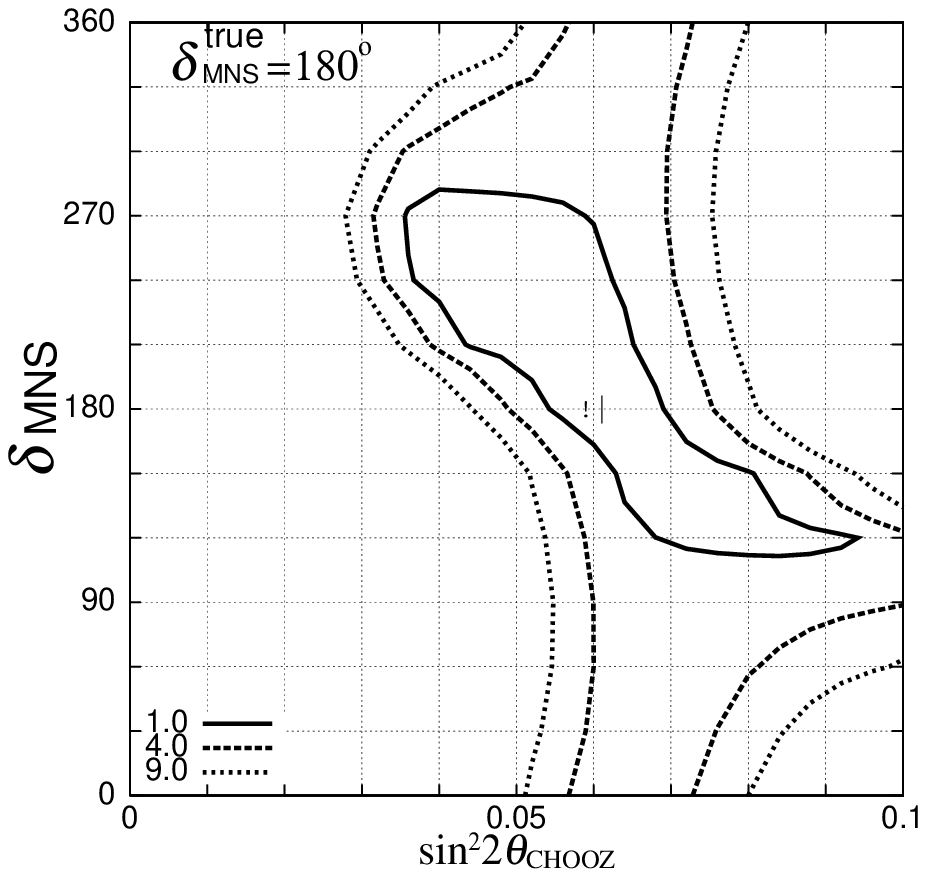}}} 
{\scalebox{0.7}{\includegraphics{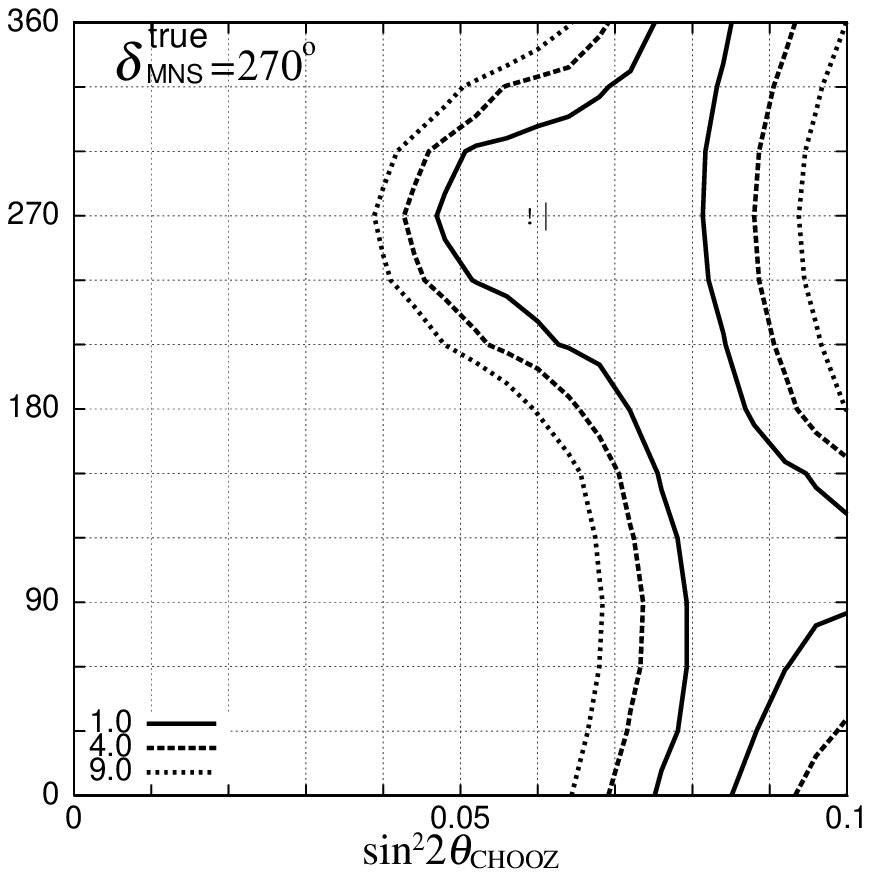}}} 
\end{center}	     
\caption{
The same as \Fgref{H2B_w_SK2}, but for $L=1,200$ km.
}
\Fglab{H2S_w_SK}
\end{figure}
 \Fgref{H2S_w_SK} shows the regions in the
$\sin^2 2 \theta_{_{\rm CHOOZ}}$ v.s. $\delta_{_{\rm MNS}}$
plane which are allowed by the VLBL experiments at $L=2,100$ km
with 500 kton$\cdot$year each for NBB ($E_{\rm peak}=5$GeV)
and NBB ($E_{\rm peak}=3$GeV), and
the LBL experiment at $L=295$ km with 100 kton$\cdot$year
for NBB ($\vev{p_\pi}=2$GeV); \eqref{ex_condition_s_B}.
All the model parameters used to calculate the expected numbers of
events and the symbols of the figures are the same as those adopted for
\Fgref{H2B_w_SK2}. The four input points at
$\delta_{_{\rm MNS}}^{true} = 0^{\circ},
90^{\circ}, 180^{\circ}, 270^{\circ}$
and $\sin^2 2\theta_{_{\rm CHOOZ}}^{true}=0.06$
are shown by the solid-circle in each plane,
and the region with $\chi^2_{min}<1$, 4, and 9
are shown by the solid, dashed, and dotted
boundaries, respectably.
 The $\chi^2$ fit has been performed by assuming the LMA scenario
with the hierarchy I,
but otherwise by allowing all the model parameters
to vary freely; see \eqref{IvsIII_fit}.

The results shown in \Fgref{H2S_w_SK} look very similar to those
of \Fgref{H2B_w_SK2} for the combination of the VLBL experiment
at $L=2,100$ km and the LBL experiment at $L=295$ km;
\eqref{ex_condition_B}.
 For $\delta_{_{\rm MNS}}^{true}=90^{\circ}$ and 270$^{\circ}$,
we cannot constrain the CP phase at all from these experiments. 
 For $\delta_{_{\rm MNS}}^{true}=0^{\circ}$ and 180$^{\circ}$,
there appear a region of $\chi^2_{min}<1$ where $\delta_{_{\rm MNS}}$
is constrained, but the preferred region covers the whole range
at $\chi^2_{min}<4$ level.
 We find that the area of the $\chi^2_{min}<1$ region is almost the 
same as that of the $L=2,100$ km experiment in \Fgref{H2B_w_SK2}
for each $\delta_{_{\rm MNS}}^{true}$ case.
We conclude that the capability of measuring
$\sin^22\theta_{_{\rm CHOOZ}}$ and $\delta_{_{\rm MNS}}^{}$
in the LMA scenario is very similar between the VLBL experiments
at $L=2,100$ km and those at $L=1,200$ km.

\begin{table}[t]
\begin{center}
\begin{tabular}{|c||c||r|r|r|r|}
\hline
\multicolumn{1}{|c}{} &  \multicolumn{1}{c||}{}
 & $\delta_{_{\rm MNS}}=0^\circ$ & $90^\circ$ & $180^\circ$ & $270^\circ$  \\
  \hline
  \hline
$\delta m^2_{_{\rm ATM}}$
& (A')
& $3.50^{+0.065}_{-0.075}$
& $3.50^{+0.075}_{-0.08}$
& $3.50^{+0.07}_{-0.08}$
& $3.50^{+0.085}_{-0.07}$
\\    
  \cline{2-6}
$(\times 10^{-3})$ (eV$^2$)
& (B')
& $3.50^{+0.065}_{-0.05}$
& $3.50^{+0.065}_{-0.075}$
& $3.50^{+0.07}_{-0.075}$
& $3.50^{+0.07}_{-0.07}$
\\
  \hline
  \hline
$ {\rm sin}^22\theta_{_{\rm ATM}} $
& (A') 
& $>0.9940$
& $>0.9940$
& $>0.9935$
& $>0.9935$ \\
  \cline{2-6}
& (B')
& $>0.9945$
& $>0.9945$
& $>0.9945$
& $>0.9945$\\
  \hline
 \end{tabular}
\end{center}
\caption{
The mean and the $1\sigma$ errors of
$\delta m^{2}_{_{\rm ATM}}$ and 
the one-sigma level bound of 
$\sin^22\theta_{_{\rm ATM}}^{true}$
when the input data are calculated for the four
LMA points (\eqref{IvsIII_true} at
$\sin^22\theta_{_{\rm CHOOZ}}^{true}$ = 0.06) 
and the fits are performed by assuming
the LMA scenarios.
(A') Results with 500 kton$\cdot$year each for NBB($E_{\rm peak}=5$ GeV) and
NBB($E_{\rm peak}=3$ GeV) at $L=1,200$ km.
(B') In addition, 100 kton$\cdot$year data from
NBB($\langle p_\pi \rangle = 2$ GeV)
at $L=295$ km are included in the fit.  
}
\tblab{atm_lma_S}
\end{table}

Finally, we show in \Tbref{atm_lma_S}
the expected accuracy of the measurements of
$\delta m^{2}_{_{\rm ATM}}$ and $\sin^22\theta_{_{\rm ATM}}^{}$.
The expected number of events are calculated for 
$\delta m^{2~true}_{_{\rm AMT}}=\numt{3.5}{-3}$eV$^2$
and
$\sin^22\theta_{_{\rm ATM}}^{true}=1.0$ in the 
LMA scenario with the hierarchy I.
The 4 LMA input points and the fit parameters are the same 
as those used for \Tbref{atm_lma}.
The mean and the one-sigma errors for fitted $\delta m^2_{_{\rm ATM}}$
and the one-sigma lower bound for $\sin^22\theta_{_{\rm ATM}}^{}$
are shown in the \Tbref{atm_lma_S}.
The results of the VLBL experiment only, \eqref{ex_condition_s_A},
are shown in the (A') row, whereas those obtained by adding the
LBL experiment at $L=295$ km, \eqref{ex_condition_s_B},
are shown in the (B') rows.
By comparing the results of \Tbref{atm_lma} and \Tbref{atm_lma_S},
we find that more precise measurements of
$\delta m^2_{_{\rm ATM}}$ and $\sin^22\theta_{_{\rm ATM}}$
can be achieved by the VLBL experiments at $L=1,200$ km.
As mentioned above, this is essentially because the two NBB's
that we choose for studying the potential of the $L=1,200$ km
experiments cover the two sides of the $P_{\nu_\mu\to\nu_\mu}=0$ node,
whereas one of the two NBB's chosen for studying the $L=2,100$ km case
sits on top of the node to minimize $N(\mu)$.
We find that the accuracy of $\delta m^2_{_{\rm ATM}}$
and $\sin^22\theta_{_{\rm CHOOZ}}$ measurements at the $L=2,100$ km
experiments can be improved, if a NBB with higher peak energy
is added.
Because higher $E_{\rm peak}$ NBB gives higher $\tau$-backgrounds
and larger $N(\mu)$, the optimal choice of the beam-type and
beam energy should be determined according to the
prime physics objectives and the details of the proposed detectors.

\section{Summary and discussions}
\clean
The KEK-JAERI joint project on HIPA (High Intensity Proton Accelerator)
\cite{HIPA} will start delivering $10^{21}$ POT (protons on target) at 
50 GeV per one operation year by the year 2007.
This compares with $ 3.85 \times 10^{19}$ POT \cite{K2K_POT} at 12 GeV
which are being provided by the KEK PS for the K2K neutrino oscillation
experiment.
The HIPA project thus provides us with an excellent opportunities of 
extending the K2K experiment, by shooting the neutrino beam from
the HIPA to SK (the Super-Kamiokande) with $L= 295$ km \cite{H2SK}.
Refinement of the measurements of the atmospheric neutrino oscillation 
parameters ($\delta m^2_{_{\rm ATM}}$ and $\sin^22\theta_{_{\rm ATM}}$)
and the first measurement of the $\nu_\mu \to \nu_e$ oscillation parameter
($\sin^22\theta_{_{\rm CHOOZ}}$) by observing the electron appearance are 
the main targets of the proposed experiment \cite{H2SK}.

In this report, we explore the possibility of extending the LBL experiments 
to a much longer base-line length ($L >$ 1000 km) as a second stage of 
the neutrino oscillation experiments using the HIPA beam.
In particular, we examine in detail the physics potential of VLBL experiments
at $L=2,100$ km, which is approximately the distance between the HIPA and
Beijing, where strong interests in constructing a huge neutrino detector
(BAND) have been expressed \cite{BAND}.
We also examine the case at $L$ = 1,200 km carefully in order to study 
the sensitivity of physics outputs on the base-line lengths of the same order.
The distance 1,200 km is approximately that between the HIPA and 
the central Korea, where strong interests in LBL neutrino oscillation
experiments are expressed.

We study the physics potential of such VLBL experiments within 
the three-neutrino model.
The three-neutrino model gives a consistent picture of all the neutrino
oscillation observations except the LSND experiment,
and has six parameters those are observables by neutrino-oscillation
phenomena:
two mass-squared differences, 
$\delta m^2_{12}=\pm\delta m_{_{\rm SOL}}^2$ and 
$\delta m^2_{13}=\pm\delta m_{_{\rm ATM}}^2$,
three angles, $\sin^22\theta_{_{\rm SOL}}^{}$,
$\sin^22\theta_{_{\rm ATM}}^{}$, and
$\sin^22\theta_{_{\rm CHOOZ}}^{}$,
and one phase, $\delta_{_{\rm MNS}}^{}$.
The main objectives of such VLBL experiments may be
summarized by the following four questions : 
\begin{enumerate}
\renewcommand{\labelenumi}{\arabic{enumi}. } 
\item  Can we distinguish the neutrino mass hierarchy cases ? 
\item Can we distinguish the solar neutrino oscillation scenarios
($\delta m^2_{_{\rm SOL}}$ and $\sin^22\theta_{_{\rm SOL}}$) ? 
\item Can we measure the two unknown parameters of the model,
$\sin^22\theta_{_{\rm CHOOZ}}$ and $\delta_{_{\rm MNS}}$ ? 
\item How much can we improve the measurements of $\delta m^2_{_{\rm ATM}}$
and $\sin^22\theta_{_{\rm ATM}}$ ? 
\end{enumerate}
In order to quantify our answers to the above questions, we make the following
simplifications for the neutrino beams and the detector capability.
\begin{enumerate} 
\renewcommand{\labelenumi}{(\roman{enumi})} 
\item  High-energy NBB's with $E_{\rm peak}= 3 \sim 6$ GeV for the VLBL 
experiments. 
\item  Low-energy NBB with $\langle p_\pi \rangle $ = 2 GeV to
represent the LBL experiment to SK. 
\item
 100 kton-level detector which is capable of detecting $\nu_\mu$ CC
and $\nu_e$ CC events almost perfectly, but is not necessarily capable 
of measuring the particle charges and hadron energies. 
\item
 Backgrounds from secondary beams and from leptonic $\tau$ decays of the
$\nu_\tau$ CC events, as well as those from $\pi^0$-$e$ misidentification
the neutral current processes are accounted for. 
\item
Common flux normalization errors of 3 $\%$ is assigned for the
high-energy
NBB's, and an independent 3 $\%$ error for the low-energy NBB. 
\item
Overall uncertainty in the matter density is assumed to be 3.3 $\%$
for each experiment.
\end{enumerate}
The detector capabilities of (iii) and (iv) are taken from those of the SK
detector
which may be common for water-$\check {\rm C}$erenkov detectors.

We find the following observations.
By assuming that the HIPA-to-SK experiment measures the atmospheric-neutrino
oscillation parameters to be centered at $\delta m^2_{_{\rm ATM}}=3.5\times
10^{-3}$
eV$^2$ and $\sin^22\theta_{_{\rm ATM}}= 1$,  and by assuming that its
cumulative
effects can be represented by 
\bea
\begin{tabular}{l}
100 kton$\cdot$year at $L=295$ km with 
NBB($\langle p_\pi \rangle=2$ GeV),
\end{tabular}
\eea
we find that VLBL experiments of
\bea
\left.
\begin{tabular}{l}
500 kton$\cdot$year at $L= 2,100$ km with NBB($E_{\rm peak}=6$ GeV) \\
500 kton$\cdot$year at $L= 2,100$ km with NBB($E_{\rm peak}=4$ GeV) 
\end{tabular}
\right\}
 \eea
can give the following answers : 
\begin{enumerate} 
\renewcommand{\labelenumi}{\arabic{enumi}. }
\item
 If the neutrino mass hierarchy I is realized in Nature, then the
hierarchy III can be
rejected at $3\sigma$ ($1\sigma$)
level if $\sin^22\theta_{_{\rm CHOOZ}} >$ 0.04
(0.02).
See \Fgref{IvsIII_B} (b). 
\item
If the LMA scenario ($\delta m^2_{_{\rm SOL}}=10\times 10^{-4}$ eV$^2$,
$\sin^22\theta_{_{\rm SOL}}=0.8$) is realized in Nature, and if
$\sin^22\theta_{_{\rm CHOOZ}} >$0.06, the SMA/VO scenarios can be rejected
at one-sigma level when $\delta_{_{\rm MNS}}$ is around $0^\circ$ or
$180^\circ$
but not at all when $\delta_{_{\rm MNS}}$ is
around $90^\circ$ or $270^\circ$.
See \Fgref{LMAvsSMA_B} (b). 
\item
If the LMA scenario ($\delta m^2_{_{\rm SOL}}=10\times10^{-5}$ eV$^2$,
$\sin^22\theta_{_{\rm SOL}}=0.8$) is realized in Nature, and if
$\sin^22\theta_{_{\rm CHOOZ}}=0.06$, then
$0.04 < \sin^22\theta_{_{\rm CHOOZ}} < 0.1$ is obtained when
$\delta_{_{\rm MNS}}$ is around $0^\circ$ or $180^\circ$,
$0.02 < \sin^22\theta_{_{\rm CHOOZ}} < 0.08$  when
$\delta_{_{\rm MNS}}$ is around $90^\circ$,
$0.045 < \sin^22\theta_{_{\rm CHOOZ}} < 0.12$ when
$\delta_{_{\rm MNS}}$ is around $270^\circ$.
$\delta_{_{\rm MNS}}$ can be constrained to local values at one-sigma level
when the true
$\delta_{_{\rm MNS}}$ is around
$0^\circ$ or $180^\circ$ but it is unconstrained when
it is around $90^\circ$ or $270^\circ$. 
See \Fgref{H2B_w_SK2}.
\item
If the LMA scenario ($\delta m^2_{_{\rm SOL}}=10\times 10^{-4}$ eV$^2$,
$\sin^22\theta_{_{\rm SOL}}=0.8$) is realized in Nature,
$\delta m^2_{_{\rm ATM}}$ is measured with the $3\%$ accuracy and 
$\sin^22\theta_{_{\rm ATM}}$ to 1$\%$ level.
See \Tbref{atm_lma}.
\end{enumerate}

Summing up, a combination of LBL experiments at $L=295$ km and VLBL
experiments at $L=2,100$ km with the NBB's from HIPA can determine 
the neutrino mass hierarchy at $3\sigma$ level if 
$\sin^22\theta_{_{\rm CHOOZ}} >$0.04, cannot distinguish the LMA scenario
from the SMA/VO scenarios well, can constrain $\sin^22\theta_{_{\rm CHOOZ}}$,
and it can also constrain $\delta_{_{\rm MNS}}$ at one-sigma level in
some cases.
The errors of the atmospheric-neutrino oscillation parameters are reduced
to $\delta (\delta m^2_{_{\rm ATM}})=\pm 0.1 \times 10^{-3}$
eV$^2$ and $\sin^22\theta_{_{\rm ATM}} > 0.991$.
Very similar results are found for a combination of $L = 295$ km and
$L=1,200$ km experiments.
See \Fgref{H2S_chi}, \Fgref{H2S_w_SK}, and
\Tbref{atm_lma_S}.

\section*{Acknowledgments}
The authors wish to thank stimulating discussions with
K.~Nakamura,
J.~Sato,
Y.F.~Wang,
K.~Whisnant,
Z.Z.~Xing,
C.G.~Yang,
J.M.~Yang,
B.L.~Young,
and
X.M.~Zhang.
The works of MA and TK are supported in part by
the Grant-in-Aid for Scientific Research from the
Ministry of Education, Culture, Sports, Science and Technology
of Japan.
KH would like to thank the core-university
program of JSPS for support.
The work of NO is supported in part by a grand from 
the US Department of Energy, DE-FG05-92ER40709.
\def\PRD#1#2#3{Phys. Rev. {\bf D#1} #2 (#3)}

\end{document}